\documentclass[12pt]{article}
\usepackage{amsmath,
			amsthm,
			amsfonts,
			amsmath,
			amssymb,
			amsfonts,
			graphicx,
			color,
			euscript,
			epstopdf,
			marvosym,
			mathtools}
\usepackage{times}
\usepackage{graphicx}
\usepackage{color}
\usepackage{multirow}
\usepackage[authoryear]{natbib}
\usepackage{rotating}
\usepackage{bbm}
\usepackage{latexsym}
\usepackage{sectsty}

\sectionfont{\fontsize{13}{15}\selectfont}

\newcommand{\captionfonts}{\normalsize}

\makeatletter  
\long\def\@makecaption#1#2{%
  \vskip\abovecaptionskip
  \sbox\@tempboxa{{\captionfonts #1: #2}}%
  \ifdim \wd\@tempboxa >\hsize
    {\captionfonts #1: #2\par}
  \else
    \hbox to\hsize{\hfil\box\@tempboxa\hfil}%
  \fi
  \vskip\belowcaptionskip}
\makeatother   
\newcommand{\e}{{\rm e}}
\newcommand{\ep}{\epsilon}
\newcommand{\bs}{\boldsymbol}
\renewcommand{\d}{{\rm d}}
\newcommand{\pd}{\partial}
\newcommand{\PP}{{\rm P}}

\newcommand{\D}{\displaystyle}

\newcommand{\mc}{\mathcal }

\definecolor{darkgreen}{rgb}{0,0.6,0}

\definecolor{violet}{rgb}{0.8,0,0.8}

\begin{document}
\hspace{13.9cm}

%\ \vspace{20mm}\\

\begin{center}
{\Large Evidence accumulation and change rate inference \\ in dynamic environments} \end{center}

\ \\
{\bf Adrian E. Radillo$^{\displaystyle 1}$, Alan Veliz-Cuba$^{\displaystyle 2}$, Kre\v{s}imir Josi\'{c}$^{\displaystyle 1, \displaystyle 3, 4*}$, and Zachary P. Kilpatrick$^{\displaystyle 1, 5,6*}$}\\
{$^{\displaystyle 1}$Department of Mathematics, University of Houston, Houston, TX 77204}\\
{$^{\displaystyle 2}$Department of Mathematics, University of Dayton, Dayton, OH 45469}\\
{$^{\displaystyle 3}$Department of Biology and Biochemistry, University of Houston, Houston, TX 77204}\\
{$^{\displaystyle 4}$Department of BioSciences, Rice University, Houston, TX 77251}\\
{$^{\displaystyle 5}$Department of Applied Mathematics, University of Colorado, Boulder, CO 80309}\\
{$^{\displaystyle 6}$Department of Physiology and Biophysics, University of Colorado School of Medicine, Aurora, CO 80045}\\
{$^{\displaystyle *}$equal contribution}\\

\ \\[-5ex]

\thispagestyle{empty}
\markboth{}{NC instructions}
\ \vspace{-0mm}\\
%
%Abstract
\begin{center} {\bf Abstract} \end{center}

In a constantly changing world, animals must account for environmental volatility when making decisions. To appropriately discount older, irrelevant information, they need to learn the rate at which the environment changes. We develop an ideal observer model capable of inferring the present state of the environment along with its rate of change. Key to this computation is an update of the posterior probability of all possible changepoint counts. This computation can be challenging, as the number of possibilities grows rapidly with time. However, we show how the computations can be simplified in the continuum limit by a moment closure approximation. The resulting low-dimensional system can be used to infer the environmental state and change rate with accuracy comparable to the ideal observer. The approximate computations can be performed by a neural network model  via a rate-correlation based plasticity rule. We thus show how optimal observers accumulate evidence in changing environments, and map this computation to reduced models which perform inference using plausible neural mechanisms.

%%%%%%%%%%%

\section{Introduction}

%{\bf Statement of problem.} 
Animals continuously make decisions in order to find food, identify mates, and avoid predators.  However, the world is seldom static.  Information that was critical yesterday may be of little value now.  Thus, when accumulating evidence to decide on a course of action, animals weight new evidence more strongly than old~\citep{pearson11}. The rate at which the world changes determines the rate at which an individual should discount previous information~\citep{deneve08,velizcuba15}. For instance, when actively tracking prey, a predator may only use visual information obtained within the last second~\citep{olberg00,portugues09}, while social insect colonies integrate evidence that can be hours to days old when deciding on a new home site~\citep{franks02}. These environmental state variables (e.g., prey location, best home site) are constantly changing, and the timescale on which the environment changes is unlikely to be known in advance. Thus, to make accurate decisions, animals must learn how rapidly their environment changes~\citep{wilson10}.
%Humans and other animals make decisions based on noisy sensory input in a changing environment~\citep{behrens07}. 
%Such decisions require that the animal take into account environmental 
%volatility when making a choice. How an environment
%fluctuates is rarely known in advance, and must therefore also be 
%learned from sensory information. 
 
%{\bf Previous models.} 
Evidence accumulators are often used to model decision processes in static and fluctuating environments~\citep{smith04,bogacz06}.  These models show how  noisy observations can be accumulated to provide a probability that one among multiple alternatives is correct~\citep{gold07,beck08}. They explain a variety of behavioral data~\citep{ratcliff08,brunton13}, and electrophysiological recordings suggest that neural activity can reflect the accumulation of evidence~\citep{huk05,kira15}. 
%\Adcomment{the following sentence introduces the adjective `normative' for the first time without any explanation of its meaning. Do we leave it as it is?}
Since normative evidence accumulation models determine the belief of an ideal observer, they also show the best way to integrate 
noisy sensory measurements, and can tell us if and how animals fail to use such information optimally~\citep{bogacz06,beck08}.
 
% {\bf More on previous models -- those more closely related. Why are they important.}
 Early decision-making models focused on decisions between two choices in a static environment~\citep{wald48,gold07}. Recent studies have extended this work to  more ecologically relevant
%\Adcomment{I prefer `ecologically'. I checked the meanings of both adverbs and they are pretty close in this sentence. I find that `ecologically' carries more the notion that organisms need to adapt their decision-making strategies to their environment.  But we can stick to `ethologically' if you prefer it} 
 situations, including multiple alternatives~\citep{churchland08,krajbich11}, multidimensional environments~\citep{Niv:2015},  
and cases where the correct choice~\citep{mcguire14,glaze15}, or context~\citep{shvartsman15}, changes in time. In these cases, normative models are more difficult to derive and analyze~\citep{wilson11}, and their dynamics are more complex. However, methods of sequential and stochastic analysis  are still useful in understanding their properties~\citep{wilson10,velizcuba15}.  

%{\bf Dynamics in case of changing environments.} 
We examine the case of a changing environment where an optimal observer discounts prior evidence at a rate determined by environmental volatility. In this work, a model performs {\em optimally} if it maximizes the likelihood of predicting the correct environmental state, given the noise in observations~\citep{bogacz06}. Experiments suggest that humans learn the rate of environmental fluctuations to make choices nearly optimally~\citep{glaze15}. During dynamic foraging experiments where the choice with the highest reward changes in time, monkeys also appear to use an evidence discounting strategy suited to the environmental change rate~\citep{sugrue04}. 

However, most previous models have assumed that the rate of change of the environment is known ahead of time to the observer~\citep{glaze15,velizcuba15}. \cite{wilson10} developed a model of an observer that infers the rate of environmental change  from observations. To do so, the observer computes a joint posterior probability of the state of the environment, the time since the last change in the environment, and a count of the number of times the environment has changed ({\em changepoint} count). With more measurements, such observers improve their estimates of the change rate, and are therefore better able to predict the environmental state. Inference of the change rate is most important when an observer makes fairly noisy measurements, and cannot determine the current state from a single observation.

We extend previous accumulator models of decision making to the case of multiple, discrete choices with asymmetric, unknown transition rates between them. We assume that the observer is primarily interested in the current state of the environment, often referred to as the {\em correct choice} in decision-making models~\citep{bogacz06}. Therefore, we show how an ideal observer can use sensory evidence to infer the rates at which the environment transitions between states, and simultaneously use these inferred rates to discount old evidence and determine the present environmental state. 
%\Adcomment{I find that the previous sentence lacks clarity because we never say, before it, \emph{what} the observer makes a `decision' about. We only mention explicitly the \emph{prediction} task from Wilson et al. in the previous paragraph,  but not quite our task of \emph{estimating} the current state of the environment. We do mention this distinction further down in the paper, but wouldn't it be worth it to clarify our task upfront?}

%{\bf How does this differ from previous models.}
Related models have been studied before~\citep{wilson10,adams07}.  However, they relied on the assumption that, after a change, the new state  does not depend on the previous state. This excludes the possibility of a finite number of states: For example, in the case of two choices, knowledge of the present state determines with complete certainty the state after a change, and the two are thus not independent. For cases with a finite number of choices our algorithm is simpler than previous ones. The observer only needs to compute a joint probability of the environmental state and the changepoint count.

%{\bf Further differences/advantages}
%We showed previously that the optimal, nonlinear model of evidence accumulation with known change rate is well approximated by a linear model~\citep{velizcuba15}.
The storage needed to implement our algorithms grows rapidly with the number of possible environmental states. However, we show that moment closure methods can be used to decrease the needed storage considerably, albeit at the expense of accuracy and the representation of higher order statistics. Nonetheless, when measurement noise is not too large, these approximations can be used to estimate the most likely transition rate, and the current state of the environment. 
This motivates a physiologically plausible neural implementation for the present computation:
We show that a Hebbian learning rule which shapes interactions between multiple neural populations representing the different choices allows a network to integrate inputs nearly optimally.
Our work therefore links statistical principles for optimal inference with stochastic neural rate models that can adapt to the environmental volatility to make near-optimal decisions in a changing environment.

%Sensory information is noisy. How to accumulate such evidence to make decisions optimally has been studied 
%extensively in static environments~\citep{bogacz06}. However, normative models of decision making in \emph{dynamic} 
%environments, where the correct choice changes in time, are not fully developed. Previous models assumed symmetry 
%in the transition rates~\citep{wilson13,glaze15}, or that the transition rates between the different choices are known 
%to the observer~\citep{velizcuba15}. The aim of this article is to develop a normative model of evidence accumulation 
%with a finite number of choices where transition rates are \emph{unknown}.  

\section{Optimal evidence accumulation for known transition rates}
\label{known}

We start by revisiting the problem of inferring the current state of the environment from a sequence of noisy observations. We assume that 
the number of  states is finite, and the state of the environment changes at times unknown to the observer.
We first review the case when the rate of these changes is known to the observer. 
In later sections, we will assume that these rates must also be learned.
Following~\cite{velizcuba15}, we derived a recursive equation for the likelihoods of the different states, and an approximating stochastic differential equation (SDE). Similar derivations were presented for decisions between two choices by \cite{deneve08} and \cite{glaze15}.

An ideal observer decides between $N$ choices, based on successive observations at 
times $t_n$ $(n=1,2,\ldots)$. We denote each possible choice by $H^i$, $(i=1,\ldots,N)$, with $H_n$ being the correct choice at time $t_n$. The \emph{transition rates} $\ep^{ij}$, $i \neq j$, correspond to the \emph{known} probabilities that the state changes between two observations:
$\ep^{ij}=\PP\left(H_n=H^{i} | H_{n-1}=H^j \right)$. The observer makes measurements, $\xi_n,$ at times $t_n$ with known conditional probability densities $f^{i}(\xi)=\PP\left(\xi_n=\xi | H_n=H^i\right)$. Here, and elsewhere, we assume that the observations are conditionally independent. We also abuse notation slightly by using $\PP (\cdot)$ to denote a probability, or the value of a probability density function, depending on the argument. We use explicit notation for the probability density function when there is a potential for confusion.

We denote by $\xi_{j:n}$ the vector of observations $(\xi_j, \ldots, \xi_n)$, and by $\PP_n(\ \cdot \ )$ the  conditional probability $\PP(\ \cdot \ |\xi_{1:n})$. 
%Similarly we denote the joint conditional probability $\PP(\ \cdot \ , \cdot \ |\xi_{1:n})$ by  $\PP_n(\ \cdot \ , \cdot \ )$. 
To make a decision, the observer can compute the index that maximizes the \emph{posterior probability}, 
$\hat{\imath}={\rm argmax}_i \; \PP_n(H_n=H^i)$. Therefore $H^{\hat{\imath}}$ is the most probable state, given 
the observations $\xi_{1:n}$.

A recursive equation for the update of each of the probabilities \mbox{$\PP_n(H_n = H^i)$} after the $n^{\text{th}}$ observation has the form~\citep{velizcuba15}
%\Adcomment{shouldn't we mention somewhere what assumptions are required for Bayes' rule to apply? Essentially this amounts to assuming that the densities $\PP (\xi_{1:n})$ never vanish, $\forall n$. Are there others?}
\begin{align}
\PP_n(H_n = H^i) \propto  f^i(\xi_n)  \sum_{j=1}^{N} \ep^{ij} \PP_{n-1}(H_{n-1}=H^j) \qquad (i=1,\ldots,N).
	\label{post_known}
\end{align} 
Thus, the transition rates, $\ep^{ij},$ provide the weights of the previous probabilities in the update equation.  Unless transition rates are large or observations very noisy, the probability $\PP_n(H_n = H^{\hat{\imath}})$ grows,
and can be used to identify the present environmental state.  However, with positive transition rates, the posterior probabilities tend to saturate at a value below unity. Strong observational evidence that contradicts an observer's current belief can cause the observer to change their belief subsequently. Such contradictory evidence typically arrives after a change in the environment. 

Following \cite{velizcuba15}, we take logarithms, $x_n^i : = \ln P_n(H_n = H^i),$ and denote by $\Delta x_{n}^i := x_{n}^i-x_{n-1}^i$ the change in log probability due to an observation at time $t_n$. Lastly, we assume the time between observations $\Delta t : = t_n - t_{n-1}$ is small, and $\ep_{\Delta t}^{ij} = \Delta t \ep^{ij} + o ( \Delta t)$ for $i \neq j$ so that dropping higher order terms yields
\begin{align*}
\Delta x_{n}^i = \ln f_{\Delta t}^i(\xi_n) + \ln \left( 1 - \sum_{j \neq i} \Delta t \ep^{ji} + \sum_{j \neq i} \Delta t \ep^{ij} \e^{x_{n-1}^j - x_{n-1}^i} \right), \hspace{6mm} i=1,...,N,
\end{align*}
where the likelihood function $f^i_{\Delta t} (\xi)$ may vary with $\Delta t$. Next, we use the approximation $\ln (1 + z) \approx z$ for $|z| \ll 1$ and replace the index $n$ by time, $t,$ to write
\begin{align*}
\Delta x_{t}^i \approx \Delta t g_{t, \Delta t}^i + \sqrt{\Delta t} W_{\Delta t}^i + \Delta t \sum_{j \neq i} \left( \ep^{ij} \e^{x_{t}^j - x_{t}^i} - \ep^{ji} \right), \hspace{6mm} i=1,...,N,
\end{align*}
where the drift $g^i_t = \frac{1}{\Delta t} {\rm E}_{\xi} \left[ \ln f^i_{\Delta t} (\xi) | H_t \right]$ is the expectation of $\ln f^i_{\Delta t}(\xi)$ over $\xi$, conditioned on the true state of the environment at time $t$, $H_t \in \{ H^1, ... , H^N \}$, and $W_{\Delta t} = ( W_{\Delta t}^1, ..., W_{\Delta t}^N )$ follows a multivariate Gaussian distribution with mean zero and covariance matrix $\Sigma_{\Delta t}$ given by
\begin{align*}
\Sigma_{\Delta t}^{ij} = \frac{1}{\Delta t} {\rm Cov}_{\xi} \left[ \ln f_{\Delta t}^i(\xi), \ln f_{\Delta t}^j (\xi) | H _t \right].
\end{align*}
Finally, taking the limit $\Delta t \to 0$, we can approximate the discrete process,  Eq.~(\ref{post_known}), with the system of SDEs:
%\Adcomment{do we want bold font $\boldsymbol{K}$ in the following equation?}
\begin{align}
\d x_t^i = g^i_t \d t + \d W^i_t + \sum_{j \neq i} \left( \ep^{ij} \e^{x_t^j - x_t^i} - \ep^{ji} \right) \d t, \hspace{6mm} i=1,...,N, \label{sdeknown}
\end{align}
where we assume the following limits hold:
\begin{align*}
g^i_t : = \lim_{\Delta t \to 0} g_{t, \Delta t}^i \hspace{6mm} \text{and} \hspace{6mm} \Sigma^{ij}_t : = \lim_{\Delta t \to 0} \Sigma_{t,\Delta t}^{ij}.
\end{align*}

The nonlinear term in Eq.~\eqref{sdeknown} implies that, in the absence of noise, the system has a stable fixed point, and  older evidence is discounted.  Such continuum models of evidence accumulation are useful because they are amenable to the methods of stochastic analysis~\citep{bogacz06}. Linearization of the SDE provides insights into the system's local dynamics~\citep{glaze15,velizcuba15}, and can be used to implement the inference process  in model neural networks~\citep{bogacz06,velizcuba15}.

We next extend this approach to the case when the observer infers the transition rates, $\ep^{ij},$  from measurements.

\section{Environments with symmetric transition rates}
\label{symmetric}
We first derive the ideal observer model when the unknown transition rates are symmetric, $\epsilon^{ij} \equiv \text{constant}$ when $j \neq i$, and  $\epsilon^{ii} := 1 - (N-1) \epsilon^{ij}$.
%\Adcomment{I dropped the variable $\ep$ to keep text short at this point, plus it did not agree with the definition given in section~\ref{S:sym-multi}}.
This simplifies the derivation, since the observer only needs to estimate a single changepoint count. The asymmetric case discussed in Section \ref{asymmetric} follows the same idea, but the derivation is more involved since the observer must estimate multiple counts.

Our problem differs from previous studies in two key ways~\citep{adams07,wilson10}: First, we assume the observer tries to identify the most likely state of the environment at time $t_n$. 
To do so the observer computes the joint conditional probability, $\PP_n(H_n,a_n),$ of the current state, $H_n$, and
the number of environmental changes, $a_n$, since beginning the observations. Previous studies focused on obtaining the 
predictive distribution,  $\PP_{n}(H_{n+1})$. The two distributions are closely related, as $\PP_n(H_{n+1})=\sum_{H_n} \PP_n(H_{n+1}|H_n)\PP_n(H_n)$.
%\Adcomment{Is this the relation?
%\[
%\PP_n(H_{n+1})=\sum_{H_n} \PP(H_{n+1}|H_n)\PP_n(H_n)
%\]
%But the terms $\PP(H_{n+1}|H_n)$ are unknown. Are we claiming that $\PP_n(H_{n+1})$ can be recovered from $\PP_n(H_n)$, or the other round?}

Second and more importantly,~\cite{adams07,wilson10} implicitly assumed that only observations since
the last changepoint provide information about the current environmental state. That is,  
if the time since the last changepoint -- the current run-length, $r_n$ -- is known to the observer, then all observations 
before that time can be discarded:
$$
\PP(H_n | \xi_{1:n},{r_n}) = \PP(H_n | \xi_{n-r_n:n}).
$$
%\Adcomment{just changed the indexing in previous equation for the last run. Let me know if you have doubts about this.}
This follows from the assumption that the state after a change is conditionally independent of the state that preceded it.  We assume that the number of environmental states is finite.  Hence  this independence assumption does not hold:  Intuitively,
if observations prior to a changepoint indicate the true state is $H^j$, then states
 $H^i, i \neq j$ are more likely after the changepoint.
  
\cite{adams07,wilson10} derive a probability update equation for
 the {\em run length},  and the number of {\em changepoints}, and use this equation
 to obtain the predictive distribution of future observations.
% This requires to keep at all times the history of all observations $\xi_{1:n}$, since there is always some nonzero probability that $r_n = n$.
 We show that it is not necessary to 
 compute run length probabilities when the number of environmental states is finite. Instead we derive a recursive equation for the joint probability of the current state, $H_n$, and number of changepoints, $a_n$. As a result, the total number of possible pairs $(H_n,a_n)$ grows as $N \cdot n$
% \Adcomment{I would say $N\cdot n$ because $0\leq a_n \leq n-1$}
 (linearly in $n$) where $N$ is the fixed number of environmental states $H^i$, rather than $n^2$ 
% \Adcomment{I am tempted to say $n^2$ here because both$0 \leq r_n, a_n \leq n-1$.
% I went through Wilson et al. once more and couldn't find these bounds explicitly stated...} 
(quadratically in $n$) as in \cite{wilson10}.\footnote{The algorithm in \cite{wilson10} requires estimating the run-length $r_n$ and changepoint count $a_n$, so the dimension of the pair $(r_n,a_n)$ grows like $n^2$.}
% \Acomment{Edited to emphasize that we do not require to keep track of the entire history of observations.}\Adcomment{yes, this is a good point. It wasn't obvious to me, but looking back at their algorithm, to compute the predictive distribution at a given timestep, they need the observations of the current run, \emph{for all potential run lenghts}, i.e. the whole history} \Kcomment{I don't think this is correct.  They have message passing algorithm, where they need to update a range of probabilities that were computed on the previous timestep.  However, to do so, they do not need to keep all observations.}
%\Zcomment{I agree with Kreso. Form Eq. (3.7), it seems $p(r_{t-1},a_{t-1};x_{1:t-1})$ is already computed and $p(x_t|x_{t-1}^{r_{t-1}})$ can be described by a finite number of suff stats, so no need to keep $x_{1:t-1}$. Changed to reflect this.}

\subsection{Symmetric 2-state process} \label{S:sym_2state_uknown}
\begin{figure}[t]
\begin{center}
\includegraphics{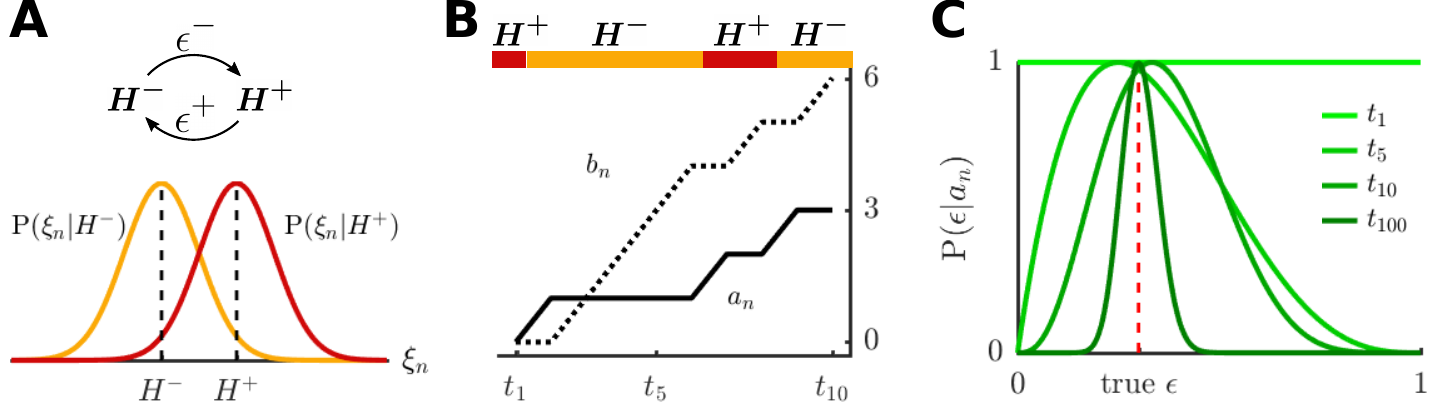}%option [width=14.65cm] could be added but not necessary as pdf already has this size
\caption{Online inference of the change rate  in a dynamic environment. (A) The environment alternates between states $H^+$ and $H^-$ with transition probabilities $\ep^+, \ep^-$. We analyze the symmetric case ($\epsilon: = \epsilon^{\pm}$) in Section \ref{S:sym_2state_uknown} and the asymmetric case ($\epsilon^+ \neq \epsilon^-$) in Section \ref{asymmetric}. The state of the environment determines $f^\pm(\xi)=\PP(\xi|H^\pm)$, which we represent as Gaussian densities.
(B) A sample path of the environment (color bar) together with the first ten values of the actual changepoint count, $a_n,$ and non-changepoint count, $b_n$.
(C) Evolution of the conditional probabilities, $\PP(\ep|a_n)$ (given by Beta distributions), corresponding to the changepoint count from panel B, until $t_n=t_{100}$. The dashed red line indicates the value of $\epsilon$  in the simulation. The densities are scaled so that each equals 1 at the mode. }
\label{fig1}
\end{center}
\end{figure}

We first derive a recursive equation for the probability of two alternatives, $H_n \in \{H^{\pm} \},$ in a changing environment, where the change process is memoryless, and the change rate, $\ep:=\PP(H_n=H^{\mp} |H_{n-1}=H^{\pm})$, is symmetric and initially unknown to the observer (See Fig.~\ref{fig1}A).
%In the present section, the setting is the following:  and the transition rates are unknown, but known  to be equal. \Adcomment{We might want to rewrite the section with the new notation $H_n\in {H^1,H^2\}$ for consistency with the other sections}
 The most probable choice given the observations up to a time, $t_n$, 
  can be obtained from the log of the posterior odds ratio $L_n=\log \left(\frac{\PP_n(H_n=H^+)}{\PP_n(H_n=H^-)}\right)$. The sign of $L_n$ indicates which option is more likely, and its magnitude indicates the strength of this evidence~\citep{bogacz06,gold07}. Old evidence should be discounted according to the inferred environmental volatility. Since this is 
  unknown, an ideal observer computes a probability distribution for the change rate, $\epsilon$ (See Fig. \ref{fig1}C), along with the
probability of environmental states. 

Let $a_n$ be the number of changepoints, and $b_n = n-1-a_n$ the count of non-changepoints  between times $t_1$ and $t_n$ $(n=1,2,\ldots)$ (See Fig.~\ref{fig1}B). The process $\{a_n\}_{n\geq 1}$ is a pure birth process with birth rate $\ep$.  The observer assumes no changes prior to the start of observation,  $\PP(a_1=0)=1$, and must make at least two observations, $\xi_1$ and $\xi_2,$ to detect a change.  

\begin{figure}
\begin{center}
\includegraphics{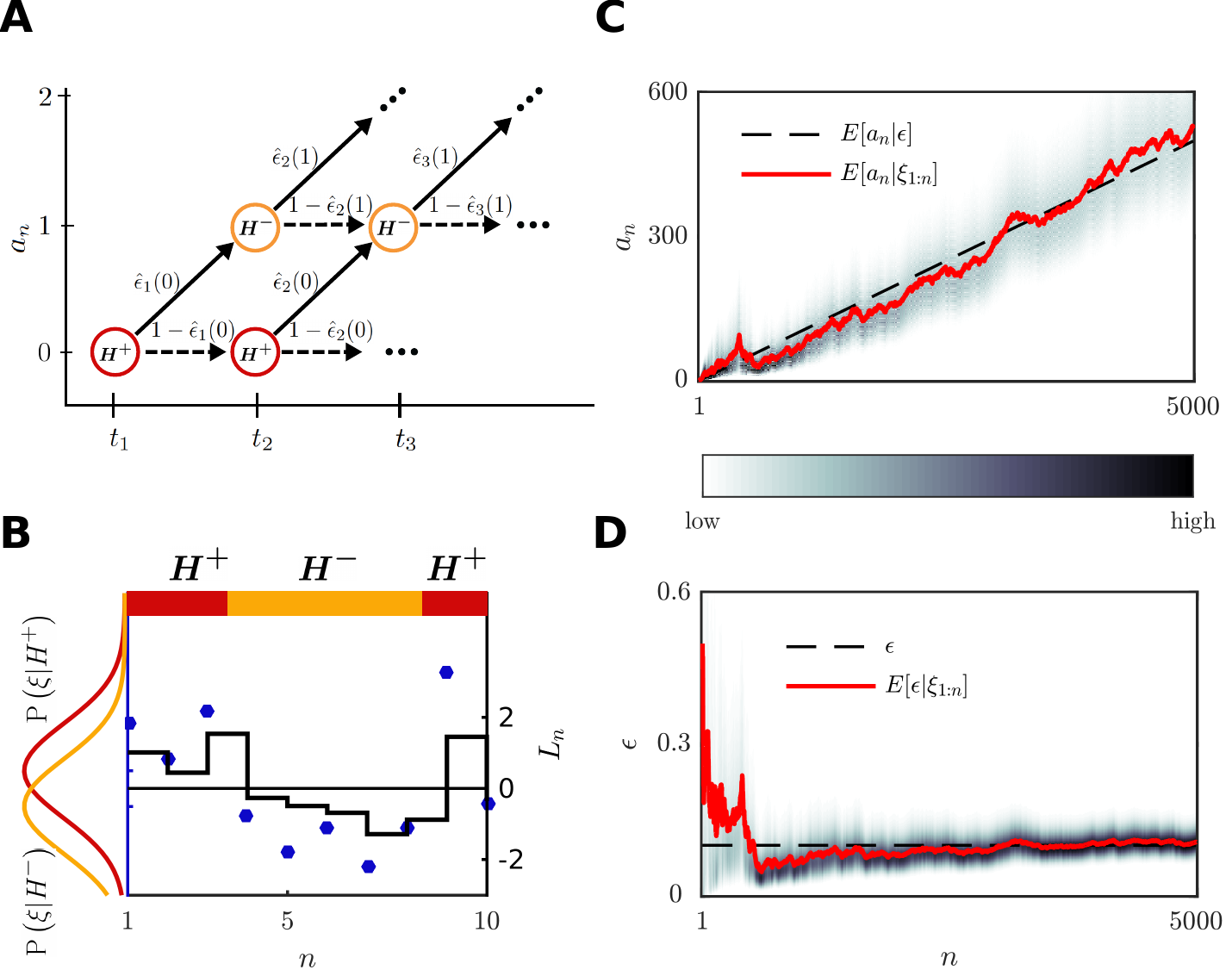}%option [width=14.65cm] could be added but not necessary as pdf already has this size
\caption{Inference of the states, $H^{\pm}$, and change rate, $\ep$. 
(A) The joint posterior probability, $\PP_n(H^\pm,a),$ is propagated along a directed graph according to Eq.~\eqref{Hptwoup_newnotation}. Only  paths corresponding to the initial condition 
$(H_1,a_1)=(H^+,0)$ are shown. 
(B) A sample sequence of environmental states (color bar, top) together with the first ten observations $\xi_1,\ldots \xi_{10}$ (blue dots), for $\ep=0.1$. Superimposed in black (right y-axis) is the log-posterior odds ratio $L_n$ as a function of time. 
(C) Evolution of the posterior over $a_n$ (gray scale). The posterior mean (red) converges to the expected number of changepoints $\ep(n-1)$ (dashed line). 
(D) Evolution of the posterior over the change rate $\ep$ (gray scale). The posterior mean (red) converges to the true value (dashed line) and the variance diminishes with the number of observations.}
\label{fig2}
\end{center}
\end{figure}

%starting at $a_1=0$ and bounded above at time $t_n$ by $n-1$. 
To develop an iterative equation for the joint conditional probability density, $\PP_n(H_n,a_n)$, given the $n$ observations $\xi_{1:n}$, we begin by marginalizing over these quantities at the time of the previous observation, $t_{n-1}$, for $n>1$ (See Appendix~\ref{2stateStart} for details):
\begin{align}
\PP_n(H_n,a_n) = \frac{\PP(\xi_{1:n-1})}{\PP(\xi_{1:n})} \PP(\xi_n | H_n) \sum_{H_{n-1} = H^\pm} 
			\sum_{a_{n-1}=0}^{n-2} \PP(H_n, a_n|H_{n-1},a_{n-1})\PP_{n-1}(H_{n-1},a_{n-1}). \label{jupdate}
\end{align}
%\Adcomment{I added the derivation of the previous equation in the appendix, as a few steps are needed}
With two choices we have the following relationships for all $n>1$:
\begin{align}
H_n=H_{n-1} \Leftrightarrow a_n=a_{n-1}, \quad \text{and} \quad H_n\neq H_{n-1} \Leftrightarrow a_n=a_{n-1}+1. \label{relationHa}
\end{align} 
The term $\PP(H_n, a_n|H_{n-1},a_{n-1})$ in Eq.~\eqref{jupdate} is therefore nonzero only if either, 
$H_{n-1} = H_n$, and $a_{n-1}=a_n$, or $H_{n-1}\neq H_n$ and $a_{n-1}=a_n-1$: If the system is in the joint state $(H_{n-1},a_{n-1})$ at  $t_{n-1}$, then at $t_n$ it can either (a) transition to $(H_n \neq H_{n-1}, a_n = a_{n-1}+1)$ or (b) remain at $(H_n = H_{n-1}, a_n = a_{n-1})$. This observation is central to the message-passing algorithm described in~\citep{adams07,wilson10}, with probability mass flowing from lower to higher values of $a$ according to a pure birth process (See Fig.~\ref{fig2}A).  We can thus simplify Eq.~\eqref{jupdate}, leaving only two terms in the double sum. Writing $\PP_n \left(H^{\pm},a\right)$ for 
$\PP_n\left(H_n=H^{\pm},a_n=a\right)$, and similarly for any conditional probabilities, we have for $n > 1$: 
%\Kcomment{Is  abbreviating present state, but not previous state confusing?  Writing both in full, i.e.$ \PP(H_n= H^{\pm}, a_n= a|H_{n-1} = H^{\pm},a_{n-1}= a)$ could be too lengthy.  Previous version was inconsistent -- see below for corrections.} 
\begin{align}
\PP_n\left(H^{\pm},a\right) =& \frac{\PP\left(\xi_{1:n-1}\right)}{\PP(\xi_{1:n})}f^{\pm} (\xi_n) 
		\left[ \PP(H^\pm,a|H_{n-1}=H^\pm,a_{n-1}=a) \cdot \PP_{n-1}\left(H^{\pm},a\right)\right. \nonumber\\
		&\qquad	\left.+\PP(H^\pm,a|H_{n-1}=H^\mp,a_{n-1}=a-1)\cdot \PP_{n-1}\left(H^{\mp},a-1\right) \right].
																	\label{jupdate2} 
\end{align}
We must also specify \emph{initial conditions} at time $t_1$, and \emph{boundary values} when $a\in \{0,n-1\}$ for these equations. At  $t_1$  we have $\PP(a_1=0)=1$. Therefore,
\begin{align} 
\PP_1(H^{\pm},0) = \frac{1}{\PP(\xi_1)}f^{\pm}(\xi_1)\PP_0(H^\pm), \label{initCond}
\end{align}
and $\PP_1(H^{\pm},a) = 0$ for $a \neq 0$. Here $\PP_0(H^\pm)$ is the prior over the two choices, which we typically take to be uniform so $\PP_0(H^+) = \PP_0(H^-)$. The probability $\PP(\xi_1)$ is unknown to the observer. However, similar to the ratio $\frac{\PP\left(\xi_{1:n-1}\right)}{\PP(\xi_{1:n})}$ in Eq.~\eqref{jupdate2}, $\PP(\xi_1)$ acts as a normalization constant and does not appear in the posterior odds ratio, $R_n$ (See Eq.~\eqref{post_odds_ratio} below).
Finally, at all future times $n > 1$, we have separate equations at the boundaries,
\begin{align}
\PP_n(H^{\pm}, 0) =& \frac{\PP\left(\xi_{1:n-1}\right)}{\PP(\xi_{1:n})}f^{\pm}(\xi_n)
		 \PP(H^\pm,0|H_{n-1}=H^\pm,a_{n-1}=0) \PP_{n-1}\left(H^{\pm} , 0\right), \label{boundary1}
\end{align}
and,
\begin{align}
\PP_n(H^{\pm},n-1) = &\frac{\PP \left( \xi_{1:n-1} \right)}{\PP(\xi_{1:n})} f^{\pm} (\xi_n)
\PP(H^\pm,n-1|H_{n-1}=H^\mp,a_{n-1}=n-2) \times \nonumber \\
&\hspace{5.5cm} \PP_{n-1}\left(H^{\mp} , n-2 \right). \label{boundary2}
\end{align}
We next compute 
$\PP(H_n, a_n|H_{n-1},a_{n-1})$ in Eq.~\eqref{jupdate}, with $n>1$, by marginalizing over all possible transition rates $\ep \in [0,1]$:
\begin{align}
\PP(H_n, a_n|H_{n-1}, a_{n-1}) = \int_0^1 \PP(H_n, a_n| \ep, H_{n-1}, a_{n-1}) \PP(\ep|H_{n-1}, a_{n-1}) \d \ep.	\label{transmarg1}
\end{align}
Note that \mbox{$\PP(\ep|H_{n-1}, a_{n-1})$} \mbox{$= \PP(\ep|a_{n-1})$,} 
so we need the distribution of $\ep$, given $a_{n-1}$ changepoints, for all $n>1$.
%or, equivalently, $b_{n-1} = n-2-a_{n-1}$ time steps with no change.
We assume that prior to any changepoint observations --- that is at time $t_1$ --- the rates follow a Beta distribution with hyperparameters $a_0, b_0>0$ (See also Sections 3.1 and 3.2 in \cite{wilson10}),
\[
\PP_0(\ep)= \beta(\ep; a_0,b_0):=\frac{\ep^{a_0-1} (1-\ep)^{b_0-1} }{B(a_0,b_0)},
\]
where $\beta$ denotes the probability density of the associated Beta distribution, and $B (x,y) :=\int_{0}^{1}\ep^{x-1}(1-\ep)^{y-1}d\ep $ is the beta function. For any $n>1$, the random variable $a_n|\ep$ follows a Binomial distribution with parameters $(n-1,\ep),$ for which the Beta distribution is a conjugate prior. The posterior over the change rate when the changepoint count is known at time $n >1$ is therefore:
\begin{align}
\ep|a_n \sim Beta(a_0+a_n,b_0+b_n). \label{betaprior}
\end{align}
For simplicity, we assume that prior to any observations, the probability over the transition rates is uniform, 
$\PP_0(\ep) = 1$ for all  $\ep \in [0,1]$, and therefore $a_0=b_0=1$ (See Fig.~\ref{fig1}C). 

We now return to Eq.~\eqref{transmarg1} and use the definition of the transition rate, $\ep$, (See Fig. \ref{fig1}) to find:
\begin{align}
\PP(H_n, a_n| \ep, H_{n-1}, a_{n-1}) = \left\{ \begin{array}{cc} 1 - \ep & H_n = H_{n-1} \ \& \ a_{n} = a_{n-1}, \\ 
												\ep & H_{n} \neq H_{n-1} \ \& \ a_{n} = a_{n-1}+1, \\ 
											0 & {\rm otherwise}. \end{array} \right. \label{reltranschange}
\end{align}
Eq.~\eqref{transmarg1} can therefore be rewritten using two integrals, depending on the values
of $(H_n,a_n)$ and $(H_{n-1},a_{n-1})$,
\begin{align}
\PP(H^\pm, a|H_{n-1}=H^\pm, a_{n-1}=a) &= \int_0^1(1-\ep) \beta(\ep; a_{n-1}+1,b_{n-1}+1)\d \ep \label{int1},
\end{align}
and similarly for $\PP(H^\pm, a|H_{n-1}=H^\mp, a_{n-1}=a-1)$.
%Here $\beta(\ep; a_{n-1}+1,b_{n-1}+1)$ denotes the probability density function of the distribution $Beta(\ep; a_{n-1}+1,b_{n-1}+1)$.

The mean of the Beta distribution, for $n>1$, can be expressed in terms of its two parameters:
\begin{align}
\hat{\ep}_{n-1}({a}_{n-1}) \coloneqq \mathbb{E}\left[\ep |a_{n-1} \right]=\frac{a_{n-1}+1}{a_{n-1}+b_{n-1}+2}. \label{meanBeta}
\end{align}
We denote this expected value by $\hat{\ep}_{n-1}({a}_{n-1})$ as it represents a point estimate of the change rate $\ep$ at time $t_{n-1}$ when the changepoint count is $a_{n-1}$, $n>1$. Since  $a_{n-1}+b_{n-1}=n-2$, we have:
\begin{align}
\hat{\ep}_{n-1}({a}_{n-1})=\frac{a_{n-1}+1}{n}. \label{epHat}
\end{align}
The expected transition rate, $\hat{\ep}_{n-1}({a}_{n-1}),$ is thus determined by  the ratio between the previous changepoint count and the number of timesteps, $n$.
%This approximately inverts the relationship between a given transition rate $\epsilon$ and the expected number of changes in $n$ timesteps $\langle a \rangle = \epsilon \cdot (n-1)$
%\Adcomment{I prefer to write $\mathbb{E}[a_n|\ep]$ for the LHS as the RHS is exactly the formula for the mean of the Binomial$(n-1,\ep)$ introduced earlier. Also, what exactly is the point of the previous sentence? Are we claiming that $\hat{\ep}_{n-1}({a}_{n-1}) \approx \mathbb{E}[a_n|\ep] $?}.
Leaving $a_0$ and $b_0$ as parameters in the prior gives $\hat{\ep}_{n-1}(a_{n-1}) = (a_{n-1} + a_0)/(n-2+a_0+b_0)$.
%\Adcomment{corrected the previous denominator}.
%One can distinguish two extreme cases that might help to picture the behavior of this estimate in the long-time limit. When $a_{n-1}$ becomes constant after a certain time, then $\hat{\epsilon}_{n-1}^{a_{n-1}} \to 0$ as $n\to \infty$. On the other hand, when $a_{n-1}$ grows linearly with $n$, then $\hat{\epsilon}_{n-1}^{a_{n-1}} \to 1$ as $n\to \infty$.\Adcomment{Zack, I tried to insert your comment but I was not sure of your goal for it, so feel free to modify the last sentences} \Kcomment{I am not sure the discussion after Eq.(11) is necessary.  Also, if $a_n$ grows linearly then the rate will be the slope, which is in general not equal to 1.}
Using the definition in Eq.~(\ref{epHat}), it follows from Eq.~\eqref{int1} that:
\begin{subequations} \label{intbis}
\begin{align}
\PP(H^\pm, a|H_{n-1}=H^\pm, a_{n-1}=a) &=1-\hat{\ep}_{n-1}(a), \label{int1bis} \\
\PP(H^\pm, a|H_{n-1}=H^\mp, a_{n-1}=a-1) &= \hat{\ep}_{n-1}(a-1). \label{int2bis}
\end{align}
\end{subequations}
Eqs.~(\ref{intbis}), which are illustrated in Fig.~\ref{fig2}A, can in turn be substituted into Eq.~\eqref{jupdate2} to yield, for all $n>1$:
\begin{align}
\PP_n\left(H^{\pm},a\right) =& \frac{\PP\left(\xi_{1:n-1}\right)}{\PP(\xi_{1:n})}f^{\pm} (\xi_n) 
		\left[ \left( 1-\hat{\ep}_{n-1}(a)\right)\cdot \PP_{n-1}\left(H^{\pm},a\right) \right. \nonumber \\
				& \hspace{3.5cm} \left.	+ \hat{\ep}_{n-1}(a-1) \cdot\PP_{n-1}\left(H^{\mp},a-1\right) \right].
																	\label{Hptwoup_newnotation} 
\end{align}
The initial conditions and boundary equations for this recursive probability update have already been described in Eqs.~(\ref{initCond}--\ref{boundary2}). 
Eq.~\eqref{Hptwoup_newnotation} is the equivalent of Eq.~(3) in~\cite{adams07}, and Eq.~(3.7) in~\cite{wilson10}. 
However, here the observer does not need to estimate the length of the interval since
the last changepoint. We demonstrate the inference process defined by Eq.~(\ref{Hptwoup_newnotation}) in Fig.~\ref{fig2}.
%\Adcomment{just moved the previous sentence which appeared further down in the text}

The observer can compute the posterior odds ratio by marginalizing over the changepoint count:
\begin{align}
R_n : & = \frac{\PP_n\left( H^+ \right)}{\PP_n\left( H^- \right)}
	=\frac{\sum_{a=0}^{n-1} \PP_n\left(H^+,a\right) }{\sum_{a=0}^{n-1}  \PP_n\left(H^-,a\right) }. \label{post_odds_ratio}
%	&=\frac{f^+(\xi_n)}{f^-(\xi_n)}\cdot 
%						\frac{\sum_{a=0}^{n-1}\left[ \left( 1-\frac{a+1}{n} \right) \cdot \PP_{n-1}\left(H^+,a\right) 
%												+\frac{a}{n} \cdot \PP_{n-1}\left(H^-,a-1\right) \right]}
%							{\sum_{a=0}^{n-1}\left[ \left( 1-\frac{a+1}{n} \right) \cdot \PP_{n-1}\left(H^-,a\right) 
%												+\frac{a}{n} \cdot \PP_{n-1}\left(H^+,a-1\right) \right]}.										
\end{align}
Here $\log(R_n)=L_n >0$ implies that $H_n = H^+$ is more likely than $H_n = H^-$ (See Fig.~\ref{fig2}B). Note that  $ \PP(\xi_{1:n-1}) / \PP(\xi_{1:n})$ and $1/\PP(\xi_1)$ need not be known to the
observer to obtain the most likely choice.

A posterior distribution of the transition rate $\ep$ can also be derived from Eq.~\eqref{Hptwoup_newnotation} by marginalizing over $(H_n,a_n)$,
\begin{align}
\PP_n(\ep) = \sum_{s = \pm} \sum_{a = 0}^{n-1} \PP(\ep|a_n=a) \PP_{n}\left(H^s,a\right), \label{rateposterior_newnotation_bis}
\end{align} 
where $\PP(\ep|a_n)$ is given by the Beta distribution prior Eq.~(\ref{betaprior}). 
% \Kcomment{I do agree that there could be a more direct way of getting to $\ep$ instead of using estimates of $\ep$ to get to $\PP_{n}\left(H^s,a\right)$, and then back to $\ep$. }
The expected rate is therefore:
\begin{align}
\bar{\epsilon} := \int_0^1 \ep \PP_n(\ep)\d \ep =\sum_{s = \pm} \sum_{a_n = 0}^{n-1}
								\int_0^1 \ep \PP(\ep|a_n) \PP_{n}(H^s,a_n) \d \ep %\nonumber \\ 
								%&= \sum_{s = \pm} \sum_{a_n = 0}^{n-1} 
  					%\frac{\int_0^1 \ep^{a_n+1} (1-\ep)^{b_n}d\ep }{B(a_n+1,b_n+1)} \PP_{n}(H^s,a_n) 
%																					\nonumber \\
%					= \sum_{s = \pm} \sum_{a_n = 0}^{n-1} 
% 					 \frac{B(a_n+2,b_n+1)}{B(a_n+1,b_n+1)} \PP_{n}(H^s,a_n) \nonumber \\
					 =\sum_{s = \pm} \sum_{a_n = 0}^{n-1} 
 							 \frac{a_n+1}{n+1} \PP_{n}(H^s,a_n).  \label{mean_posterior} 
\end{align}
Explicit knowledge of the transition rate, $\epsilon,$ is not used in the inference process described by Eq.~(\ref{Hptwoup_newnotation}). However, computing it allows us to evaluate how the observer's estimate converges to the true transition rate (See Fig.~\ref{fig2}D). We will also relate this estimate  to the coupling strength between neural populations in the model described in Section \ref{neuralpop}. 

\begin{figure}
\begin{center} 
\includegraphics{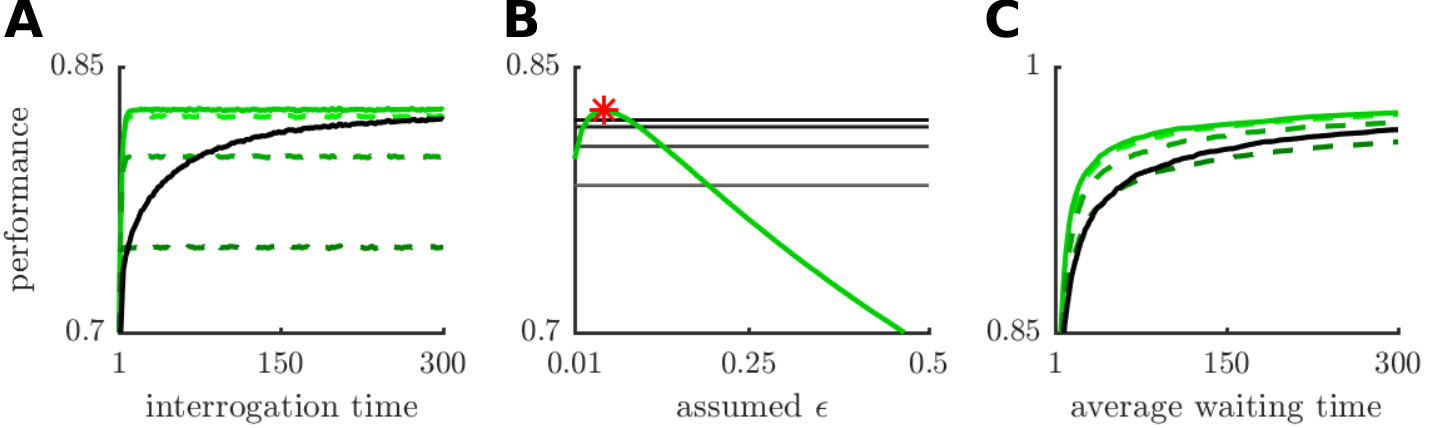}%option [width=14.65cm] could be added but not necessary as pdf already has this size
\end{center}
\caption{The performance of the inference algorithm. 
(A)  Performance under the interrogation paradigm measured as the percentage of correct responses  at the interrogation time. Here and in the next panel $\epsilon=0.05,$ and SNR$=1$. The black curve represents the performance of an ideal observer who infers the change rate from measurements. The green curves, represent the performance of observers that assume a fixed change rate (0.3, 0.15, 0.05, 0.03 from darker to lighter,  see Eq.~(\ref{post_known})).  The solid green line corresponds to an observer who assumes the true rate, dashed lines to erroneous rates. 
%As the interrogation time increases, the ideal observer who is agnostic about the change rate 
% reaches an accuracy that gets closer and closer \Adcomment{asymptotically closer?}
 %\Acomment{Will they saturate to the same value? I checked similar figures in Slack and it wasn't clear what ``closer and closer'' means exactly.} 
% to the one from the ideal observer who knows the exact change rate $\ep$. Each performance curve that corresponds to an assumed rate saturates at its own ceiling value, which depends on the error made in the rate, on the SNR and on the value of the true $\ep$. 
(B)  The green curve represents the performance at interrogation time $t_{300}$ of an observer that assumes a fixed change rate. The red star marks the maximum of this curve, corresponding to the true change rate $\ep = 0.05$. The horizontal black curves represent the performance   at times $t_{40}, t_{100}, t_{200},t_{300}$ (from bottom to top) of the observer that learns the change rate. 
(C) The accuracy as a function of the average threshold hitting time in the free response protocol. Here $\ep=0.1,$ and SNR=$0.75$. See Appendix \ref{freemethod} for details on numerical simulations.
%\Kcomment{Need to rephrase following sentence.}
%We used the same $400$ threshold values for all curves, ranging from $0$ to $3.89$ and simulations were stopped at $t_{5,000}$.
%Each accuracy value $y$ corresponds uniquely to a threshold $\theta$ for the decision variable $L_n$. Decisions are made based on the sign of $L_n$, whenever $|L_n|\geq \theta$ for the first time, resulting in $\PP(Correct)=y$. Here again, as the desired level of accuracy increases, the performance of the agnostic ideal observer (black) gets closer
%to the performance of the ideal observer who knows the true transition rate $\ep$. 
See also Fig.~3 in~\cite{velizcuba15}.
}
\label{fig3}
\end{figure}

We conjecture that when measurements are noisy, the variance of the distribution $\PP_n(\ep)$ does not converge to a point mass at the
true rate, $\epsilon,$ in the limit of infinitely many observations, $n \to \infty$, \emph{i.e.} the estimate of $\ep$ is not consistent. As we have shown, to infer the
rate we need to infer the parameter of a Bernoulli variable. It is easy to show that the posterior over this parameter converges to a point mass at the actual rate value if the probability of misclassifying the state is known to the observer~\citep{djuric2000}.  However, when the misclassification probability is not known, the variance of the posterior remains positive even in the limit of infinitely many observations.  In our case, when measurements are noisy, the observer does not know the exact number of changepoints at finite time.  Hence, the observer does not know exactly how to weight previous observations to make an inference about the current state.  As a result, the probability of misclassifying the current state may not be known. 
We conjecture that this implies that even in the limit  $n \to \infty$ the posterior over $\ep$ has positive variance (See Fig. \ref{fig2}D). 

%\Zcomment{Adrian, can you provide a paragraph of text here, describing and interpreting the findings of Fig. 3? Start by describing the procedure used to probe accuracy, and then provide a theory as to the rate at which the unknown rate model approaches the accuracy of the known model. You may want to port some of this from the figure caption.}
In Fig.~\ref{fig3}, we compare the performance of this algorithm in three cases: when the observer knows the true rate (point mass prior over the true rate $\epsilon$); when the observer assumes a wrong rate (point mass prior over an erroneous $\epsilon$); and when the observer learns the rate from measurements (flat prior over $\epsilon$). 
We define performance as the probability of a correct decision. 

Under the interrogation protocol, the observer infers the state of the environment at a fixed time.  As expected, performance increases with interrogation time, and is highest if the observer uses the true rate (See Fig.~\ref{fig3}A, also Eq.~(\ref{post_known}) above). Performance plateaus quickly when the observer assumes a fixed rate, and more slowly if the rate is learned.  The performance of observers that learn the rate slowly increases toward that of observers who know the true rate.  In Fig. \ref{fig3}B, we present the performance of the unknown-rate algorithm at 4 different times ($t_{40}, t_{100}, t_{200},t_{300}$) and compare it to the asymptotic values with different assumed rates (green curves). 

Note, an observer that assumes an incorrect change rate can still perform near optimally (e.g., curve for 0.03 in Fig.~\ref{fig3}A), especially when the signal-to-noise ratio (SNR) is quite high. The SNR is the difference in means of the likelihoods divided by their common standard deviation. Change rate inference is more effective at lower SNR values, in which case multiple observations are needed for an accurate estimate of the present state. However, at very low SNR values the observer will not be able to substantially reduce uncertainty about the change rate, resulting in high uncertainty about the state.

%\Kcomment{I suggest removing this paragraph, or shortening it considerably.}
%We see that the observer is presented with at least two alternatives when the rate is unknown: One, costly, is to assume a uniform prior distribution over the change rate (unknown-rate algorithm); the other one, cost effective, is to guess the wrong change rate and apply the known-rate algorithm. If either the error in the assumed change rate is small, or the signal-to-noise ratio (SNR) is low, the observer is likely to perform better with the cost-effective strategy. On the other hand, the observer should apply the unknown-algorithm for an intermediate range of SNR values or if she is likely to make a big error in her guess. We define the SNR as the ratio of the difference between the means of the likelihoods to their common standard deviation.
%In the unknown-rate algorithm, for very low SNR -- that is, when the observations are dominated by noise -- the observer has  little chance to accrue her certainty about the number of occurred changepoints. In turn, this prevents the accrual of certainty about the change rate and performance is limited (in this algorithm, high uncertainty about the change rate goes hand in hand with high uncertainty about the current state\Adcomment{I am not sure about this. Kreso, could you clarify this part if it is appropriate please?}). When the SNR is high, the current state can be accurately read-off from the observations and inferring the change rate adds little value to the performance. In this case, the observer will be better-off assuming a wrong $\ep$.

In the free response protocol, the observer makes a decision when the log-odds ratio reaches a predefined threshold.  In Fig.~\ref{fig3}C, we present simulation results for this protocol in a format similar to Fig.~\ref{fig3}A, with empirical performance as a function of average hitting time. Each performance level corresponds to unique log-odds threshold. Similar to the interrogation protocol (Fig. \ref{fig3}A), performance of the free response protocol saturates much more quickly for an observer that fixes their change rate estimate than one that infers this rate over time.

%\Kcomment{Why is the last equation important? I think it is interesting, but will the observer want to know the 
%rate? Also, is there a more direct way to get it?} 

%%%%%%%---------------------------------------- symmetric multistate - unknown rates ------------------------------------------------------------
\subsection{Symmetric multistate process}
\label{S:sym-multi}

We next consider evidence accumulation in an environment with an arbitrary number of states, $\{ H^1, H^2, ... , H^N\}$, with symmetric transition probabilities, $\ep^{ij}\equiv \text{constant}$, 
whenever $i\neq j$. 
We define $\ep:=(N-1)\ep^{ij}$ for any $i \neq j$, so that the probability of remaining in the same state becomes $\epsilon^{ii} = 1 -  \epsilon$, for all $i = 1,...,N$.
The  symmetry in transition rates means that an observer still only needs to track the total number of changepoints, $a_n,$ as in Section~\ref{S:sym_2state_uknown}. 

Eqs.~(\ref{jupdate}-\ref{relationHa}) remain valid with $N$ possible choices, $\{H^1,\ldots,H^N\}$. When $n>1$, the double sum in Eq.~\eqref{jupdate} simplifies to:
\begin{align*}
\PP_n\left(H^{i},a\right) =& \frac{\PP\left(\xi_{1:n-1}\right)}{\PP(\xi_{1:n})}f^{i} (\xi_n) 
		\left[ \PP \left(H^i,a|H_{n-1}=H^i,a_{n-1}=a\right) \cdot \PP_{n-1}\left(H^{i},a\right)\phantom{\sum_{j\neq i}}\right. \nonumber\\
		&\qquad	\left.+\sum_{j\neq i}\PP\left(H^i,a|H_{n-1}=H^j,a_{n-1}=a-1\right)\cdot \PP_{n-1}\left(H^{j},a-1\right) \right].
%																	\label{jupdate3} 
\end{align*}
As in Section~\ref{S:sym_2state_uknown}, we have 
$\PP_1(H^{i},0) = f^{i}(\xi_1)\PP_0(H^i)/\PP(\xi_1)$ and $\PP_1(H^{i},a_1) = 0$ for $a_1 \neq 0$, where $\PP_0(H^i)$ describes the observer's belief prior to any observations. 
At all future times, $n > 1$, we have at the boundaries for all $i=1,\ldots,N$:
\begin{align*}
\PP_n(H^{i}, 0) =& \frac{\PP\left(\xi_{1:n-1}\right)}{\PP(\xi_{1:n})}f^{i}(\xi_n)
		 \PP(H^i,0|H_{n-1}=H^i,a_{n-1}=0) \PP_{n-1}\left(H^{i} , 0\right), 						
\end{align*}
and,
\begin{align*}
\PP_n(H^{i},n-1) = &\frac{\PP \left( \xi_{1:n-1} \right)}{\PP(\xi_{1:n})} f^{i} (\xi_n)
\sum_{j\neq i}\PP(H^i,n-1|H_{n-1}=H^j,a_{n-1}=n-2)\PP_{n-1}\left(H^{j} , n-2 \right).
\end{align*}

Eq.~\eqref{transmarg1} remains unchanged and we still have $\PP(\ep|H_{n-1},a_{n-1})=\PP(\ep|a_{n-1})$. Furthermore, assuming a  Beta prior on the change rate,  Eq.~\eqref{betaprior} remains valid, and Eq.~\eqref{reltranschange} is replaced by:
\begin{align*}
\PP(H_n, a_n| \ep, H_{n-1}, a_{n-1}) = 
						\left\{ \begin{array}{cc} 1 - \ep & H_n = H_{n-1} \ \& \ a_{n} = a_{n-1}, \\ 
										\ep/(N-1) & H_{n} \neq H_{n-1} \ \& \ a_{n} = a_{n-1}+1, \\ 
											0 & {\rm otherwise}. \end{array} \right.
%											\label{reltranschange_multi}
\end{align*}
The integral from Eq.~\eqref{transmarg1} gives, once again, the mean of the Beta distribution, $\hat{\ep}_{n-1}(a),$ defined in Eqs.~(\ref{meanBeta}-\ref{epHat}). As in Section~\ref{S:sym_2state_uknown}, $\hat{\ep}_{n-1}(a_{n-1})$ is a point estimate of the change rate $\ep$ at time $t_{n-1}$ when the  changepoint count is $a_{n-1}$. We have, 
\begin{align}
\PP(H_n, a_n| H_{n-1}, a_{n-1}) &=\left\{ \begin{array}{cc} 
										\D 1- \hat{\ep}_{n-1}(a_{n}) & H_n = H_{n-1} \ \& \ a_{n} = a_{n-1}, \\
										\D \hat{\ep}_{n-1}(a_{n}-1)/(N-1) 
 														& H_{n} \neq H_{n-1} \ \& \ a_{n} = a_{n-1}+1, \\ 
										0
 														& {\rm otherwise}, \end{array} \right.
\label{transavg_multi}
\end{align}
and the main probability update equation is now:
%As in the previous section, we must specify appropriate boundary and initial conditions. By employing Eq.~(\ref{transavg_multi}) in our update equation, we again need only account for prior change point probabilities for $a_n$ and $a_n-1$ in the equation for $P_n(H_n,a_n)$. Setting $\PP_0(H^i) := \PP (H_1 = H^i)$, we have the iterative equation
%We define $\hat{\ep}_n(a_n):=\frac{a_n+1}{n}$ and 
%Eq.~\eqref{Hptwoup_newnotation} becomes:
\begin{align*}
\PP_n\left(H^i,a\right) = \frac{\PP\left(\xi_{1:n-1}\right)}{\PP(\xi_{1:n})}f^i(\xi_n) 
		  \left[ \left( 1- \hat{\ep}_{n-1}(a_n) \right) \cdot \PP_{n-1}\left(H^i,a\right) 
								+\frac{\hat{\ep}_{n-1}(a_n-1)}{N-1} \sum_{j\neq i} \PP_{n-1}\left(H^j,a-1\right) \right]. 
%																	\label{Hptwoup_multi} 
\end{align*}

The observer can infer the most likely state of the environments, by computing the index that maximizes the {\em posterior probability}, marginalizing over all changepoint counts,
\begin{align*}
\hat{\imath} &={\rm argmax}_i \PP_n(H^i) ={\rm argmax}_i \left( \sum_{a=0}^{n-1}\PP_n\left(H^i,a\right) \right).
\end{align*}
%The choice $H^{\hat{\iota}}$ corresponds to the most probable state, given the observations $\xi_{1:n}$. 
The observer can also compute the posterior probability $\PP_n(\ep)$ of the transition rate $\ep$  by marginalizing over all states $H_n$ and changepoint counts $a_n,$ as in Eq.~(\ref{rateposterior_newnotation_bis}). Furthermore, a point estimate of  $\epsilon$ is given by  the mean of the posterior  after marginalizing,  as in Eq.~\eqref{mean_posterior}.

\section{Environments with asymmetric transition rates}
\label{asymmetric}

In this section, we depart  from the framework of \cite{adams07}, and \cite{wilson10}, and  consider unequal transition rates between states. This includes the possibility that some transitions are not allowed. We  consider an arbitrary number, $N,$ of states with unknown transition rates, $\ep^{ij},$ between them. The  switching process between the states is again memoryless,   so that $H_n$ is a stationary, discrete-time Markov chain with finite state space, 
$\Omega:=\{H^1,\ldots,H^N\}$. We write the (unknown) transition matrix for this chain as a left stochastic matrix,
\begin{align*}
\bs{\ep}:=\begin{pmatrix} \ep^{11} & \ldots & \ep^{1N}  \\  
						\vdots & \ddots & \vdots \\
						\ep^{N1} & \ldots & \ep^{NN}  \end{pmatrix},		
%																		\label{trans_matrix}
\end{align*} 
where $\ep^{ij}=\PP (H_n=H^i|H_{n-1}=H^j)$, with $i,j\in \{1,\ldots,N\}$. 
We denote by $\bs{\ep}^{\cdot i}$ the $i$-th column of the matrix $\bs{\ep}$, and similarly for other matrices. Each such column  sums to $1$.
We define the changepoint counts matrix at time $t_n$ as,
\begin{align*}
	\bs{a}_n:= 
							\begin{pmatrix} a_n^{11} & \ldots & a_n^{1N}  \\  
							\vdots & \ddots & \vdots \\
							a_n^{N1} & \ldots & a_n^{NN}  \end{pmatrix}, 
%							\label{cp_matrix}
\end{align*}
where $a_n^{ij}$ is the number of transitions from state $j$ to state $i$ up to time $t_n$. There can be a maximum of $n-1$ transitions at time $t_n$.
%\Kcomment{Why can't $m = 1$?}
%Denote by $\bs{a}_n^{\cdot i}$ the $i$-th column of $\bs{a}_n$.
%The following table illustrates 
%the first five states of a hypothetical 3-state environment and the corresponding values of ${\bf a}_n$:
%\begin{center}
%{\renewcommand{\arraystretch}{1.5}
%\renewcommand{\tabcolsep}{0.2cm}
%\begin{tabular}{|c|c|c|c|c|c|c|}
%\hline
%$n$ & $1$& $2$& $3$& $4$& $5$\\
%\hline
%$H_n$ & $H^1$& $H^1$& $H^2$& $H^3$& $H^3$ \\
%\hline
%${\bf a}_n$ &\rule{0pt}{9ex}$\begin{pmatrix}0&0 & 0\\0 & 0&0 \\ 0&0 &0\end{pmatrix}$& 
%			$\begin{pmatrix}1&0 & 0\\0 & 0&0 \\ 0&0 &0\end{pmatrix}$&
%			$\begin{pmatrix}1&0 & 0\\1 & 0&0 \\ 0&0 &0\end{pmatrix}$& 
%			$\begin{pmatrix}1&0 & 0\\1 & 0&0 \\ 0&1 &0\end{pmatrix}$& 
%			$\begin{pmatrix}1&0 & 0\\1 & 0&0 \\ 0&1 &1\end{pmatrix}$ \\
%\hline
%\end{tabular}} 
%\end{center}
For a fixed $n\geq 1$,  all entries in $\bs{a}_n$ are nonnegative and  sum to $n-1$, \emph{i.e.} $\sum_{i,j} a_n^{ij} = n-1 $. 
As in the symmetric case, the changepoint matrix at time $t_1$ must be the zero matrix,  $\bs{a}_1=\bs{0}$.

\begin{figure}[t]
\begin{center} 
\includegraphics{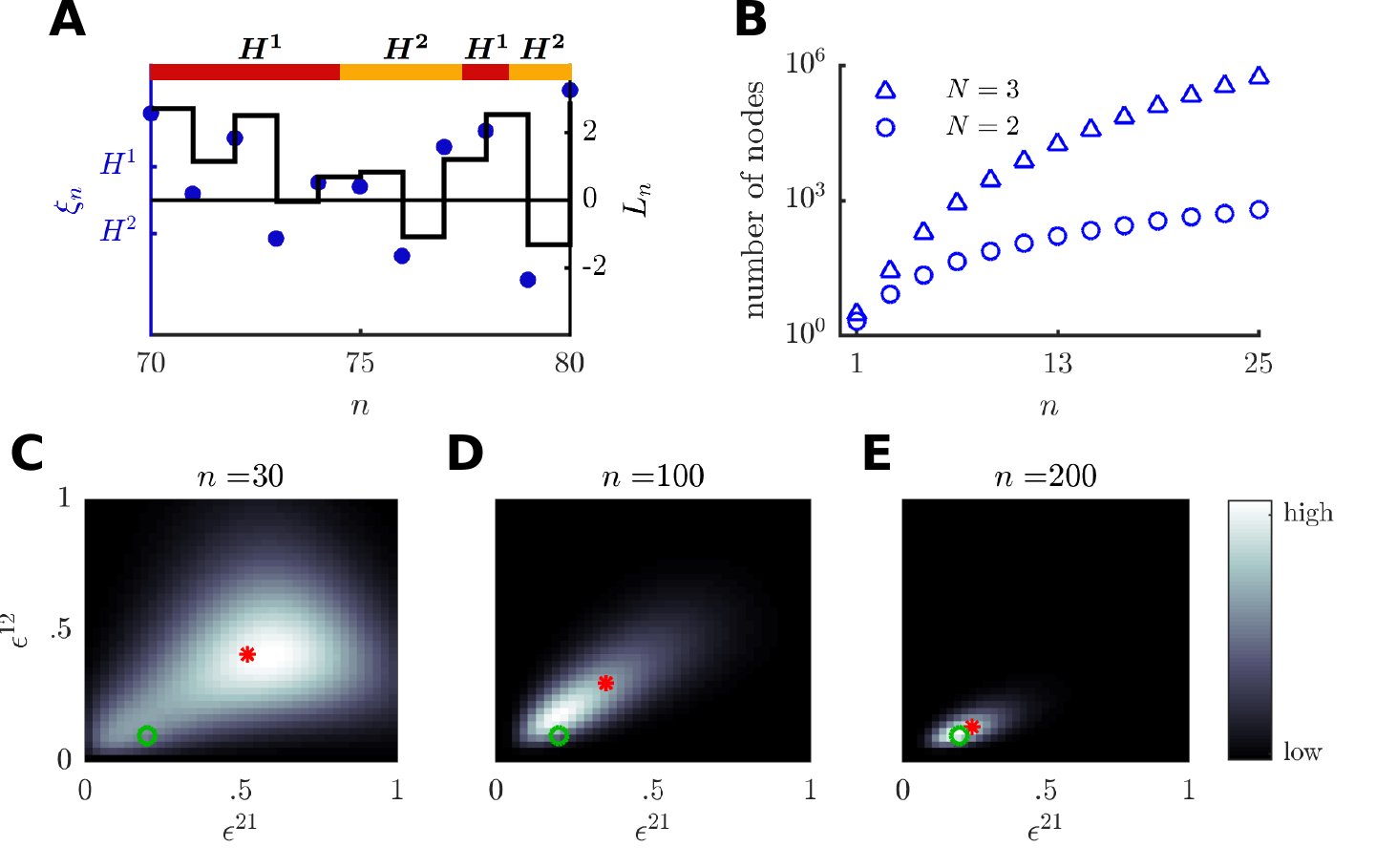} %option [width=14.65cm] could be added but not necessary as pdf already has this size
\end{center}
\caption{Evidence accumulation and change rates inference in a 2-state asymmetric system. (A) Sample path (color bar, top) of the environment between times $t_{70}$ and 
$t_{80}$ (same simulation as in panels C-E) with corresponding observations (blue dots), and log-posterior odds ratio (black step function). Here and in panels C-E, $(\ep^{21},\ep^{12})=(0.2,0.1)$, SNR$=1.4$. (B) The number of allowable changepoint matrices as a function of observation number, $n$, for $N = 2$ (blue circles), and $N = 3$ (blue triangles).  (C)-(E) Color plots (gray scale) of the joint density, $\PP_n\left(\ep^{21},\ep^{12} \right),$ with mean value (red star) approaching the true transition rates (green circle).}
\label{fig4}
\end{figure}

We will show that our inference algorithm assigns positive probability only to changepoint matrices that correspond to possible transition paths between the states $\{H^1,\ldots,H^N\}$. Many nonnegative integer matrices with entries that sum to $n-1$ are not possible changepoint matrices $\bs{a}_n$. A combinatorial argument shows that when $N=2$,  the number of possible pairs, $(H_n,\bs{a}_n),$ grows quadratically with the number of steps, $n,$ to leading order.  It can also be shown that the growth is polynomial for  $N>2$, although we do not know the growth rate in general (See Fig.~\ref{fig4}B). 
An ideal observer has to assign a probability of each of these states which is much more demanding than
in the symmetric rate case where the number of possible states grows linearly in $n$.
%\Zcomment{Add more if we figure out $N=3$ case. Then just say we expect the trend $n^N$ continues for $N>3$, so the algorithm could be implemented for smaller $N$ and $n$.} \Kcomment{I have added a note.  I think this is sufficient. Please check.}
%This means, for example, an entry $\bs{a}_n^{ij}$ can only be nonzero if the state $H^j$ has been visited. %Eqs.~\eqref{eq1_unknown}--\eqref{eq5_unknown} remain valid if we replace ${\bf w}_n$ by $\bs{a}_n$. 

We next derive an iterative equation for  $\PP_n(H_n,\bs{a}_n)$, the joint 
probability of the state $H_n$, and an allowable combination of the $N(N-1)$ changepoint counts (off-diagonal terms of $\bs{a}_n$), and $N$ non-changepoint counts (diagonal terms of $\bs{a}_n$).
%\Kcomment{Previous version of above was not precise.}
The derivation is similar to the symmetric case:  For $n>1$, we first marginalize over  $H_{n-1}$ and $\bs{a}_{n-1}$,
\begin{align*}
\PP_n( H_n , \bs{a}_n ) = \frac{1}{\PP ( \xi_{1:n} )} \sum_{H_{n-1}, \bs{a}_{n-1} } \PP ( \xi_{1:n} | H_n, H_{n-1} , \bs{a}_n, \bs{a}_{n-1})\PP\left(H_n,H_{n-1},\bs{a}_n,\bs{a}_{n-1}\right),
\end{align*}
where the sum is over all $H_{n-1} \in \{ H^1, ..., H^N \}$ and possible values of the  changepoint matrix, $\bs{a}_{n-1}$.

Using
$
\PP ( H_n, H_{n-1}, \bs{a}_n, \bs{a}_{n-1} )  = \PP ( H_n , \bs{a}_n | H_{n-1} , \bs{a}_{n-1} ) \PP ( H_{n-1} , \bs{a}_{n-1} ),
$
and applying Bayes' rule to write
\begin{align*}
\PP ( \xi_{1:n-1} | H_{n-1} , \bs{a}_{n-1}) \PP ( H_{n-1} , \bs{a}_{n-1} ) = \PP ( H_{n-1}, \bs{a}_{n-1} | \xi_{1:n-1} ) \PP ( \xi_{1:n-1}),
\end{align*}
gives
\begin{align}\label{eq:asymmetrix0}
\PP_n ( H_n, \bs{a}_n) = \frac{\PP ( \xi_{1:n-1})}{\PP ( \xi_{1:n})} \PP ( \xi_n | H_n) \sum_{H_{n-1}, \bs{a}_{n-1}} \PP_{n-1} ( H_{n-1} , \bs{a}_{n-1} ) \PP ( H_n, \bs{a}_n | H_{n-1} , \bs{a}_{n-1} ).
\end{align}
%Again, note that the sum is over all possible states $H_{n-1}$ and changepoint matrices, $\bs{a}_{n-1}$.

We compute the conditional probability $\PP ( H_n , \bs{a}_n | H_{n-1} , \bs{a}_{n-1} )$ by marginalizing over all possible transition matrices, $\bs{\epsilon}$. To do so, we relate the probabilities of $\bs{\epsilon}$ and $\bs{a}$. 
Note that if the observer assumes the columns $\bs{\ep}^{\cdot j}$ 
are independent prior to any observations, then the exit rates conditioned on the changepoint counts, $\bs{\ep}^{\cdot j}|\bs{a}_n^{\cdot j}$, are 
independent for all states, $j=1,\ldots,N$. 

To motivate the derivation we first consider a single state,  $j=1,$ and assume that the environmental state has
been observed perfectly over $T>1$ timesteps, but the transition rates are unknown. Therefore, all $\bs{a}_n^{\cdot 1}$ are known to the observer $(1\leq n \leq T)$, but the $\bs{\ep}^{\cdot 1}$ are not. 
The state of the system at time $n+1$, given that it was in state $H^1$ at time $n,$ is a categorical random 
variable, and $\PP(H_{n+1}=H^i | H_{n}=H^1)=\bs{\ep}^{i1}$, for $1 \leq n \leq T-1$. The observed transitions $H^1\mapsto H^i$ are independent samples from a categorical distribution with unknown parameters $\ep^{\cdot 1}$.

% The observed states of the environment 
%are therefore samples from a categorical distribution with unknown parameters.   
The conjugate prior to the categorical distribution is the Dirichlet distribution, and we therefore use it 
as a prior on the changepoint probabilities. For simplicity we again assume a flat prior over $\bs{\ep}^{\cdot1}$, that is $\PP(\bs{\ep}^{\cdot 1})=\chi_S(\bs{\ep}^{\cdot 1})$, where $\chi_S$ is the indicator function on the standard $(N-1)$-simplex, $S$.  

Denote by $D$ the sequence of states that the environment transitioned to at time $n+1$ whenever it was in 
state $H^1$ at time $n$, for all $1 \leq n \leq T-1$. Therefore $D$ is a sequence of states from the set $\{H^1, \ldots, H^N\}$.
By definition,
$\PP(D|\bs{\ep}^{\cdot 1})=\prod_{i=1}^N\left( \bs{\ep}^{i1}\right)^{\sum_{n=1}^{T-1}\chi (H_{n+1} = H^i, H_n = H^1)}$, where $\chi(H_{n+1} = H^i, H_n = H^1)$ is the indicator function, which is unity only when  $H_{n+1}=H^i$ and $H_n = H^1$ and zero otherwise. 
Equivalently, we can write $\PP(\bs{a}_T^{\cdot 1}|\bs{\ep}^{\cdot 1})=\prod_{i=1}^N\left( \bs{\ep}^{i1}\right)^{\bs{a}_T^{i1}}$, since $\bs{a}_T^{i1} = \sum_{n=1}^{T-1}  \chi(H_{n+1} = H^i, H_n = H^1)$. 
For general $n >1$, the posterior distribution for the transition probabilities $\bs{\epsilon}^{\cdot 1}$ given the changepoint vector $\bs{a}_n^{\cdot 1}$ is then
%Adcomment{in the following, I slightly changed the notation for the normalization constant. We might even drop it all together and use the $\propto$ symbol and say that the Dirichlet density is the only one satisfying  this equation}
\begin{align*}
\PP ( \bs{\epsilon}^{\cdot 1} | \bs{a}_n^{\cdot 1} ) = 
\frac{\Gamma \left( \sum_{i=1}^N (a_n^{i 1} + 1)  \right)}{\prod_{i=1}^N \Gamma( a_n^{i 1} + 1 )} \prod_{i=1}^N \left( \epsilon^{i1} \right)^{a_n^{i1}} 
%\frac{\Gamma \left( \sum_{i=1}^N (\bs{a}_n^{\cdot 1} + \bs{1})^i  \right)}{\prod_{i=1}^N \Gamma( \bs{a}_n^{\cdot 1} + \bs{1} )^i} \prod_{i=1}^N \left( \bs{\epsilon}^{i1} \right)^{\bs{a}_n^{i1}} 
= dir( \bs{\epsilon}^{\cdot 1}; \bs{a}_n^{\cdot 1} + \bs{1}).
\end{align*}
%\begin{align}
%\PP(\bs{\ep}^{\cdot 1}|D) =\PP(\bs{\ep}^{\cdot 1}|\bs{a}_n^{\cdot 1})
%							 \propto \prod_{i=1}^N\left( \ep^{i1}\right)^{a_n^{i1}} 
%						\sim Dir(\bs{a}_n^{\cdot 1}+1), \label{posterior_Dirichlet}
%\end{align}
Here $\bs{1}=(1,...,1)^T$, so $\bs{a}_n^{\cdot 1}+\bs{1}$ should be interpreted as the vector with entries $(\bs{a}_n^{i1}+1)_{i=1}^N$, $\Gamma (x)$ is the gamma function,
and
$dir( \bs{\epsilon}^{\cdot 1}; \bs{a}_n^{\cdot 1} + \bs{1})$  the probability density function of the  $N$-dimensional Dirichlet distribution, $Dir(\bs{a}_n^{\cdot 1}+\bs{1})$. 

The same argument applies to all initial states, $H^j$, $j \in \{1,\ldots, N\}.$  
We assume that the transition rates are conditionally independent, so that
\begin{align}
\PP(\bs{\ep}|\bs{a}_n) &=\prod_{j=1}^N dir(\bs{\ep}^{\cdot j};\bs{a}_n^{\cdot j}+\bs{1}) = \prod_{j=1}^N\frac{\Gamma \left( \sum_{i=1}^N ( a_n^{ij} + 1)  \right)}{\prod_{i=1}^N \Gamma( (a_n^{ij} + 1) )} \prod_{k=1}^N \left( \epsilon^{kj} \right)^{a_n^{kj}}. \label{prior_Dirichlet}
\end{align}
%where $\bs{a}_n +\bs{1}$ is the matrix $(a_n^{ij} + 1)_{i,j=1}^N$.

Using this observation, the transition probability between two states can be computed by marginalizing over all possible transition matrices, $\bs{\epsilon},$ conditioned on $\bs{a}_{n-1}$,
%Eq.~\eqref{transmarg_pm1},~\eqref{transmarg_pm2} now read,
\begin{align}
{\rm P}(H_n,\bs{a}_n|H_{n-1}, \bs{a}_{n-1}) 
	=&\int_\mc{M} {\rm P}(H_n, \bs{a}_n| \bs{\ep}, H_{n-1}, 
			\bs{a}_{n-1}) {\rm P}(\bs{\ep}|H_{n-1}, \bs{a}_{n-1}) \d \bs{\ep} \nonumber \\
	=&\int_S\cdots \int_S {\rm P}(H_n, \bs{a}_n|\bs{\ep}^{\cdot 1},\ldots,\bs{\ep}^{\cdot N}, 
															H_{n-1}, \bs{a}_{n-1}) \label{transmarg_multi2} \\ 
		&\hspace{-5mm} \vspace{.1cm} \times dir(\bs{\ep}^{\cdot 1};\bs{a}_{n-1}^{\cdot 1}+1) 
				\times \cdots \times dir(\bs{\ep}^{\cdot N};\bs{a}_{n-1}^{\cdot N}+1)
						\d \bs{\ep}^{\cdot 1}\cdots \d \bs{\ep}^{\cdot N},\nonumber 
\end{align}
where $\mc{M}$ represents the space of all $N \times N$ left stochastic matrices and  $S$ is the $N-1$ dimensional simplex of $\bs{\epsilon}^{\cdot j} \in [0,1]^N$ such that $\sum_{i=1}^N \bs{\epsilon}^{ij} = 1$. 

Let $\bs{\delta}^{ij}$ be the $N\times N$ matrix containing a $1$ as its $ij$-th entry, and $0$ everywhere else. For all $i,j \in \{1,\ldots,N\}$ we have
\begin{align}
\PP(H_n=H^i, \bs{a}_n| \bs{\ep}, H_{n-1}=H^j, \bs{a}_{n-1})=
	\left\{ \begin{array}{cl} \ep^{ij}  & \text{if }\bs{a}_{n}=\bs{a}_{n-1}+\bs{\delta}^{ij},\\
					  0 & {\rm otherwise}. \end{array} \right. \label{reltranschange_multi_asymm}
\end{align}
Implicit in Eq.~(\ref{reltranschange_multi_asymm}) is the requirement that the environment must have been in the state $H_{n-1} = H^j$ in order for the transition $H^j \mapsto H^i$ to have occurred between $t_{n-1}$ and $t_n$. This will ensure that the changepoint matrices $\bs{a}_n$ that are assigned nonzero probability  correspond to admissible paths through the  states $\{ H^1,..., H^N\}$.  %\Zcomment{Added the previous comment, explaining how $\bs{a}_n$ is kept consistent.} 
Applying Eq.~(\ref{reltranschange_multi_asymm}), we  can compute the integrals in Eq.~\eqref{transmarg_multi2} for all pairs $(i,j)$. We let $\hat{\ep}^{ij}_{n-1}(\bs{a}_{n-1})  : = \PP(H_n=H^i,\bs{a}_n=\bs{a}_{n-1}+\bs{\delta}^{ij}|H_{n-1}=H^j,\bs{a}_{n-1})$ 
to simplify notation, and find
%Following a similar procedure as in Eq.~\eqref{transavg_pm1} we can compute the integrals in Eq.~\eqref{transmarg_multi2} for each pair $(i,j)$ to get:
\begin{align}
\hat{\ep}^{ij}_{n-1}(\bs{a}_{n-1}) 
	&= \int_S\cdots \int_S  \ep^{ij} \prod_{k=1}^N dir(\bs{\ep}^{\cdot k};\bs{a}_{n-1}^{\cdot k}+1)
								\d \bs{\ep}^{\cdot 1}\cdots \d \bs{\ep}^{\cdot N} \nonumber \\
	&=\int_S  \ep^{ij} dir(\bs{\ep}^{\cdot j};\bs{a}_{n-1}^{\cdot j}+1)\d\bs{\ep}^{\cdot j}
		\prod_{k\neq j} \int_S dir(\bs{\ep}^{\cdot k};\bs{a}_{n-1}^{\cdot k}+1)
												\d \bs{\ep}^{\cdot k} \nonumber \\
	&=\int_S  \ep^{ij} dir(\bs{\ep}^{\cdot j};\bs{a}_{n-1}^{\cdot j}+1) \d \bs{\ep}^{\cdot j} 
%						=\frac{a_{n-1}^{ij}+1}{\sum_{k=1}^N(a_{n-1}^{kj}+1)}
																			=\frac{a_{n-1}^{ij} + 1}{N+\sum_{k=1}^Na_{n-1}^{kj}}.	 \label{transavg_multi_asymm}
%	&=\frac{a_{n}^{ij}}{N-1+\sum_{k=1}^Na_{n}^{kj}} \label{transavg_multi_asymm}
\end{align}
As in the point estimate of the rate $\hat{\ep}_{n-1}(a_{n-1})$  in Eq.~(\ref{epHat}), each  $\hat{\ep}^{ij}_{n-1}(\bs{a}_{n-1})$ is a ratio containing the number of $H^j \mapsto H^i$ transitions in the numerator, and the total number of transitions out of the $j$th state in the denominator. Thus, the estimated transition rate  $\hat{\ep}^{ij}_{n-1}(\bs{a}_{n-1})$ increases with the number of transitions $H^j \mapsto H^i$ in a given interval $\{ 1,...,n \}$.  Furthermore, each column sums to unity:
\begin{align*}
\sum_{i=1}^N \hat{\ep}^{ij}_{n-1}(\bs{a}_{n-1})  =  \frac{\sum_{i=1}^N \left( a_{n-1}^{ij} + 1 \right) }{N+\sum_{k=1}^Na_{n-1}^{kj}} = \frac{N + \sum_{i=1}^N a_{n-1}^{ij}  }{N+\sum_{k=1}^Na_{n-1}^{kj}}  = 1,
\end{align*}
so the point estimates $\hat{\ep}^{ij}_{n-1}(\bs{a}_{n-1}) $ for the transition rates out of each state $j$ do provide an empirical probability mass function along each column. However, as in the symmetric case, these estimates are biased toward the interior of the domain. This is a consequence of the hyperparameters we have chosen for our prior density, $dir(\bs{\ep} ; \bs{a}_0 + \bs{1} )$.
%\Adcomment{Will the previous comment be clear to the reader? Did we mention it in the symmetric section?}
%\Kcomment{Good point. We should mention this previously, and also why they are biased.  I changed ``skewed'' which typically applies to densities, to ``biased'' which is used for estimates. Please check.} \Zcomment{I also added a sentence stating this arises due to our choice of hyperparameters. The bias need not be there if other hyperparameters are chosen.}
%Show an example of the accumulation of the transition matrix, given a set of transitions. We will apply this to a two state system with asymmetric rates. Perhaps we could also consider a three state system?

Therefore, for $n >1$, the probability update equation in the case of asymmetric transition rates (Eq.~\eqref{eq:asymmetrix0}) is given by,
\begin{align}
\PP_n(H_n=H^i,\bs{a}_n)=\frac{\PP(\xi_{1:n-1})}{\PP(\xi_{1:n})}f^i(\xi_n) 
			\sum_{j=1}^N \hat{\ep}^{ij}_{n-1}(\bs{a}_n - \bs{\delta}^{ij}) \PP_{n-1}\left(H_{n-1}=H^j,\bs{a}_n - \bs{\delta}^{ij} \right).		\label{multi_update}
\end{align}
The point estimates of the transition rates, $\hat{\ep}^{ij}_{n-1}(\bs{a}_{n-1} = \bs{a}_n - \bs{\delta}^{ij}) $, are defined in  Eq.~(\ref{transavg_multi_asymm}). 
As before,
% $\PP(H^i, \bs{a} ) = 0$ for any $a^{ij} <0$, $\PP_0(H^i,\bs{a}) = 0$ for any $a^{ij} >0$, and $\PP_0(H^i,0) = \PP_0(H^i)$. In addition, 
$\PP_1(H^i,\bs{a}_1=\bs{0}) = f^i(\xi_1)\PP_0(H^i)/ \PP ( \xi_1)$ and $\PP_1(H^i, \bs{a}_1) = 0$ for any $\bs{a}_1 \neq \bs{0}$. At future times, it is only possible to obtain changepoint matrices $\bs{a}_n$ whose entries sum to $\sum_{i,j} a_n^{ij} = n-1$, the changepoint matrices $\bs{a}_n$ and $\bs{a}_{n-1}$ must be related as $\bs{a}_n = \bs{a}_{n-1} + \bs{\delta}^{ij},$ as noted in Eq.~(\ref{reltranschange_multi_asymm}). This considerably reduces the number of terms in the sum in Eq.~(\ref{multi_update}).

The observer can find the most likely state of the environment by  maximizing the posterior probability after marginalizing over the changepoint counts $\bs{a}_n$, 
\begin{align*}
\hat{\imath} &={\rm argmax}_i \PP_n(H^i) ={\rm argmax}_i \left( \sum_{\bs{a}_n} \PP_n\left(H^i, \bs{a}_n\right) \right).
\end{align*}
The transition rate matrix can also be computed by marginalizing across all possible states, $H_n,$ and changepoint count matrices, $\bs{a}_n$,
\begin{align*}
\PP_n( \bs{\epsilon} ) = \sum_{s=1}^N \sum_{\bs{a}_n} \PP ( \bs{\epsilon} | \bs{a}_n) \PP_n( H^s, \bs{a}_n),
\end{align*}
where $\PP( \bs{\epsilon} | \bs{a}_n)$ is the product of probability density functions, $dir ( \bs{\epsilon}^{\cdot j}; \bs{a}_n^{\cdot j} + 1),$ given in Eq.~(\ref{prior_Dirichlet}). The mean of this distribution is given by 
\begin{align}
\bar{\bs{\epsilon}} &= \int_\mc{M} \bs{\epsilon} \PP_n( \bs{\epsilon}) d \bs{\epsilon} =  \sum_{s=1}^N \sum_{\bs{a}_n} \PP_n( H^s, \bs{a}_n) \int_\mc{M} \bs{\epsilon} \PP ( \bs{\epsilon} | \bs{a}_n) \d \bs{\epsilon} \nonumber \\
&= \sum_{s=1}^N \sum_{\bs{a}_n} \PP_n( H^s, \bs{a}_n) \bs{E}(\bs{a}_n),  \label{asym_transest}
\end{align}
where $\bs{E}(\bs{a}_n)^{ij} = \hat{\ep}^{ij}_n(\bs{a}_n) =  \mathbb{E}\left[\ep^{ij} |  \bs{a}_n \right]$  defined in Eq.~(\ref{transavg_multi_asymm}), is a conditional expectation over each possible changepoint matrix $\bs{a}_n$.

Eq.~(\ref{multi_update}) is easier to interpret when $N=2$. Using Eq.~(\ref{transavg_multi_asymm}), we find
\begin{align*}
\hat{\epsilon}^{21}_{n-1} (\bs{a}_{n-1}) = \frac{a_{n-1}^{21}+1}{2 + a_{n-1}^{21} + a_{n-1}^{11}}, \ \ \ 
\hat{\epsilon}^{12}_{n-1} (\bs{a}_{n-1}) = \frac{a_{n-1}^{12}+1}{2 + a_{n-1}^{12} + a_{n-1}^{22}},
\end{align*}
and we can express $\hat{\epsilon}^{11}_{n-1} (\bs{a}_{n-1}) = 1 - \hat{\epsilon}^{21}_{n-1} (\bs{a}_{n-1}) $ and $\hat{\epsilon}^{22}_{n-1} (\bs{a}_{n-1}) = 1 - \hat{\epsilon}^{12}_{n-1}(\bs{a}_{n-1}) $.
Expanding the sum in Eq.~(\ref{multi_update}), we have
\begin{subequations} \label{asymupdate}
\begin{align}
\PP_n\left(H^1,\bs{a}_n\right)&=\frac{\PP(\xi_{1:n-1})}{\PP(\xi_{1:n})}f^1(\xi_n) \left[ \hat{\ep}^{11}_{n-1}(\bs{a}_n-\bs{\delta}^{11}) \PP_{n-1}\left(H^1,\bs{a}_n-\bs{\delta}^{11} \right) \right.  \nonumber \\
& \left.  \hspace{40mm} + \hat{\ep}^{12}_{n-1}(\bs{a}_n-\bs{\delta}^{12}) \PP_{n-1} \left(H^2,\bs{a}_n-\bs{\delta}^{12}\right) \right], \label{H1_asymupdate} \\
\PP_n\left(H^2,\bs{a}_n\right)&=\frac{\PP(\xi_{1:n-1})}{\PP(\xi_{1:n})}f^2(\xi_n) \left[ \hat{\ep}^{22}_{n-1}(\bs{a}_n-\bs{\delta}^{22}) \PP_{n-1}\left(H^2,\bs{a}_{n}-\bs{\delta}^{22}\right) \right.  \nonumber \\
& \left.  \hspace{40mm} + \hat{\ep}^{21}_{n-1}(\bs{a}_n-\bs{\delta}^{21}) \PP_{n-1} \left(H^1,\bs{a}_{n}-\bs{\delta}^{21}\right) \right].	\label{H2_asymupdate}	
\end{align}
\end{subequations}
The boundary and initial conditions will be given as above, and the mean inferred transition matrix is given by Eq.~(\ref{asym_transest}). Importantly, the inference process described by Eqs.~(\ref{asymupdate}) allows for both asymmetric changepoint matrices, $\bs{a}_n,$ and inferred transition rate matrices $\bs{E}(\bs{a}_n)$, unlike the process in Eq.~(\ref{Hptwoup_newnotation}). However, the variance of the posteriors over the rates will decrease more slowly, as fewer transitions out of each particular state will be observed. 

This algorithm can be used to infer unequal transition rates as shown in Fig.~\ref{fig4}: Panels C through E show that the mode of the joint posterior distribution, $\PP_n(\ep^{21},\ep^{12}),$ approaches the correct rates, while its variance decreases. As in Section~\ref{S:sym_2state_uknown} we conjecture that this joint density does not converge to a point mass at the true rate values unless the SNR is infinite.

\section{Continuum limits and stochastic differential equation models}
\label{sdemodel}
\begin{figure}
\begin{center} \includegraphics[width=14.5cm]{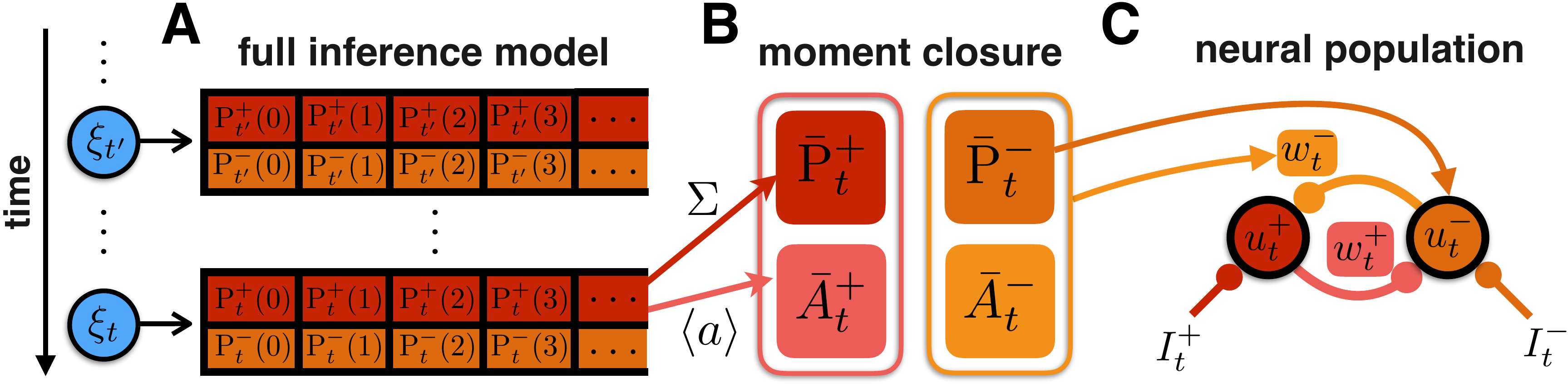} \end{center}
\caption{Schematic showing the reduction of the full inference model, Eq.~(\ref{fulllikeSDE}), for a two state ($H^{\pm}$) symmetric environment ($\ep = \ep^{\pm}$) carried out in Sections \ref{sdemodel} and \ref{neuralpop}. (A) Observations $\xi_t$ arrive continuously in time, and are used to update the probabilities, $\PP_t^{\pm}(a)$ that the environment is in state $H^{\pm}$ after $a$ changepoints. (B) Red and pink arrows from panels A to B represent, respectively, the summation and averaging of $\PP_t^{\pm}(a)$ over $a$ to obtain Eq.~(\ref{likeSDE}) for the zeroth $\bar{\PP}_t^{\pm}$ and first $\bar{A}_t^{\pm}$ moments in Section \ref{momhier}. Arrows from $\PP_t^-(a)$ have been omitted for clarity. (C) Moment equations are converted to a neural population model, Eq.~(\ref{plasticnet}), by assigning the probabilities to population variables, $\bar{\PP}_t^{\pm} \mapsto u_t^{\pm}$, and the ratio of first and zeroth moments to synaptic weights, $\bar{A}_t^{\pm}/ \bar{\PP}_t^{\pm} \mapsto w_t^{\pm}$. Orange arrows from B demonstrate this mapping. The transition rate $\ep$ is learned via changes in the weights $w_t^{\pm}$. If the observer assumes or knows the rate $\ep$ ahead of time, the weights remain fixed.}
\label{roadmap}
\end{figure}

We next derive continuum limits of the discrete probability update equations for the symmetric  case discussed in Section \ref{symmetric}. We assume that observers make measurements rapidly, so we can derive a stochastic differential equation (SDE) that models the update of an ideal observer's belief~\citep{gold07}. SDEs are generally easier to analyze than their discrete counterparts~\citep{gardiner04}. For example,
response times can be studied by examining mean first passage times of log-likelihood ratios~\citep{bogacz06}, or log-likelihoods~\citep{mcmillen06}, which is much easier done in the continuum limit~\citep{Redner:2001}. For simplicity, we begin with an analysis of the two state process, and then extend our results to the multistate case. The full inference model (Fig. \ref{roadmap}A), in the two state case, can be reduced using moment closure to truncate the resulting infinite system of SDEs to an approximate finite system (Fig. \ref{roadmap}B). This both saves computation time and suggests a potential  mechanism for learning the rate $\ep$ of environmental change. We map this approximation to a neural population model in Section \ref{neuralpop} (Fig. \ref{roadmap}C). This model consists of populations that track the environmental state and synaptic weights that learn the transition rate $\ep$.

\subsection{Derivation of the continuum limit} 
\label{sde2state}

\noindent
{\bf Two-state symmetric process.} We first assume that the state of the environment, $\{H_t\}$, is  a homogeneous \emph{continuous-time} Markov chain with state space $\{H^+,H^-\}$. The probability of transitions between the two states is symmetric, and  given by $\PP(H_{t+\Delta t}=H^\pm|H_{t}=H^\mp)=\epsilon \Delta t +o(\Delta t)$, where $0 \leq \epsilon<\infty$. The number of changepoints, $a_t,$ up to  time $t$ is a Poisson process with rate $\epsilon$. An observer  infers the present state from a sequence of observations, $\xi_{1:n}$, made at equally spaced times,
%\Adcomment{could we add a footnote here explaining why this is not a strict requirement. It is not clear to me...}
$t_{1:n},$ with $\Delta t = t_j - t_{j-1}$.\footnote{Equal spacing $\Delta t = t_j - t_{j-1}$ is not necessary for all $j=2, \ldots, n$, but it does allow for a more concise derivation of the continuum limit. Irregular spacings would require a more careful selection of the scaling of the log-likelihoods $ \ln f^{\pm} (\xi )$.}
%\Adcomment{could settle here what happens at $t_0$? Do we have $t_0=0$? If yes, is there an observation at $t_0$?}.
Each observation, $\xi_n,$ has probability $f^{\pm}_{\Delta t}(\xi_n) : = {\rm Pr}(\xi_n | H^{\pm})$ (See \cite{velizcuba15} for more details). We again use the notation $\PP_n(H^{\pm} ,a)=\PP(H_{t_n}=H^{\pm},a_{t_n}=a | \xi_{1:n})$
%\Adcomment{could we write $\PP_n(H^\pm,a)=\PP(H_{t_n}=H^\pm,a_{t_n}=a | \xi_{1:n})$ as in Eq.~\eqref{Hptwoup_newnotation}? Note the typo in the condition $\xi_n$ in main text...I assume it is a typo}
where $t_n$ is the time of the $n^\text{th}$ observation.  
%\Kcomment{I think this needs to be expanded.  Up to this point we have not really discussed observations, and when they are made.  I suggest using the same terminology as in the SIAM paper.}
 
As in the previous sections, an estimate of the rate parameter, $\epsilon,$ is obtained from the posterior distribution over the changepoint count, $a_{t_n}$, at  the time of the $n^\text{th}$ observation, $t_n$. %\Kcomment{We actually abused notation here. We first said that we use $a_t$ as the changepoint count, but then go back to $a_n$.  I state here what we mean.} 
For simplicity, we assume  a Gamma prior with parameters $\alpha$ and $\beta$ over $\epsilon$, so that $\epsilon \sim Gamma(\alpha , \beta)$.  By assumption the changepoint count follows a Poisson distribution with parameter $\epsilon t_n $, so that $\PP(a_{t_n}=a|\epsilon)= (\epsilon t_n)^{a} \e^{-\epsilon t_n}/a!$. Therefore, once $a_n$ changepoints have been observed, we have the posterior distribution  $\epsilon|a_n \sim Gamma( a_n + \alpha, t_n + \beta)$, that is,
\begin{align}
{\rm P}(\epsilon|a_n) = \frac{(t_n+\beta)^{a_n+\alpha}\epsilon^{a_n+\alpha - 1} \e^{-\epsilon (t_n+\beta)}}{\Gamma(a_n+\alpha)}.  \label{betaprior_cts_SDE}
\end{align}
%Therefore, if we write
%\begin{align}
%{\rm P}(H_n, a_n|H_{n-1}, a_{n-1}) = \int_0^\infty {\rm P}(H_n, a_n| \epsilon, H_{n-1}, a_{n-1}) {\rm P}(\epsilon|H_{n-1}, a_{n-1}) d\epsilon,  \label{transmarg1_SDE}
%\end{align}
%then we can use the fact that we need only know the changepoint count to determine our prior $ {\rm P}(\epsilon|H_{n-1}, a_{n-1}) = {\rm P}(\epsilon|a_{n-1})$, so
%\Kcomment{This is wrong, and a bit unclear. The parameter of the Poisson distribution is $\lambda = \epsilon t_n$.  Thus the distribution $\PP(a_{t_n}=a|\epsilon)= (\epsilon t_n)^{a} e^{-\epsilon t_n})/a!$ is correct.
%The conjugate on $\lambda$ is a Gamma distribution.  What was unspecified here is what hyperparameters
%we chose on the prior.  If we choose the improper prior, and set $\alpha =  \beta = 0$, then the
%posterior on $\lambda$ will be $\lambda | a_n = \epsilon t_n | a_n \sim Gamma( \lambda; a_n, 1)$.
%I think this gives $\epsilon | a_n \sim Gamma( \epsilon; a_n, t_n)$ where the posterior hyperparameters are
%just $a_n$ and $t_n$.  The density is then ${\rm P}(\epsilon|a_n) = t_n^{a_n} \epsilon^{a_n - 1}   e^{\epsilon t_n} /  (a_n - 1)!$.  I think this changes the derivation below.}

We can substitute Eq.~(\ref{betaprior_cts_SDE}) into Eq.~(\ref{transmarg1}) describing the probability of transitions between time $t_{n-1}$ and $t_n$ to find 
\begin{align}
{\rm P}(H_n, a_n|H_{n-1}, a_{n-1}) =  
\int_0^\infty {\rm P}(H_n, a_n| \epsilon, H  _{n-1}, a_{n-1}) \gamma(\epsilon; a_{n-1}+\alpha,t_{n-1}+\beta)
d\epsilon,  \label{transmarg2_SDE}
\end{align}
where $\gamma(\epsilon; \alpha, \beta) = \beta^\alpha \epsilon^{\alpha-1}\e^{-\epsilon \beta}/\Gamma(\alpha)$ is the density of the Gamma distribution.
%\Zcomment{Note, the prefactor here is the reciprocal of that given in Wilson et al (2010) in Eq. (3.8). Is this a typo on their part or am I wrong?} \Adcomment{It seems to me that you are right and that Wilson et al. made a typo.}
Using the definition of the transition rate $\epsilon$, we can relate it to the first conditional probability in the integral of Eq.~(\ref{transmarg2_SDE}) via
\begin{align}
{\rm P}(H_n, a_n| \epsilon, H_{n-1}, a_{n-1}) = \left\{ \begin{array}{cc} 
1 -  \epsilon \Delta t & H_n = H_{n-1} \ \& \ a_{n} = a_{n-1}, 
\\
\epsilon  \Delta t & H_{n} \neq H_{n-1} \ \& \ a_{n} = a_{n-1}+1, 
\\
 0 & {\rm otherwise}. \end{array} \right. \label{reltranschange_cts}
\end{align}
We have dropped the $o(\Delta t)$ terms as we are interested in the limit $\Delta t\rightarrow 0$.
%\Adcomment{Equation \eqref{reltranschange_cts} is an approximation up to $o(\Delta t)$. This is because the probability in the case $H_{n} \neq H_{n-1} \ \& \ a_{n} = a_{n-1}+1$ is the probability of exactly one changepoint occuring in the interval $\Delta t$ for a Poisson process. The exact expression is $\epsilon\Delta t \cdot e^{-\epsilon_r\Delta t}$. We should make sure that this approximation does not influence our passage to the limit below when $\Delta t \to 0$}

Using Eq.~(\ref{reltranschange_cts}) and properties of the Gamma distribution we can evaluate the integral in Eq.~(\ref{transmarg2_SDE}) to obtain
\begin{align}
{\rm P}(H_n, a_n|H_{n-1}, a_{n-1}) 
&= \left\{ \begin{array}{cc} 	
1-\Delta t \frac{a_{n}+\alpha}{t_{n-1}+\beta}
& H_n = H_{n-1} \ \& \ a_{n} = a_{n-1} 
\\
\Delta t \frac{a_{n}+\alpha-1}{t_{n-1}+\beta}
& H_{n} \neq H_{n-1} \ \& \ a_{n} = a_{n-1}+1 
\\
 0 & {\rm otherwise.} \end{array} \right. 
\label{integral_betas}
\end{align}
We can use Eq.~\eqref{integral_betas} in the update equation, Eq.~(\ref{jupdate}), to obtain the probabilities of $(H_n,a_n)$ given observations $\xi_{1:n}$. As before, only terms involving  $a_n-1$ and $a_n$ remain in the sum for $n \geq 1$. Using the same notational convention as in previous sections, we obtain,
%Writing ${\rm P}_n(H^\pm,a)$ for ${\rm P}(H_n=H^\pm,a_n=a|\xi_{1:n})$ we have for $H^\pm$ and $n \geq 1$:
\begin{align}
{\rm P}_{n}(H^\pm,a) = \frac{{\rm P}(\xi_{1:n-1})}{{\rm P}(\xi_{1:n})}f^\pm_{\Delta t}(\xi_n) 
&\left[ 
\left( 1 - \Delta t \frac{a+\alpha}{t_{n-1}+\beta} \right) {\rm P}_{n-1}\left(H^\pm,a\right) \right. \nonumber \\
&+ \left. \Delta t \frac{a+\alpha-1}{t_{n-1}+\beta} {\rm P}_{n-1}\left(H^\mp,a-1\right) \right]. \label{Hptwoup_newnotation_SDE} 
\end{align}
%The initial and boundary conditions are given by: $\PP_n(H^{\pm},a)=0$ for $a<0$, $\PP_0(H^{\pm},a)=0$ for $a>0$, and $\PP_0(H^{\pm},0)=\PP_0(H^{\pm})$.
%\Kcomment{Relate to the earlier symmetric case. The equations are similar, and differ only in the point estimate of the transition probability at this point.} \Zcomment{Why do the transition rates have denominator one-off from the discrete case? $(a+1)/n$ versus $(a+1)/(n-1)$ here?}
Note, Eq.~(\ref{Hptwoup_newnotation_SDE}) is similar to the update Eq.~(\ref{Hptwoup_newnotation}) we derived in Section \ref{symmetric}, with the time index replaced by $t_{n-1}/\Delta t$ up to the $\beta$ term.
%\Adcomment{up to the $\beta$ term. Does this ratio $t_{n-1}/\Delta t$ have a particular meaning?}
Also, since we have used a Gamma instead of a Beta distribution as a prior, the point estimate of the transition rate is slightly different (See Eq.~\eqref{epHat}). As in the discrete time case, a point estimate of the transition rate is required even before the first changepoint can be observed. We therefore cannot use an improper prior, as the rate point estimate would be undefined.

To take the limit of Eq.~(\ref{Hptwoup_newnotation_SDE}) as $\Delta t\rightarrow 0$ we proceed as in \cite{bogacz06} and \cite{velizcuba15}, working with logarithms of the probabilities. 
Dividing Eq.~\eqref{Hptwoup_newnotation_SDE} by ${\rm P}_{n-1}(H^\pm,a)$, taking logarithms of both sides, and using the notation $x^{\pm}_{t_n}(a):=\ln {\rm P}_{n}(H^\pm,a)$, we obtain, \footnote{Note, we drop the $\ln \left[ \PP(\xi_{1:n-1})/ \PP (\xi_{1:n}) \right]$ term below since it is common to all evolution equations. For determining the most likely option, only the  relative magnitudes of the log-likelihoods are important. In numerical simulations, we normalize to account for this discrepancy.}
%\Adcomment{strictly speaking, the $\propto$ sign in the following 2 equations is wrong, as we don't multiply by a constant but add one, since logs have been taken...}
\begin{align*}
\Delta x^{\pm}_{t_n}(a) \propto \ln f^\pm_{\Delta t} (\xi_n) +
\ln \left[ 
 1 - \Delta t \frac{a+\alpha}{t_{n-1}+\beta} 
 + \Delta t \frac{a+\alpha-1}{t_{n-1}+\beta} 
 \e^{x^{\mp}_{t_{n-1}}(a-1)-x^{\pm}_{t_{n-1}}(a)}
 \right].
% \label{Hptwoup_newnotation_log} 
\end{align*}
Using the approximation $\ln (1+z)\approx z$ for small $z$ yields
\begin{align*}
\Delta x^{\pm}_{t_n}(a) \propto \ln f^\pm_{\Delta t}(\xi_n) +
  \Delta t \left( 
 \frac{a+\alpha-1}{t_{n-1}+\beta}  \e^{x^{\mp}_{t_{n-1}}(a-1)-x^{\pm}_{t_{n-1}}(a)}
 -\frac{a+\alpha}{t_{n-1}+\beta} \right).
% \label{log_discrete} 
\end{align*}
Since the proportionality constant is equal for all $a$, we drop it in the SDE for the log-likelihood $x_t$, (See~\cite{velizcuba15} for the details of the derivation)
%\Kcomment{I suggest using a subscript in $x_t$.  This is more usual in SDEs. Changed it below.}
\begin{align}
\d x_t^{\pm}(a) = g^{\pm}_t \d t  + \d W_t^{\pm} +
  \left( 
 \frac{a+\alpha-1}{t+\beta}  \e^{x_t^{\mp}(a-1)-x_t^{\pm}(a)}
 -\frac{a+\alpha}{t+\beta} \right) \d t,
 \label{eq:SDE} 
\end{align}
where $g^{\pm}_t=\lim_{\Delta t \to 0} \frac{1}{\Delta t} {\rm E}_{\xi}[\ln f^{\pm}_{\Delta t} (\xi)|H_t]$ and $W^{i}$ satisfies $\langle W_t^i W_t^j \rangle=\Sigma^{ij}_t \cdot t$ with $\Sigma^{ij}_t= \lim_{\Delta t \to 0} \frac{1}{\Delta t} {\rm Cov}_{\xi}[\ln f^{i}_{\Delta t} (\xi),\ln f^{j}_{\Delta t} (\xi)|H_t]$ for $i,j \in \{ + , - \}$.
%\Adcomment{is it really $Cov(\ln f^\pm ,\ln f^\pm)$?} 
%\Kcomment{Before we had $\langle W_t^\pm W_t^\pm \rangle=\Sigma^{\pm\pm}(t) \cdot t$.  I think the last $t$ should be a $\delta(t)$. Please confirm. Same equations appear below.} \Zcomment{Not sure about this. For instance, $\langle W_t^+ W_t^+ \rangle = \sigma_1^2 t$, since this is just the variance of a Wiener process.}

Note that Eq.~\eqref{eq:SDE} is an infinite set of differential equations, one for each pair $(H^\pm, a)$, $a\in\mathbb{Z}_{\geq 0}$.
The initial conditions  at $t=0$ are given by $x^{\pm}(a)=\ln \PP_{0}(H^{\pm}, a)$.
To be consistent with the prior over the rate, $\epsilon$, we can choose 
a Poisson prior over $a$ with mean, $\alpha$, \emph{i.e.} $\displaystyle {\rm P}_{0}(a):=\frac{\alpha^a \e^{-\alpha}}{a!}$.
%Alternatively, we can  use the  geometric distribution over the count, $\displaystyle {\rm P}_{0}(a):= (1 - p)^{a}p$, where $0\leq p\leq 1$. For 
%$p$ close to 1, this distribution will be concentrated at $a = 0$\Adcomment{what will be the difference of choosing one approach versus the other?} \Kcomment{Maybe just drop the discussion of these other priors since we never use them.}. 
%Alternatively, we can define ${\rm P}_0(a|\epsilon):=\frac{(\epsilon \beta)^a e^{-\epsilon \beta}}{a!}$ and obtain
%\begin{align*}
%{\rm P}_0(a):= \int_{0}^\infty {\rm P}_0(a|\epsilon) {\rm P} (\epsilon)d\epsilon = \int_0^\infty \frac{(\epsilon \beta)^a e^{-\epsilon \beta}}{a!} \frac{\beta^\alpha \epsilon^{\alpha-1}e^{-\epsilon \beta}}{\Gamma(\alpha)}d\epsilon = \frac{\Gamma(a+\alpha)}{2^{a+\alpha} a! \Gamma(\alpha)}.
%\end{align*}
The initial conditions for Eq.~(\ref{eq:SDE}) are given by $x^\pm = \ln {\rm P}_0(H^\pm,a)=\ln \left[ {\rm P}_0(H^\pm) {\rm P}_0(a) \right ]$. Note also that Eq.~(\ref{eq:SDE}) at the boundary $a=0$ is a special case. Since at $a=0$ there is no influx of probability from $a-1,$ Eq.~(\ref{eq:SDE}) reduces to 
$$
 \d x_t^{\pm}(0) = g^{\pm}_t \d t  + \d W^{\pm} + \left( (\alpha -1) \e^{-x_t^{\pm}(0)} - \alpha \right) \frac{\d t}{t + \beta}.
 $$
%\Kcomment{Should we mention that the equation at the boundary $a = 0$ is special since there is no influx of probability from $a-1$?}

Lastly, note that we can obtain evolution equations for the likelihoods, $\PP^{\pm}_t(a) = \PP(H_t = H^{\pm}, a),$ by applying the change of variables $\PP^{\pm}_t(a) = \e^{x^{\pm}_t(a)}$. It\^{o}'s  change of coordinates rules~\citep{gardiner04} imply that  Eq.~(\ref{eq:SDE}) is equivalent to
%\Adcomment{could one derive these first and then derive the log forms? I wonder if the detour prob -> log -> prob was necessary }
%\Acomment{It was needed to go from $ln f$ to $g dt + dW$ using the ``standard'' techniques. It would be less clear without taking logs first.}
\begin{align}
\d \PP^{\pm}_t(a) &= \PP^{\pm}_t(a) \left[ \left( g^{\pm}_t + \frac{1}{2} \right) \d t + \d W^{\pm}_t \right] \nonumber \\ & \hspace{3.5cm} + \left[ \frac{a + \alpha - 1}{t + \beta} \PP^{\mp}_t(a-1) - \frac{a+\alpha}{t + \beta} \PP^{\pm}_t(a) \right] \d t,  \label{fulllikeSDE}
\end{align}
where now initial conditions at $t=0$ are simply $\PP^{\pm}_0(a) = \PP_0(H^{\pm},a) = \PP_0(H^{\pm}) \PP_0(a)$. We will compare the full system, Eq.~(\ref{fulllikeSDE}), with an approximation using a moment expansion in Section \ref{momhier} (See also Fig. \ref{roadmap}). \\
\vspace{-4mm}

\noindent
{\bf Two states with asymmetric rates.} Next we consider the case where the state of the environment, $\{ H_t \}$, is still a continuous-time Markov chain with state space $\{ H^1, H^2 \}$, but the probabilities of transition between the two states are asymmetric: $\PP(H_{t + \Delta t} = H^{i} | H_t = H^{j} ) = \epsilon^{ij} \Delta t + o ( \Delta t )$, $i \neq j$, where $\epsilon^{12} \neq \epsilon^{21}$.
%\Adcomment{everything before this and after section 5.2 uses the $\pm$ notation so I believe that we should stick to it for the 2-state cases}
Thus, we must separately enumerate changepoints, $a^{12}_t$ and $a^{21}_t$, to obtain an estimate of the rates $\epsilon^{12}$ and $\epsilon^{21}$. In addition, we will rescale the enumeration of non-changepoints so that $a^{jj}_t = a^{jj} \Delta t$, in anticipation of the divergence of $a^{jj}$  as $\Delta t \to 0$. This will mean the total {\em dwell time}, $a^{jj}_t,$ will be continuous, while the changepoint count will be  discrete. The quantities $a^{ij}_t$ are then placed into a $2 \times 2$ matrix, $\bs{A}_t=(a_t^{ij}) \in \mathbb{R}^{2 \times 2}$, where $a^{ij} \in \mathbb{Z}_{\geq 0}$ for $i \neq j$ and $a^{jj} \in \mathbb{R}^*$. Note that if the number of changepoints, $a^{ij}_t$, and the total dwell time in a state, $a^{jj}_t$, were known, the change rate could be estimated as $\widetilde{\epsilon}^{ \ ij} = a^{ij}_t/a^{jj}_t$.
%\Kcomment{I changed the previous sentence. I am not sure that this is actually a Poisson process - For example, if rates are the same (but unknown), you would count every second event as a transition.  The inter-event times (actual times), would be Gamma distributed then, since they are sums of two exponentially distributed random variables.}

As before, we will estimate the rate parameters, $\epsilon^{ij}$, using the posterior probability of the changepoint matrix, $\bs{a}_t$. We assume Gamma priors on each rate,  so that $\epsilon^{ij} \sim Gamma ( \alpha_j, \beta_j)$. By assumption the changepoint count, $a^{ij}_t,$ follows a Poisson distribution with parameter $ \epsilon^{ij} a^{jj}_t$, so that $\PP ( a^{ij}_t = a | \epsilon^{ij} a^{jj}_t) = \left( \epsilon^{ij} a^{jj}_t \right)^a \e^{- \epsilon^{ij} a^{jj}_t} / a!$. Therefore, once $a^{ij}_t$ changepoints have been observed along with the dwell time $a^{jj}_t$, we have the posterior distribution $\epsilon^{ij} | (a^{ij}_t, a^{jj}_t) \sim Gamma ( a^{ij}_t + \alpha_j, a^{jj}_t + \beta_j)$, so
\begin{align}
\PP ( \epsilon^{ij} | a^{ij}_t, a^{jj}_t ) = \frac{(a^{jj}_t + \beta_j)^{a^{ij}_t + \alpha_j} \left( \epsilon^{ij} \right)^{a^{ij}_t + \alpha_j - 1} \e^{- \epsilon^{ij} (a^{jj}_t + \beta_j)}}{\Gamma (a^{ij}_t + \alpha_j)}.  \label{pepsasym}
\end{align}

We now derive the continuum limit of Eq.~(\ref{asymupdate}). One key step of the derivation is the application of a change of variables to the changepoint matrix $\bs{a}$, where we replace the non-changepoint counts with dwell times $t^j$, defined as $t^i_{\Delta t} : = \Delta t a^{ii}_{\Delta t}$ for $\Delta t = t_n - t_{n-1}$. This is necessary, due to the divergence of $a^{ii}_{\Delta t}$ as $\Delta t \to 0$. In the limit $\Delta t\rightarrow 0$, the modified changepoint matrix becomes
\begin{align*}
\bs{A} = \left( \begin{array}{cc} t^1 & a^{12} \\ a^{21} & t^2 \end{array} \right),
\end{align*}
where $a^{ij} \in \mathbb{Z}^*$ is the changepoint count from $H^j \mapsto H^i$, while $t^i \in \mathbb{R}^*$ is the dwell time in state $H^i$. Thus, taking logarithms, linearizing, and taking the limit $\Delta t \to 0$, we obtain the following system of stochastic partial differential equations (SPDEs) for the log-likelihoods, $x^j_t (\bs{A})= \ln {\rm P}_{n}(H^\pm, \bs{A})$ :
\begin{subequations} \label{sdeasymmetric}
\begin{align}
\d x_t^1 ( \bs{A} ) = g^1_t \d t + \d W_t^1 + \left( \frac{a^{12} + \alpha_2 -1}{t^2 + \beta_2} \e^{x_t^2 ( \bs{A} - \bs{\delta}^{12}) - x_t^1(\bs{A})} - \frac{a^{21} + \alpha_1}{t^1 + \beta_1} - \frac{\pd x_t^1}{\pd t^1}\right) \d t, \\
\d x_t^2 ( \bs{A} ) = g^2_t \d t + \d W_t^2 + \left( \frac{a^{21} + \alpha_1 -1}{t^1 + \beta_1} \e^{x_t^1 ( \bs{A} - \bs{\delta}^{21}) - x_t^2(\bs{A})} - \frac{a^{12} + \alpha_2}{t^2 + \beta_2} - \frac{\pd x_t^2}{\pd t^2} \right) \d t,
\end{align}
\end{subequations}
where the drift, $g^i_t,$ and noise, $W^i_t,$ are defined as before (for details, see Appendix \ref{climasym}). Note that the flux terms, $\displaystyle \frac{\pd x_t^i}{\pd t^i},$ account for the flow of probability to longer dwell times $t^i$. For example, the SPDE for $x^1_t$ has a flux term for the linear increase of the dwell time $t^{1}$, since this represents the environment remaining in state $H^1$. These flux terms simply propagate the probability densities $\e^{x_t^i(\bs{A})}=\PP_t\left(H^i,\bs{A}\right)$ % \Adcomment{$=\ln P_n\left(H^\pm,\bs{A}\right)$ or $=\PP_t^j(\bs{A})$ would be clearer, although $=\PP_t^j(\bs{A})$ is only introduced a bit further} 
over the space $(t^1,t^2)$, causing no net change in the probability of residing in either state $H^i$: $\PP_t^{i} = \int_0^{\infty} \int_0^{\infty} \e^{x_t^i(\bs{A})} \d t^1 \d t^2$.
%This is a continuum analog of the discrete exchange of probability in the terms involving the changepoint counts $a^{ij}$\Adcomment{I don't understand this sentence}.

Eq.~(\ref{sdeasymmetric}) generalizes  Eq.~\eqref{asymupdate} as an infinite set of SPDEs, indexed by the discrete variables $(H^j,a^{12}, a^{21})$ where $a^{12}, a^{21} \in \mathbb{Z}_{\geq 0}$. Each SPDE is over the space $(t^1,t^2)$, and it is always true that $t^1 + t^2 = t$.
%\Kcomment{I think this condition does not need to be imposed, but is satisfied automatically, no?}
%\Acomment{Is it also indexed by the cont. variables  $t^1,t^2$ (with the condition that $t^1+t^2=t$)?}
Initial conditions at $t=0$ are given by $x^j(\bs{A}) = \ln \left[ \PP_0(H^j) \cdot \PP_0(\bs{A}) \right]$. For consistency with the prior on the rates, $\epsilon^{ij}$, we choose a Poisson prior over the changepoint counts $a^{ij}$, $i \neq j$, and a Dirac delta distribution prior over the dwell times $t^i$,
\begin{align}
\PP_0(\bs{A}) = \frac{\alpha_1^{a^{21}} \e^{-\alpha_1}}{a^{21}!}  \frac{\alpha_2^{a^{12}} \e^{-\alpha_2}}{a^{12}!} \delta ( t^1 - \beta_1) \delta ( t^2 - \beta_2). \label{asymprior}
\end{align}
As before, Eq.~(\ref{sdeasymmetric}) at the boundaries $a^{12} = 0$ and $a^{21} = 0$ is a special case, since there will be no influx of probability from $a^{12} -1$ or $a^{21} - 1$.

As in the symmetric case, we can convert Eq.~(\ref{sdeasymmetric}) to equations describing the evolution of the likelihoods $\PP^i_t(\bs{A}) = \PP (H_t = H^i, \bs{A})$. Applying the change of variables $\PP_t^i(\bs{A}) = \e^{x_t^i(\bs{A})}$, we find
\begin{subequations} \label{likesdeasym}
\begin{align}
\d \PP_t^1 (\bs{A}) =& \PP_t^1(\bs{A}) \left[ \left( g^1_t + \frac{1}{2} \right) \d t + \d W_t^1 \right] \nonumber \\
& + \left[ \frac{a^{12} + \alpha_2 - 1}{t^2 + \beta_2} \PP_t^2(\bs{A} - \bs{\delta}^{12}) - \frac{a^{21} + \alpha_1}{t^1 + \beta_1} \PP_t^1(\bs{A}) - \frac{\pd \PP_t^1(\bs{A})}{\pd t^1} \right] \d t \\
\d \PP_t^2 (\bs{a}) =& \PP_t^2(\bs{a}) \left[ \left( g^2_t + \frac{1}{2} \right) \d t + \d W_t^2 \right] \nonumber \\
& + \left[ \frac{a^{21} + \alpha_1 - 1}{t^1 + \beta_1} \PP_t^1(\bs{A} - \bs{\delta}^{21}) - \frac{a^{12} + \alpha_2}{t^2 + \beta_2} \PP_t^2(\bs{A}) - \frac{\pd \PP_t^2(\bs{A})}{\pd t^2} \right] \d t
\end{align}
\end{subequations}
where now initial conditions at $t=0$ are $\PP_0^i(\bs{A}) = \PP_0(H^i) \PP_0(\bs{A})$. \\
\vspace{-4mm}

\noindent
{\bf Multiple states with symmetric rates.} The continuum limit in the 
case of $N$ states, $\{H^1,\ldots,H^N\}$, with symmetric transition rates can be derived as with $N = 2$ (See Appendix \ref{manysym} for details).
Again, denote the transition probabilities by $\PP(H_{t+\Delta t}=H^i|H_t=H^j)=\epsilon^{ij} \Delta t +o(\Delta t)$, 
and the rate of switching from one to any other state by $\epsilon= (N-1) \epsilon^{ij} $. 

Assuming again a Gamma prior on the transition rate, $\epsilon \sim Gamma(\alpha , \beta),$
and introducing $x^i_{t_n}(a):=\ln {\rm P}_{n}(H^i,a)$, we obtain the SDE 
\begin{align}
{\rm d} x^i_t(a) = g^i_t {\rm d} t   + {\rm d} W_t^i +
  \left( 
 \frac{a+\alpha -1 }{(N-1)(t+\beta)}  \sum_{j\neq i} \e^{x_t^j(a-1)-x_t^i(a)}
 -\frac{a+\alpha}{t+\beta} \right) {\rm d} t ,
 \label{eq:SDEmulti} 
\end{align}
where $g^i_t=\lim_{\Delta t \to 0} \frac{1}{\Delta t} {\rm E}_{\xi}[\ln f^i_{\Delta t}(\xi)|H_t]$ and $W^i$ satisfies $\langle W^iW^j \rangle=\Sigma^{ij}_t\cdot t$ with $\Sigma^{ij}_t=\lim_{\Delta t \to 0} \frac{1}{\Delta t} {\rm Cov}_{\xi}[\ln f^{i}_{\Delta t} (\xi),\ln f^{j}_{\Delta t} (\xi)|H_t]$. 

Eq.~\eqref{eq:SDEmulti} is again an infinite set of stochastic differential equations, 
one for each pair $(H^i, a)$, \mbox{$i \in {1, \ldots, N}$,} $a  \in\mathbb{Z}_{\geq 0}$. 
We have some freedom in choosing  initial conditions at $t=0$. For example, since  $x^i(a)=\ln \PP_{0}(H^i, a)$, we can
use the Poisson distribution discussed in the case of two states.

The posterior over the transition rate, $\epsilon,$ is 
\[
\PP_n(\epsilon)=\sum_{s = 1}^N \sum_{a_n = 0}^\infty 
\PP(\epsilon|a_n)\PP_n(H^s,a_n),
\]
where $\PP(\epsilon|a_n)$ is the Gamma distribution given by Eq.~(\ref{betaprior_cts_SDE}). Similar to Eq.~(\ref{mean_posterior}),
the expected rate is
\begin{align*}
\bar{\epsilon} := \int_0^\infty \ep \PP_n(\ep) {\rm d} \ep =\sum_{s = 1}^N \sum_{a_n = 0}^{\infty}
\int_0^\infty \ep \PP(\ep|a_n) \PP_{n}(H^s,a_n) {\rm d} \ep 
=\sum_{s = 1}^N \sum_{a_n = 0}^\infty
\frac{a_n+\alpha}{t_n+\beta} \PP_{n}(H^s,a_n).
%\label{mean_posterior_SDE} 
\end{align*}
An equivalent argument can be used to obtain the posterior over the rates in the asymmetric case with $N$ states.

\subsection{Moment hierarchy for the 2-state process}
\label{momhier}
In the previous section, we approximated the evolution of the joint probabilities of environmental states and changepoint counts. The result, in the symmetric case, was an infinite set of SDEs, one for each combination of state and changepoint values $(H^i, a)$. However, an observer is mainly concerned with the current state of the environment. The changepoint count is important for this inference, but may not be of direct interest itself. We next derive simpler, approximate models that do not track the entire joint distribution over all changepoint counts, but only essential aspects of this distribution. We do so by deriving a hierarchy of iterative equations for the moments of the distribution of changepoint counts, $a\in\mathbb{Z}_{\geq 0}$, focusing specifically on the two state symmetric case.

Our goal in deriving moment equations is to have a low-dimensional, and reasonably tractable, system of SDEs. Similar to previous studies of sequential  decision making algorithms~\citep{bogacz06}, such low-dimensional systems can be used to inform neurophysiologically relevant population rate models of the evidence accumulation process. To begin, we consider the infinite system of SDEs given in the two state symmetric case, Eq.~(\ref{fulllikeSDE}). Our reduction then proceeds by computing the SDEs associated with the lower order (0th, 1st, and 2nd) moments over the changepoint count $a$:%\Adcomment{should there be $^\pm$ in the RHS below?}
\begin{align}
\bar{\PP}_t^{\pm} = \sum_{a \in \mathbb{Z}_{\geq 0}} \PP^\pm_t(a),
 \ \ 
 \bar{a}_t^{\pm} = \sum_{a \in \mathbb{Z}_{\geq 0}} (a + \alpha) \PP^\pm_t(a), 
 \ \ 
 \bar{b}_t^{\pm} = \sum_{a \in \mathbb{Z}_{\geq 0}} (a + \alpha)^2 \PP^\pm_t(a). \label{momentdef}
\end{align}
We denote the moments using bars ($\bar{b}_t^{\pm}$). Below, when we discuss cumulants, we will represent them using hats $\left(\hat{b}_t^{\pm}\right)$. Note that the ``0th'' moments are the marginal probabilities of $H^+$ and $H^-$.

We begin by summing Eq.~(\ref{fulllikeSDE}) over all $a \in \mathbb{Z}_{\geq 0}$ and applying Eq.~(\ref{momentdef}) to find this generates an SDE for the evolution of the moments $\bar{\PP}_t^{\pm}$ given
\begin{align}
\d \bar{\PP}_t^{\pm} = \bar{\PP}_t^{\pm} \left[ \left( g^{\pm}_t + \frac{1}{2} \right) \d t + \d W_t^{\pm} \right] + \frac{1}{t + \beta} \left[ \bar{a}_t^{\mp} - \bar{a}_t^{\pm} \right] \d t. \label{momsde1}
\end{align}
where we have used the fact that
\begin{align*}
\overline{(a-1)}_t^{\pm} = \sum_{a=1}^{\infty} (a + \alpha -1) \PP_t^{\pm}(a-1) = \sum_{a=0}^{\infty} (a+ \alpha) \PP_t^{\pm}(a) = \bar{a}_t^{\pm}.
\end{align*}
The SDE given by Eq.~(\ref{momsde1}) for the zeroth moment, $\bar{\PP}_t^{\pm},$ depends on the first moment, $\bar{a}_n^{\pm}$. We can determine values for the first moment by either obtaining the next SDE in the moment hierarchy, or assuming a reasonable functional form for $\bar{a}_t^{\pm}$. For instance, if the transition rate $\epsilon$ is known we can assume $\bar{a}_t^{\pm} : = (t + \beta) \epsilon \bar{\PP}_t^{\pm} + {\mc O}(1)$, so that $\bar{a}_t^{+} + \bar{a}_t^{-}$ is approximately the mean of the counting process with rate $\ep$.
%\Adcomment{is there any intuition behind this?}
In this case, the continuum limit $t \to \infty$ of Eq.~(\ref{momsde1}) becomes
\begin{align*}
\d \bar{\PP}_t^{\pm} = \bar{\PP}_t^{\pm} \left[ \left( g^{\pm}_t + \frac{1}{2} \right) \d t + \d W_t^{\pm} \right] +  \epsilon \cdot \left[  \bar{\PP}_t^{\mp} - \bar{\PP}_t^{\pm} \right] \d t. 
%\label{momsde2}
\end{align*}
As expected, this is the two state version of Eq.~(\ref{sdeknown}), with known rate, $\ep$. However, if the observer has no prior knowledge of the rate, $\epsilon$, then $\bar{a}_t^{\pm}$ should evolve towards $(t+ \beta) \epsilon \bar{\PP}_t^{\pm}$ at a rate that depends on the noisiness of observations.

To obtain an equation for $\bar{a}_t^{\pm}$ we multiply Eq.~(\ref{fulllikeSDE}) by $(a + \alpha)$ and sum to yield,
\begin{align}
\d \bar{a}_t^{\pm} = \bar{a}_t^{\pm} \left[ \left( g^{\pm}_t + \frac{1}{2} \right) \d t + \d W_t^{\pm} \right] + \frac{1}{t + \beta} \left[ \bar{a}_t^{\mp} + \bar{b}_t^{\mp} - \bar{b}_t^{\pm} \right] \d t.  \label{momsde3}
\end{align}
This equation relates the zeroth, first, and second moments, $\bar{\PP}_t^{\pm}$, $\bar{a}_t^{\pm}$, and $\bar{b}_t^{\pm}$. Again, we require an expression for the next moment, $\bar{b}_t^{\pm}$, to close the system of Eqs.~(\ref{momsde1},  \ref{momsde3}). We could obtain an equation for $\bar{b}_t^{\pm}$ by multiplying Eq.~(\ref{fulllikeSDE}) by $(a + \alpha)^2$ and summing. However, as is typical with moment hierarchies, we would not be able to close the system as equations for subsequent moments will depend on successively higher moments~\citep{socha07,kuehn16}. To close the equations for $\bar{\PP}_t^{\pm}, \bar{a}_t^{\pm}, ...$ we can truncate: One possibility is cumulant-neglect~\citep{whittle57,socha07}, which assumes all cumulants above a given order grow more slowly than the moment itself and can thus be ignored. 
This allows one to express the highest order moment as a function of the lower order moments, since a moment is an algebraic function of its associated cumulant and lower moments.
%In the following we use hats to distinguish cumulants $(\hat{b}_t^{\pm})$, whereas bars still denote moments ($\bar{b}_t^{\pm}$).
For instance, neglecting the second cumulant $\hat{b}_t^{\pm} \approx 0$ allows us to approximate the second moment as $\bar{b}_t^{\pm} = \hat{b}_t^{\pm} + \left( \bar{a}_t^{\pm} \right)^2 \approx  \left( \bar{a}_t^{\pm} \right)^2$.\footnote{Note, we have used a hat to distinguish cumulants $(\hat{b}_t^{\pm})$, whereas bars still denote moments ($\bar{b}_t^{\pm}$).}
%Neglecting the third cumulant $\hat{c}_n^{\pm} \approx 0$ leads to an approximation for the third moment, $\overline{c}_n^{\pm} = \hat{c}_n^{\pm} + 3 \overline{b}_n^{\pm} \overline{a}_n^{\pm} - 2 \left( \overline{a}_n^{\pm} \right)^3 \approx 3 \overline{b}_n^{\pm} \overline{a}_n^{\pm} - 2 \left( \overline{a}_n^{\pm} \right)^3$.
%\Kcomment{Maybe state beforehand that we will denote moments with bars, and cumulants with hats.}

\begin{figure}
\vspace{-1cm}
\begin{center} \includegraphics[width=12cm]{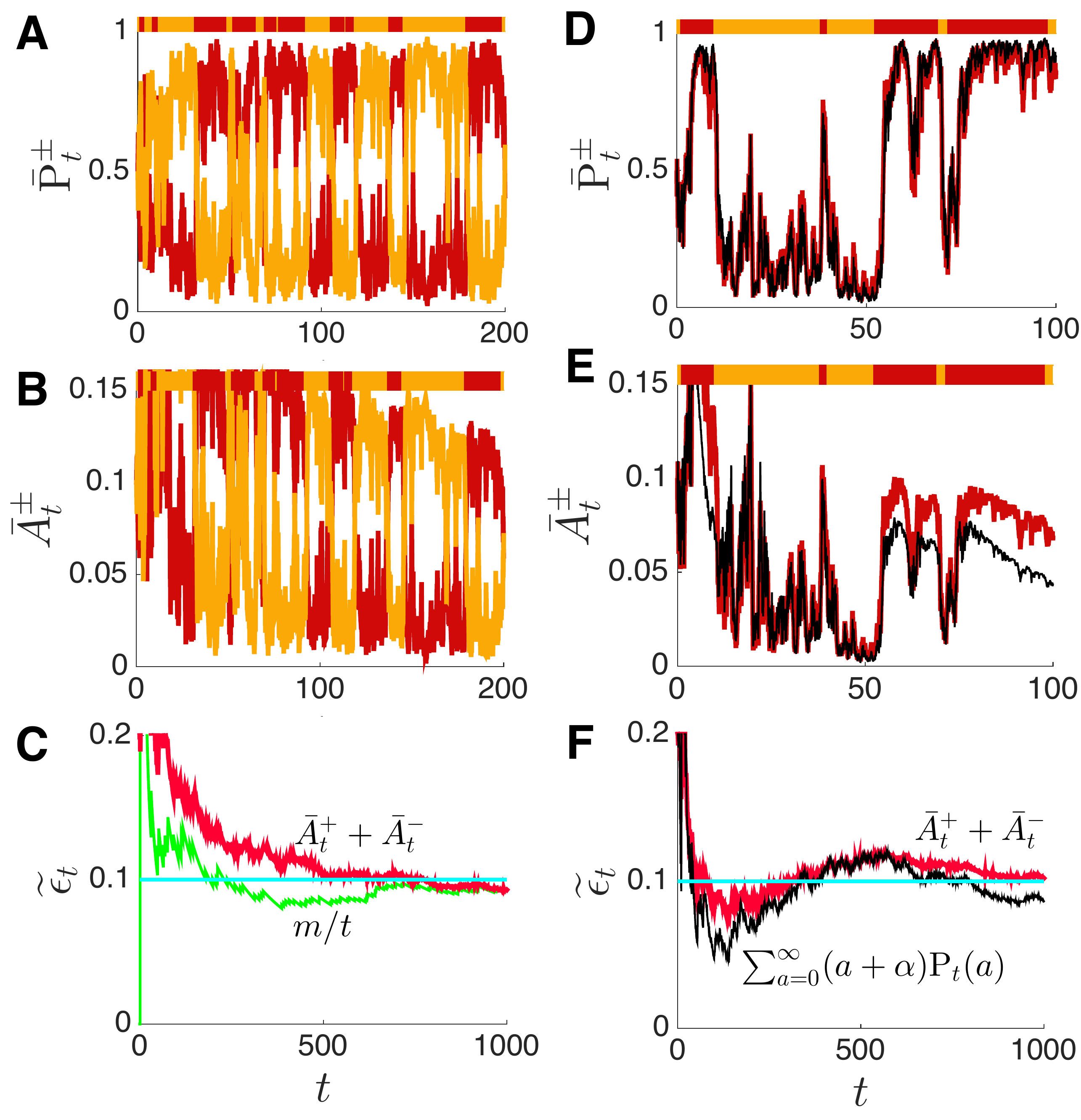} \end{center}
\vspace{-5mm}
\caption{The dynamics of the first two moments, as approximated by Eq.~(\ref{likeSDE}). (A) The probabilities $\bar{\PP}_t^{\pm}$  track the present state of the environment (bar above), switching with rate $\ep = 0.1$, and approach the stationary densities around the equilibria determined by the dichotomous drift terms $g^{\pm}_t$. (B) The first moments, $\bar{A}^{\pm}_t,$ also switch with the environmental state and alternate between the neighborhoods of two points. (C) The sum $\widetilde{\ep}_t : = \bar{A}^+_t + \bar{A}_t^-$ provides a running point estimate of the environmental transition rate, $\ep,$ as shown in Eq.~(\ref{epmomest}). The estimate is determined by the actual changepoints, and noisily tracks $m/t$, where $m$ is the actual number of changepoints.
%\Adcomment{why not $a$?}\Acomment{I think with $m$ we refer to the actual changepoint count and $a$ is a variable that ``spans'' $\mathbb{Z}_{\geq 0}$.}
(D,E,F) Same as A,B,C, but the moment simulations are compared with numerical simulations of the full system of SDEs given by Eq.~(\ref{fulllikeSDE}).
%\Kcomment{How can you simulate infinitely many SDEs?} \Zcomment{Removed infinite, and explained in the Appendix.}
(D) Thick red line is $\bar{\PP}_t^{+}$ from Eq.~(\ref{likeSDE}) and thin black line is $\sum_{a=0}^{\infty} \PP_t^{+}(a)$ using Eq.~(\ref{fulllikeSDE}); (E) Thick red line is $\bar{A}_t^{+}$ from Eq.~(\ref{likeSDE}) and thin black line is $\sum_{a=0}^{\infty} (a + \alpha) \PP_t^{+}(a)$ using Eq.~(\ref{fulllikeSDE});  (F) Estimates of $\widetilde{\ep}_t$ using Eq.~(\ref{likeSDE}) and Eq.~(\ref{fulllikeSDE}). Details about the simulation method, initial conditions, and parameters are provided in Appendix~\ref{simappend}.}
\label{fig5}
\end{figure}

Applying cumulant-neglect to the second moment, $\bar{b}_t^{\pm},$ in Eqs.~(\ref{momsde1}--\ref{momsde3}), using the change of variables, $\bar{A}_t^{\pm} = \bar{a}_t^{\pm}/(t+ \beta)$, and the fact that $\d \bar{A}^{\pm}_t = \left[ (t+ \beta) \d \bar{a}_t^{\pm} - \bar{a}_t^{\pm} \d t \right]/(t+\beta)^2,$ we obtain a closed system of equations for the zeroth and first moments,
\begin{subequations} \label{likeSDE}
\begin{align}
\d \bar{\PP}^{\pm}_t &= \bar{\PP}^{\pm}_t  \left[ \left( g^{\pm}_t + \frac{1}{2} \right) \d t + \d W^{\pm} \right] + \left[ \bar{A}^{\mp}_t - \bar{A}^{\pm}_t \right] \d t \label{likeSDEa} \\
\d \bar{A}^{\pm}_t &=  \bar{A}^{\pm}_t \left[  \left( g^{\pm}_t + \frac{1}{2} \right) \d t + \d W^{\pm} \right] +  \left( \bar{A}^{\mp}_t - \bar{A}^{\pm}_t \right) \left( \frac{1}{t + \beta} + \bar{A}^{\mp}_t + \bar{A}^{\pm}_t \right) \d t. \label{likeSDEb}
\end{align}
\end{subequations}
Here initial conditions are given by $\bar{\PP}_0^{\pm} : = \PP_0(H^{\pm})$ and $\displaystyle \bar{A}_0^{\pm} = \alpha \PP_0(H^{\pm})/ \beta$. We show in Appendix~\ref{momentconsist} that Eq.~(\ref{likeSDE}) is also consistent with Eq.~(\ref{sdeknown}), which holds in the case of  two states and known rate $\epsilon$. Trajectories of Eq.~(\ref{likeSDE}) are shown in Fig. \ref{fig5}. Note that both $\bar{\PP}^{\pm}_t$ and $\bar{A}^{\pm}_t$ tend to increase when $g^{\pm}_t$ is the maximal drift rate, {\em i.e.} when $H^{\pm}$ is the true environmental state. Thus, we expect that when $\bar{\PP}^{\pm}$ is high (close to unity) then $\bar{A}^{\pm}_t$ will tend to be larger than $\bar{A}^{\mp}_t$. 

%\Kcomment{The following paragraph is crucial, as it really motivates the implementation as well. I therefore separated it out. Maybe flesh out a bit more?}  
Immediately after a changepoint (where the maximal drift rate $g^{\pm}_t$ changes), there is an additional contribution to the increase of $\bar{A}^{\pm}_t$ due to the $(\bar{A}^{\mp}_t - \bar{A}^{\pm}_t)$ term. It is this brief burst of additional input to the subsequently dominant variable that generates the counting process, enumerating the changepoints. For instance, when a $H^+ \mapsto H^-$ switch occurs, an increase in $\bar{A}^-_t$ will temporarily be driven both by the drift term, $g^-_t,$ and the nonlinear term involving $(\bar{A}^+_t - \bar{A}^-_t)$. The burst of input generated by the nonlinear term in Eq.~(\ref{likeSDEb}) has an amplitude that decays nonautonomously with time. In fact, it can be shown that when the signal-to-noise ratio of the system is quite high, the variables $\bar{A}^{\pm}_t \approx (m+\alpha)/(t+\beta)$, which is effectively the true changepoint count $m$ divided by elapsed time as modified by the prior.

We can also obtain a point-estimate of the transition rate of the environment, which we define as $\widetilde{\epsilon}_t:= \bar{A}_t^+ + \bar{A}_t^-$, since the following relations hold:
\begin{align}
\bar{A}^+_t + \bar{A}^-_t &= \frac{1}{t + \beta} \sum_{a \in \mathbb{Z}_{\geq 0}} (a + \alpha) \left[  \PP(H^+, a|\xi_t) +\PP(H^-, a|\xi_t) \right] \nonumber \\
&= \sum_{a \in \mathbb{Z}_{\geq 0}} \frac{(a + \alpha)}{t + \beta}  \PP(a|\xi_t) = \int_0^{\infty} \epsilon \sum_{a \in \mathbb{Z}_{\geq 0}}  \PP(\epsilon|a) \PP(a|\xi_t)  \d \epsilon.  \label{epmomest}
\end{align}
This estimate is an average over the distribution of possible changepoint counts, $a,$ given the observations, $\xi_t.$ Here $\PP(\epsilon|a)$ is  a Gamma distribution with parameters $\alpha$ and $\beta$. In Fig. \ref{fig5}C we compare this approximation, $\widetilde{\epsilon}_t$, with the true change rate $\epsilon$ and the running estimate $m/t$, obtained from the actual number of changepoints, $m$. 

In Fig. \ref{fig5}D,E these approximations are compared to Eq.~(\ref{fulllikeSDE}), the full SDE  giving the distribution over all changepoint counts, $a$. Notice that the first moments $\bar{A}_t^{\pm}$ are overestimates of the true average, $\sum_{a=0}^{\infty} (a + \alpha) \PP_t^{\pm}(a)/(t+\beta),$ obtained from Eq.~(\ref{fulllikeSDE}). We expect this is due to the fact that the moment equations, Eq.~(\ref{likeSDE}), tend to overcount the number of changepoints. Fluctuations lead to an increase in the number of events whereby $\bar{A}_t^+$ and $\bar{A}_t^-$ exchange dominance $\left(\bar{A}_t^+ = \bar{A}_t^-\right)$, which will lead to a burst of input to one of the variables $\bar{A}_t^{\pm}$. As a consequence, the transition rate tends to be overestimated by Eq.~(\ref{likeSDE}) compared to Eq.~(\ref{fulllikeSDE}).

%Lastly, we can also consider the extreme case of small noise ($\langle W^{ij} \rangle  = \Sigma^{ij}(t) t$, $\Sigma^{ij} \ll 1$). In the singular limit $\Sigma^{ij} \to 0, \forall t$, we can reduce Eq.~(\ref{likeSDE}) considerably since ${\mathcal P}^{\pm}(t) \in \{ 0,1 \}$ for $t>0$. This reduces Eq.~(\ref{likeSDEa}) to ${\mathcal P}^{\pm}(t) = H(g^{\pm}(t) - g^{\mp}(t))$; e.g., ${\mathcal P}^+(t)=1$ when $g^{+}(t) > g^-(t)$ so ${\mathcal P}^-(t) = 0$. Similarly, when $g^{+}(t) > g^-(t)$, ${\mathcal A}^-(t) = 0$ and
%\begin{align}
%\frac{\d {\mc A}^{+}(t)}{\d t} &= {\mc A}^{+}(t)\left[  g^{+}(t) + \frac{1}{2}  -  \frac{1}{t + \beta} - {\mc A}^{+} (t) \right], 
%\end{align}
%so we can apply the fact that ${\mc P}^+(t)=1$ and $\d {\mc P}^+(t) =0$

In sum, while the inference  approximation given by Eq.~(\ref{likeSDE})  does not provide an  estimate of the variance, it does provide  insight into the computations needed to estimate the changepoint count and transition probability. Transitions increment the running estimate of the changepoint count, and this increment decays over time, inversely with the total observation time $t$. Similar equations for the moments can be obtained in the case of asymmetric transition rates, or more than two choices using Eq.~\eqref{likesdeasym} and Eq.~\eqref{eq:SDEmulti} respectively, although we omit their derivations here.

\section{Learning transition rate in neural populations with plasticity}
\label{neuralpop}
Models of decision making often consist of mutually inhibitory neural populations with finely tuned synaptic weights~\citep{machens05,mcmillen06,wong07}. For instance,  many models of evidence integration in two alternative choice tasks
assume that synaptic connectivity is tuned so that the full system exhibits line attractor dynamics  in the absence of inputs. Such networks integrate inputs perfectly, and maintain this integrated information in memory after the inputs are removed. However,  in changing environments optimal evidence integration should be leaky, since older information becomes irrelevant for the present decision~\citep{deneve08,glaze15}. 

We previously showed that optimal integration in changing environments can be implemented by mutually excitatory neural populations~\citep{velizcuba15}. Instead of a line attractor, the resulting dynamical systems contain globally attracting fixed points.
Such leaky integrators maintain a limited memory of their inputs on a  timescale determined by the frequency of environmental changes. However, in this previous work we assumed that the rates of the environmental changes were known to the observer.
Here, we show that when these rates are not known a priori, a plastic neuronal network is capable of learning and implicitly representing them through coupling strengths between neural populations.

\noindent
{\bf Symmetric environment.} We begin with Eq.~(\ref{likeSDE}), the leading order equations for the likelihood  $\bar{\PP}^{\pm}_t$ and change rate variables $\bar{A}^{\pm}_t$  derived in Section~\ref{momhier}.
%\Kcomment{I think there is a change in notation here -- these should be roman, not script.  Replace with $\bar{\PP}_t$?}
We interpret the likelihoods as neural population activity variables $u^{\pm}_t : = \bar{\PP}^{\pm}_t$, reflecting a common modeling assumption that two populations  receive separate streams of input associated with evidence for either choice~\citep{bogacz06}. Next, we define a new variable $w^{\pm}_t: = \bar{A}^{\pm}_t/ \bar{\PP}^{\pm}_t$, which represents the synaptic weight between these neural populations (See Fig. \ref{roadmap}C). In particular, $w^{\pm}_t$ represents the strength of coupling from $u^{\pm}_t$ to $u^{\mp}_t$ as well as the local inhibitory coupling within $u^{\pm}_t$. Applying this change of variables to Eq.~(\ref{likeSDE}), we derive a set of equations for the population activities, $u^{\pm}_t,$ and their associated synaptic weights, $w^{\pm}_t$:
%\Kcomment{Rewrote this in standard SDE form.}
\begin{subequations} \label{rateSDE} 
\begin{align}
\d u_t^{\pm} &= u_t^{\pm} \left[ \left( g^{\pm}_t + \frac{1}{2} \right) \d t + \d W_t^{\pm} \right] + \left[ w_t^{\mp} u_t^{\mp} - w_t^{\pm} u_t^{\pm} \right] \d t \label{rateSDEa} \\
\d w_t^{\pm}  &= - \left[ w_t^{\pm} + w_t^{\mp} - \frac{w_t^{\mp}}{u_t^{\pm}} \right] \left[ \frac{1}{t + \beta} +  u_t^{\mp} (w_t^{\mp} - w_t^{\pm}) \right] \d t .  \label{rateSDEb}
\end{align}
\end{subequations}
This is a neural population model with a rate-correlation based plasticity rule~\citep{miller1994,Pfister2006}. Each neural population $u_t^{\pm}$ impacts its neighboring population via mutual excitation as in \cite{velizcuba15}. Note that each population in Eq.~(\ref{rateSDE}) is also locally affected by self-inhibition, whose weight evolves according to the same dynamics as the excitatory weights between populations. We expect that such dynamics could arise as the quasi-static approximation of a network with separate excitatory and inhibitory populations, but we save such analyses for future work. We can interpret the non-autonomous term, $1/(t+ \beta)$, as modeling the dynamics of a chemical  agent involved in the plasticity process whose availability decays over time. Simple chemical degradation kinetics for a concentration $C_t$ yield such a function when
\begin{align}
\d C_t = -C^2 \d t, \ \ \ C(0) = 1/\beta \ \ \ \  \Rightarrow \ \ \ \ \ C_t = \frac{1}{t + \beta}.  \label{chemeq}
\end{align}
We briefly analyze the model, Eq.~(\ref{rateSDE}), by considering the limit of no observation-noise. That is, we assume $g^{\pm}_t \to \pm \infty$ when $H_t = H^{+}$ and $\Sigma^{++} \to 0$, where $\langle W^{+}_t W^{+}_t \rangle = \Sigma^{++}_t \cdot t$, and analogous relations hold when $H_t = H^{-}$. As a result, when $H_t = H^+$, then $u^+_t \to 1$ and $u^-_t \to 0$, which we demonstrate in Appendix \ref{lownoiselim}. Plugging the expressions $u^+_t = 1$ and $u^-_t = 0$ into Eq.~(\ref{rateSDEb}) for $w^+_t$, we find
\begin{align}
\d w^+_t = - \left[ \frac{1}{t + \beta} \right] w^+_t \d t. \label{wpdom}
\end{align}
Next, we write Eq.~(\ref{rateSDEb}) for $w^-_t$ in the form
\begin{align*}
u^-_t \d w_t^{-}  &= - \left[ u^-_t w_t^{-} + u_t^- w_t^{+} - w^+_t \right] \left[ \frac{1}{t + \beta} +  u_t^{+} (w_t^{+} - w_t^{-}) \right] \d t,
\end{align*}
so by plugging in $u^+_t = 1$ and $u^-_t = 0$, we find $0 = w^+_t \left[ \frac{1}{t+ \beta} + w^+_t - w^-_t \right]$, which, when $w^+_t \neq 0$, simplifies to
\begin{align}
w^-_t = w^+_t + \frac{1}{t + \beta}. \label{wmsup}
\end{align}
An analogous pair of equations holds when $H_t = H^-$ and thus $u^-_t \to 1$ and $u^+_t \to 0$. Solving Eq.~(\ref{wpdom},\ref{wmsup}) and their $H_t = H^-$ counterparts iteratively, we find that in the limit of no observation-noise (e.g., $g^{\pm}_t \to \pm \infty$ when $H_t = H^{\pm}$),
\begin{align}
w^{\pm}_t = \frac{w_0 \beta + m}{t + \beta}, \ \ \ \  \  \ \ w^{\mp}_t = \frac{w_0 \beta + m + 1}{t + \beta}, \label{solveforw}
\end{align}
where $H_t = H^{\pm}$ and $w_0 : = w^{j}(0)$ for $H(0) = H^j$, so $w_0$ constitutes the initial estimate of the change rate of the environment. Here, $m$ is the number of changepoints in the time series $H_t$ during the time interval $[0,t]$. Eq.~(\ref{solveforw}) can be re-expressed in the form of a rate-based plasticity rule
\begin{align}
\d w^{\pm}_t  = \left[ \delta( u^+_t - u^-_t )  - w^{\pm}_t \right] \cdot C_t \, {\rm d} t,  \label{plastic1}
\end{align}
where $\delta (u)$ is the Dirac delta distribution, along with Eq.~(\ref{chemeq}) for the concentration decay of the  agent $C_t$. Note, that the non-negative term $\delta (u^+_t - u^-_t)$ in Eq.~(\ref{plastic1}) results in long term potentiation (LTP) of both synaptic weights, $w^{\pm}_t,$ whenever the neural activities, $u^{\pm}_t,$ are both high, {\em i.e.} when their values cross  at $u^{\pm}_t = 0.5$. During such changes, the weights $w^{\pm}_t$ are incremented. Outside of these transient switching epochs, there is long term depression (LTD) of the synaptic weights $w^{\pm}_t$ modeled by the term $(- w^{\pm}_t)$.
%In the noise-free limit the solution of these equations are given in Eq.~(\ref{solveforw}).  \Kcomment{Last sentence repeats what was said above.  Remove?}

\begin{figure}
\begin{center} \includegraphics[width=14.5cm]{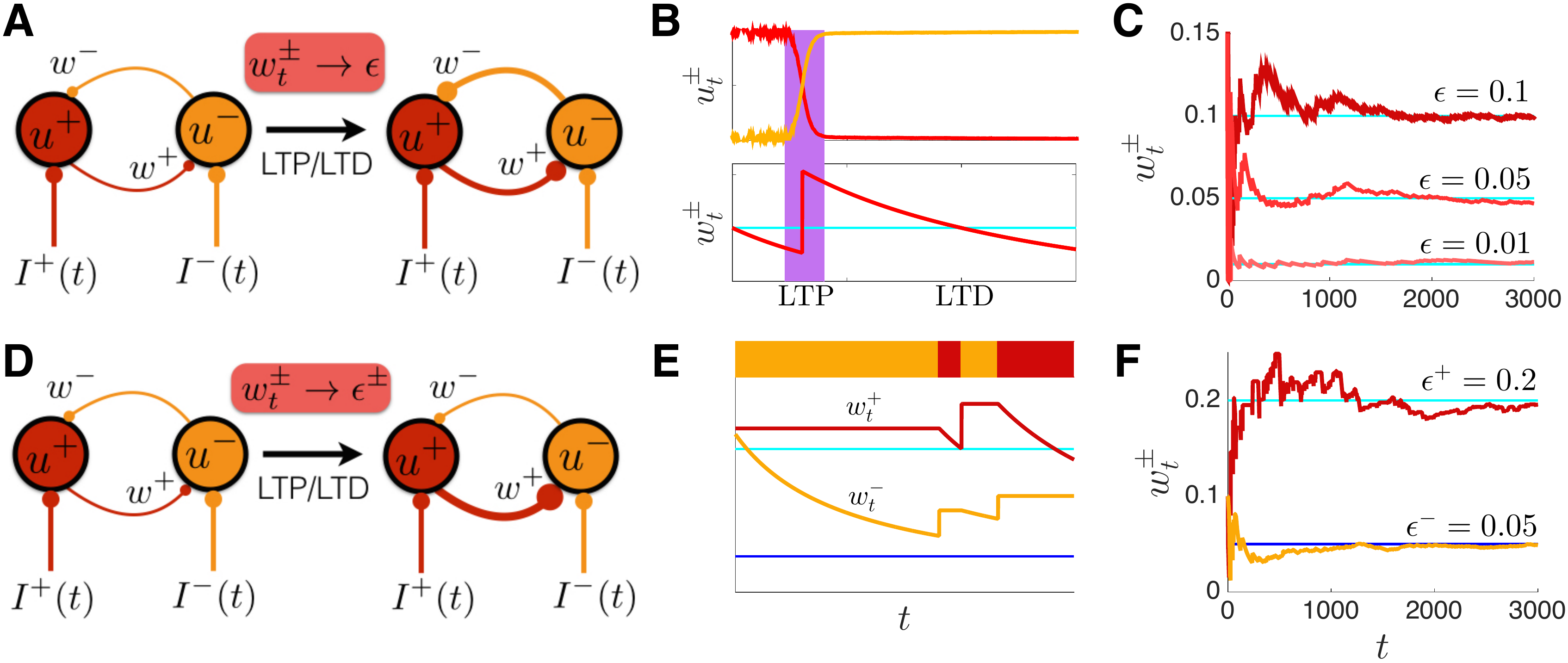} \end{center}
\caption{Neural network model with plasticity, inferring the current state $H_t$ and rates $\ep^{\pm}$ of environmental change. (A) Schematic showing the synaptic weight $w^{\pm}$ from neural population $u^{\pm} \mapsto u^{\mp}$ evolving through long term potentiation (LTP) and long term depression (LTD) to match the environment's rate of change, $\ep^{\pm} : = \ep$. (B) When the neural populations exchange dominance, their activity levels $u^{\pm}$ are both transiently high. As a result, both synaptic weights, $w^{\pm},$ increase via LTP. When only one population is active, both weights decay via LTD, as described by Eq.~(\ref{plasticnet2}). (C) Inference of the rate, $\ep,$ via long term plasticity of the weights for $\ep = 0.01,0.05,0.1$. Though the signal-to-noise ratio is finite (See Appendix \ref{simappend}), the weights in the network described by Eq.~(\ref{plasticnet}) converge to the actual rate, $\ep$. (D) Schematic showing the evolution of  weights $w^{\pm}_t$ when rates are asymmetric, $\ep^+ > \ep^-$, so that $w^+_{\infty} > w^-_{\infty}$. The network is described by Eq.~(\ref{asymweight2}). (E) Only the weight $w_t^{\pm}$ decays through LTD when population $u^{\pm}_t$ is active, and only the weight $w_t^{\pm}$ is potentiated through LTP when dominance switches from $u^{\pm}_t$ to $u^{\mp}_t$, as in Eq.~(\ref{asymweight2}).  (F) Network weights $w^{\pm}_t$ converge to the asymmetric rates, $\ep^{\pm}$. See Appendix \ref{simappend} for  details about the simulations.}
\label{fig6}
\end{figure}

We schematize (Fig. \ref{fig6}A) and simulate (Fig. \ref{fig6}B,C) the resulting plastic neural population network:
\begin{subequations} \label{plasticnet}
\begin{align}
\d u^{\pm}_t &= u^{\pm}_t \left[ I^{\pm}_t\d t + \d W^{\pm}_t \right] + \left[ w^{\mp}_t u^{\mp}_t - w^{\pm}_t u^{\pm}_t \right] \d t \label{plasticnet1} \\
\d w^{\pm}_t&= \left[ \delta( u^+_t - u^-_t )  - w^{\pm}_t \right] \cdot C_t \, \d t. \label{plasticnet2}
\end{align}
\end{subequations}
The constant $1/2$ has been absorbed into the population input, so that $I^{\pm}_t = g^{\pm}_t + 1/2$. It is important to note that Eq.~(\ref{plasticnet}) only performs optimal inference in the limit of no observation-noise. Perturbing away from this case, we expect the performance to be sub-optimal. However, as can be seen in Fig. \ref{fig6}C, the correct change rate is approximated reasonably well. 

Many previous neural population models of evidence accumulation assume that neural activity represents log-probabilities or log-likelihoods~\citep{bogacz06,mcmillen06,velizcuba15}. In our model the rate variables, $u^{\pm}_t,$ represent the probability that the environment is in state $H^{\pm}$. This particular form of the population model leads dynamical equations which are consistent with an accepted rate-correlation based plasticity rule~\citep{miller1994,Pfister2006}. Using log probabilities would lead to  models that contain exponential functions of the rate~\citep{velizcuba15}, which are less common. In addition, since probabilities can assume a finite range of values, we required that $u_t^{\pm} \in [0,1]$. Using log probabilities would require that we use a semi-infinite range, $(- \infty, 0],$ or that we truncate. Note also that the inputs $I^{\pm}_t$ and noise $\d W^{\pm}_t$ are gain-modulated using the population rates $u^{\pm}_t$. Gain-modulating circuits have been identified in many sensory areas~\citep{salinas96}, and recent studies suggest evidence-accumulating circuits may also modulate input in a history-dependent way~\citep{wyart12}.

Eq.~(\ref{plasticnet}), thus models evidence accumulation in a symmetrically changing environment when the change rate, $\ep,$ is not known a priori. The model is based on the recursive equation for the joint probability of the environmental state, $H^{\pm},$ and changepoint count, $a$,  derived in Section \ref{S:sym_2state_uknown}. We obtained a tractable model by first passing to the continuum limit,  and then applying a moment closure approximation to reduce the dimension of the resulting equations. Obtaining the low-dimensional approximation in Eq.~(\ref{likeSDE}) was crucial to obtaining a neural population model that approximates state inference.  We next extend this model to the case of asymmetric rates of change. \\
\vspace{-4mm}

\noindent
{\bf Asymmetric environment.} The continuum limit of the inference process in an asymmetric environment, Eq.~(\ref{likesdeasym}), provides  several pieces of information we can use to identify an approximate neural population model. First, under the assumption of large signal-to-noise ratios, the synaptic weights should evolve to reflect the number of detected changepoints, rescaled by the amount of time spent in each state
\begin{align}
w^+_t = \frac{w^+_0 \beta^+ + m^+_t}{t^+ + \beta^+}, \ \ \ \ \  \  \ w^-_t = \frac{w^-_0 \beta^- + m^-_t}{t^- + \beta^-},  \label{asymweights}
\end{align}
where $w_0^{\pm} : = w^{\pm}(0)$ are the network's initial estimates of the change rates $\ep^{\pm}$, $m^{\pm}$ is the true number of changepoints $H^{\pm} \mapsto H^{\mp}$ during the time interval $[0,t]$, and $t^{\pm}$ is the total length of time spent in the state $H^{\pm}$.\footnote{We use the notation $H^{\pm}$ for the two states here, for convenience and consistency with Eq.~(\ref{plasticnet}). Similarly, we use $\ep^{\pm}$ and $t^{\pm}$ rather than the numerical notation of the asymmetric case in Section \ref{sde2state}.} Second, the flux term in Eq.~(\ref{likesdeasym}) indicates that a memory process is needed to store the estimated time $t^{\pm}$ spent in each state $H^{\pm}$. This can be accomplished by modifying  Eq.~(\ref{chemeq}) for the plasticity agent, so that  each  $C^{\pm}_t$  decays only when the neural population of origin, $u^{\pm}_t$, is active. Thus we obtain the pair of equations:
\begin{align*}
\d C^{\pm}_t &= - H(u^{\pm}_t - \theta) \left[ C^{\pm}_t \right]^2. 
%\label{asymchem}
\end{align*}
Expressing Eq.~(\ref{asymweights}) as a system of equations for the synaptic weights, $w^{\pm}_t,$ yields:
\begin{align}
\d w^{\pm}_t &= H(u^{\pm}_{t - \tau} - \theta) \left[ \delta (u^{+}_t - u^{-}_t) - w^{\pm}_t \right] \cdot C^{\pm}_t \, \d t.  \label{asymweight2}
\end{align}
Here the function $H(u^{\pm}(t-\tau)-\theta)$ for $\theta \geq 0.5,$ and $\tau >0$ enforces the requirement that the population $u^{\pm}_t$ must have a high rate of activity prior to the LTP event. Thus, to learn asymmetric weights, there should be a small delay $\tau$ accounting for the time it takes for the presynaptic firing rate to trigger the plasticity process~\citep{gutig2003}. We demonstrate the performance of the network whose weights evolve according to Eq.~(\ref{asymweight2}) in Fig. \ref{fig6}D,E. Note, the network with weights evolving according to Eq.~(\ref{asymweight2}) can still infer symmetric transition rates $\ep^{\pm} = \ep$, but it will do so at half the rate of the network Eq.~(\ref{plasticnet}). This is due to the fact that Eq.~(\ref{asymweight2}) counts changepoints and dwell times of each state $H^{\pm}$ separately.

We have thus shown that the recursive update equations for the state probability in a dynamic environment lead to plausible neural network models that approximate the same inference. Previous neural network models of decision making have tended to interpret population rates as a representation of posterior probability~\citep{bogacz06,beck08}. We have shown that the synaptic weight between populations can represent the change rate of the environment. As a result, standard rate-correlation models of plasticity can be used to implement the change rate inference process.

\section{Discussion}

Evidence integration models have a long history in neuroscience~\citep{ratcliff08}. 
These normative models conform with behavioral observations across species~\citep{brunton13},
and have been used to explain the neural activity that underpins decisions~\citep{gold07}. However,
animals make decisions in an environment that is seldom static~\citep{portugues09}. The relevance of available information, the accessibility, and the payoff of different choices can all fluctuate. It is thus important to extend evidence accumulation models to such cases.

We have shown how ideal observers accumulate evidence to make decisions when there are multiple, discrete choices, and the correct choice changes in time. We assumed that the rates of transition between environmental states are initially unknown to the observer.  An ideal observer must therefore integrate information from measurements to concurrently estimate both the transition rates and the current state of the environment. Importantly, these two inference processes are coupled: Knowledge of the rate allows the observer to appropriately discount older information to infer the present state, while knowledge of transitions between states is in turn necessary to infer the rate. 

Inference when all transition rates are identical is straightforward to implement in resulting models. An ideal observer only needs to track the probability of the environmental state and the total changepoint count, regardless of the states between which the change occurred.
%\Adcomment{as personal curiosity, are there cases where the animal may be interested in the rate but not in the current state? If yes, would it be possible to infer the rate without using $H_n$?}
However, when the transition rates are asymmetric, the resulting models are more complex. In this case, an ideal observer must estimate a matrix of changepoint counts, distinguished by the starting and ending states. The number of possible matrices grows polynomially with the number of observations. This computation is difficult to implement, and we do not suggest that 
animals make inferences about environmental variability in this way. 
However, understanding the ideal inference process allowed us to identify its most important features.
In turn, we derived tractable approximations and plausible neural implementations, whose performance compared well with that of an ideal observer (Fig. \ref{fig5}D,E,F). We believe humans and other animals do generally implement approximate strategies when they need to infer such rates~\citep{lange09}.
Ideal observer models  allow us to understand what inferences can be made with the available information, which assumptions of the observer are important (e.g., assuming an incorrect transition rate does not always have a  large impact on performance), and how such inferences could be approximated in networks of the brain and other biological computers.

In many naturally occurring decisions like foraging, mate selection, and home-site choice, animals simply need to identify the best alternative rather than the rate of environmental change~\citep{johnson13}. Therefore, rapid approximations, or a guess of the environmental change rate may provide better initial performance than learning the rate, which could be slow. Moreover, it appears that when measurements are noisy, rates cannot be learned precisely even in the limit of infinite observations.  Thus, learning the rate may only improve performance when noise is too high for single measurements to determine the correct alternative, but sufficiently low to make rate inference possible. It is within this range of parameters that we expect to be able to distinguish the performance of our normative model from that of different approximations. We plan to carry out such a systematic comparison of model performance in future work. There is evidence that humans adjust their rate of evidence-discounting, based on the actual change rate of the environment~\citep{glaze15}. However, further psychophysical studies are needed to identify whether subjects use heuristic strategies to learn or something close to the normative models we derived here.

A number of related models have been developed previously~\citep{wilson10,adams07}.  The present model is somewhat different, as a finite number of choices implies that the present environmental state is dependent on the previous state. As a result, we found it was more efficient to implement an update equation that estimated the present environmental state and the changepoints, rather than the time in the present state. 

%\Kcomment{I suggest removing following paragraph.  We already said this above.}
%Because our models take the form of one-step, iterative equations, it is straightforward to derive the continuum limit, assuming observations are made with high frequency. An advantage of the continuum limit is that it allows one to employ standard methods in stochastic analysis~\citep{gardiner04}. For instance, we could use linearization to obtain approximate models whose performance could be compared with the exact model. Furthermore, we have used moment closure methods~\citep{socha07} to produce low-dimensional nonlinear approximations of the full models. These reduced models could give insight into computations animals may actually apply in dynamic environments.

Several of the assumptions we have made in this study could be modified to extend our analysis to more general situations. For instance, we have assumed that the observer's eventual choice does not affect the environment. However, in many natural situations changes in the environment are a consequence of the observer's actions~\citep{Cisek:2014ei}. In more realistic situations it is likely that there is a sequence of actions leading to an ultimate decision, and each action can influence the information available to the observer. An animal making a foraging decision in a group collects more evidence once it moves toward a particular food patch, but it may also draw other members with it, changing the subsequent availability of food there~\citep{petit09}. Thus including a sequence of actions, and their impact on the available information and the environment would be necessary in a realistic model. Another possibility is that changes to the environment are non-Markovian and/or involve multiple timescales. Extending our ideal observer models to estimate such change statistics might require derivation of multi-step update equations. In such cases, we expect the truncations we have applied in this work would be useful for identifying tractable approximations of the optimal inference process.

Optimal models of evidence accumulation are useful both as baselines to compare to performance in psychophysical experiments, and starting points for identifying plausible neuronal network implementations. Our core contribution here has been to present a general model of evidence accumulation in a dynamic environment, when an observer has no prior knowledge of the rate of change. An unavoidable feature of these models is that the number of variables the observer must track grows as more observations are made, and growth is more rapid in asymmetric environments with multiple environmental states. This motivated our development of continuum approximations and low-dimensional moment equations for the optimal models, which suggest more plausible neural computations. We hope this work will foster future theoretical studies that will extend this framework, as well as experiments that could validate the models herein. To fully understand the neural mechanisms of evidence accumulation, we must account for the wide variety of conditions that organisms encounter when making decisions.

\subsection*{Acknowledgments} Funding was provided by NSF-DMS-1517629 (AER, KJ, and ZPK); NSF-DMS-1311755 (ZPK); and NSF/NIGMS-R01GM104974 (KJ).

\section*{Appendix} 
\subsection{Two-state system with unknown symmetric rate}
\label{2stateStart}
We show how to derive Eq.~\eqref{jupdate} from the main text.
Bayes' rule and the law of total probability first yield:
\begin{align*}
{\rm P}_n(H_n,a_n) = \frac{1}{{\rm P}(\xi_{1:n})} \sum_{H_{n-1} = H^\pm} \sum_{a_{n-1} = 0}^{n-2} 
					{\rm P}(\xi_{1:n}|H_n, H_{n-1}, a_n , a_{n-1}) {\rm P}(H_n, H_{n-1}, a_n, a_{n-1}).
%					\label{marg1}
\end{align*}
Using the conditional independence of observations, 
\begin{align*}
 {\rm P}(\xi_{1:n}|H_n, H_{n-1},a_n, a_{n-1}) = {\rm P}(\xi_n | H_n) {\rm P}(\xi_{1:n-1}|H_{n-1},a_{n-1}),
\end{align*}
we find that,
\begin{align*}
%\label{E:intermediate}
{\rm P}_n(H_n,a_n) = \frac{1}{{\rm P}(\xi_{1:n})} \sum_{H_{n-1} = H^\pm} \sum_{a_{n-1} = 0}^{n-2} 
				{\rm P}(\xi_n | H_n) {\rm P}(\xi_{1:n-1}|H_{n-1},a_{n-1}) {\rm P}(H_n, H_{n-1}, a_n, a_{n-1}).
\end{align*}
Furthermore, we can use the definition of conditional probability to write,
\begin{align*}
{\rm P}(H_n, H_{n-1}, a_n, a_{n-1}) = {\rm P}(H_n, a_n|H_{n-1},a_{n-1}) {\rm P}(H_{n-1},a_{n-1}),
\end{align*}
and Bayes' rule also implies,
\begin{align*}
{\rm P}(\xi_{1:n-1}| H_{n-1},a_{n-1}) {\rm P}(H_{n-1},a_{n-1}) = {\rm P}_{n-1}(H_{n-1},a_{n-1}) {\rm P}(\xi_{1:n-1}).
\end{align*}
Hence, we derive Eq.~\eqref{jupdate} from the main text,
\begin{align*}
{\rm P}_n(H_n,a_n) = \frac{{\rm P}(\xi_{1:n-1})}{{\rm P}(\xi_{1:n})} {\rm P}(\xi_n | H_n) \sum_{H_{n-1} = H^\pm} 
			\sum_{a_{n-1}=0}^{n-2} {\rm P}_{n-1}(H_{n-1},a_{n-1}) {\rm P}(H_n, a_n|H_{n-1},a_{n-1}).
%			\label{jupdateApp}
\end{align*}

\subsection{Numerical methods for free response protocol}
\label{freemethod}
The free response protocol is simulated by evolving the update Eq.~(\ref{Hptwoup_newnotation}) and subsequently computing the log likelihood ratio $L_n : = \log (R_n)$ using Eq.~(\ref{post_odds_ratio}) at each timestep $n$. Each point along the curves in Fig. \ref{fig3}C corresponds to an average waiting time and average performance corresponding to a threshold value $\theta$ over 100,000 simulations. For each value of $\theta$, the simulation is terminated when $|L_n|>\theta$ and the choice is given by the sign of $L_n$. To avoid excessively long simulations, we removed any that lasted longer than $n=5000$, but we found changing this upper bound did not affect averages considerably. There were 400 values of $\theta$ chosen, discretizing the interval from $\theta =0$ to $\theta = 3.89$.

\subsection{Continuum limit for two states with asymmetric rates}
\label{climasym}

We begin by considering Eq.~(\ref{H1_asymupdate}), which provides an update of the probability of being in state $H^1$ after $n$ observations, given the specific changepoint matrix $\bs{a}$:
\begin{align}
\PP_n(H^1,\bs{a})&= {\mc F}_{n,\Delta t}^1 \left[ \left( 1 - \frac{a^{21} + 1}{1 + a^{21} + a^{11}} \right) \PP_{n-1}\left(H^1,\bs{a}-\bs{\delta}^{11} \right) \right. \label{Pn1asym} \\
& \hspace{3cm} \left. + \frac{a^{12}}{1 + a^{12} + a^{22}} \PP_{n-1} \left(H^2,\bs{a}-\bs{\delta}^{12}\right) \right],  \nonumber
\end{align}
where we have defined ${\mc F}_{n,\Delta t}^1 = \frac{\PP(\xi_{1:n-1})}{\PP(\xi_{1:n})}f^1_{\Delta t}(\xi_n)$. Subsequently, we divide by $\PP_{n-1}(H^1, \bs{a})$ and take the logarithm to find:
\begin{align*}
\Delta x_n^1(\bs{a}) &= \ln {\mc F}_{n, \Delta t}^1 + \ln \left[ \left( 1 - \frac{a^{21} + 1}{1 + a^{21} + a^{11}} \right) \e^{x_{n-1}^1 (\bs{a}-\bs{\delta}^{11} ) - x_{n-1}^1( \bs{a})}  \right. \\
& \hspace{4cm} \left. + \frac{a^{12}}{1 + a^{12} + a^{22}} \e^{x_{n-1}^2 (\bs{a}-\bs{\delta}^{12}) - x_{n-1}^1(\bs{a})} \right].
\end{align*}
Now, as we are taking the continuum limit, we consider  $\PP(H_{t + \Delta t} = H^{\pm} | H_t  = H^{\mp}) =  \ep^{ij} \Delta t + o ( \Delta t)$, where $0 \leq \ep^{ij} < \infty$. Given a $\mathbb{R}^{2 \times 2}$ transition rate matrix with off-diagonal entries $\ep^{ij} \Delta t + o ( \Delta t)$ and diagonal entries $1- \ep^{ij} \Delta t + o ( \Delta t)$, the expected changepoint and non-changepoint counts after $n$ timesteps will be $\langle a^{ij}_{\Delta t} \rangle = \Delta t \ep^{12} \ep^{21} n/(\ep^{12}+\ep^{21})$ and $\langle a^{ii}_{\Delta t} \rangle = \ep^{ij} n/(\ep^{12}+\ep^{21})$. Thus, while the changepoint counts $a^{ij}_{\Delta t}$ scale with $\Delta t$, the non-changepoint counts do not. In the continuum limit, we will choose a time $t := n \Delta t$ and take $\Delta t \to 0$ while keeping $t$ constant, so the number of timesteps diverges like $n = t/(\Delta t)$ for a fixed time $t$. While the expected changepoint counts $\langle a^{ij}_{\Delta t} \rangle$ thus remain fixed, the non-changepoint counts $\langle a^{ii}_{\Delta t} \rangle$ will diverge as $\left( \Delta t \right)^{-1}$, suggesting we should rescale non-changepoint counts to the absolute dwell time $t^i_{\Delta t} = \Delta t a^{ii}_{\Delta t}$. The expected value of the dwell times is then finite in this limit $\lim_{\Delta t \to 0} \langle t^{i}_{\Delta t} \rangle = \ep^{ij} t /(\ep^{12}+\ep^{21})$. Performing this change of variables, we then define the changepoint matrix as involving changepoint counts $a^{ij}$ on the off-diagonal and dwell times along the diagonal: $\bs{A} = \left( \begin{array}{cc} t^1 & a^{12} \\ a^{21} & t^2 \end{array} \right)$ so an increment of non-changepoint count $a^{ii}$ now takes the form $\bs{A} + \Delta t \bs{\delta}^{ii}$. As such, we can now expand:
\begin{align*}
\e^{x^i(\bs{a} - \bs{\delta}^{ii})} = \e^{x^i(\bs{A} - \Delta t \bs{\delta}^{ii})} = \e^{x^i(\bs{A})} - \Delta t\e^{x^i(\bs{A})} \frac{\pd x^i( \bs{A})}{\pd t^i} + {\mc O}\left( \left( \Delta t \right)^2 \right),
\end{align*}
via application of the chain rule and noting $ \d t^i/ \d a^{ii} = \Delta t$. Note, we cannot perform such an expansion in $\Delta t$ to $x^i(\bs{A} - \bs{\delta}^{ij})$, since perturbations to the matrix $\bs{A}$ in this case are ${\mc O}(1)$. Truncating Eq.~(\ref{Pn1asym}) to terms of ${\mc O}(\Delta t)$ and incorporating the Poisson-delta prior Eq.~(\ref{asymprior}), we find the discrete update equation becomes:
\begin{align*}
\Delta x_n^1(\bs{A}) = \ln {\mc F}_{n,\Delta t}^1 + \Delta t \cdot \left[ \frac{a^{12} + \alpha_2 -1}{t^2 + \beta_2} \e^{x_{n-1}^2 (\bs{A}-\bs{\delta}^{12}) - x_{n-1}^1(\bs{A})} - \frac{a^{21} + \alpha_1}{t^1 + \beta_1} - \frac{\pd x^1_{n-1}( \bs{A})}{\pd t^1}  \right].
\end{align*}
Lastly, upon taking the continuum limit $\Delta t \to  0$, we find that
\begin{align*}
\d x_t^1(\bs{A}) = \left[ g^1_t \d t + \d W^1_t \right] + \left[ \frac{a^{12} + \alpha_2 -1}{t^2 + \beta_2} \e^{x_t^2 (\bs{A}-\bs{\delta}^{12}) - x_t^1(\bs{A})} - \frac{a^{21} + \alpha_1}{t^1 + \beta_1} - \frac{\pd x^1_t( \bs{A})}{\pd t^1}  \right] \d t,
\end{align*}
where the statistics of the drift $g^1_t$ and noise $\d W^1_t$ are analogous to those given after Eq.~(\ref{eq:SDE}), only the transition rates of $H_t$ from $H^j \mapsto H^i$ are now $\ep^{ij}$. Note, due to the flux term $\displaystyle \frac{\pd x^1_t( \bs{A})}{\pd t^1}$ and continuum values for $t^j \in \mathbb{R}^*$, this is a stochastic partial differential equation (SPDE). An analogous SPDE can be derived for $x_t^2(\bs{A})$ in the same way.

\subsection{Continuum limit with multiple states and symmetric rates}
\label{manysym}

The derivation parallels that with two states.  
To obtain the continuum limit, we use the generalized version of Eq.~(\ref{jupdate}), 
\begin{align}
{\rm P}_n(H_n,a_n) = \frac{{\rm P}(\xi_{1:n-1})}{{\rm P}(\xi_{1:n})} {\rm P}(\xi_n | H_n) \sum_{H_{n-1}} \sum_{a_{n-1}=0}^{\infty} {\rm P}_{n-1}(H_{n-1},a_{n-1}) {\rm P}(H_n, a_n|H_{n-1},a_{n-1}). \label{jupdate_SDEmulti}
\end{align} 

Assuming again a Gamma prior on the transition rate, $\epsilon \sim Gamma(\alpha , \beta)$,  and following the derivations of Eq.~\eqref{transavg_multi} and \eqref{integral_betas}, we obtain
\begin{align}
{\rm P}(H_n, a_n|H_{n-1}, a_{n-1}) &  =
\left\{ \begin{array}{cc} 	
1-\Delta t \frac{a_{n}+\alpha}{t_{n-1}+\beta}
& H_n = H_{n-1} \ \& \ a_{n} = a_{n-1} 
\\
\Delta t \frac{a_{n}+\alpha-1}{(N-1)(t_{n-1}+\beta)}
& H_{n} \neq H_{n-1} \ \& \ a_{n} = a_{n-1}+1 
\\
 0 & {\rm otherwise.} \end{array} \right. 
\label{integral_betas_multi}
\end{align}

Using Eq.~(\ref{integral_betas_multi}) in Eq.~\eqref{jupdate_SDEmulti} yields
\begin{align*}
{\rm P}_{n}(H^i,a) = &  \frac{{\rm P}(\xi_{1:n-1})}{{\rm P}(\xi_{1:n})}f^i_{\Delta t}(\xi_n) 
\left[ 
\left( 1 - \Delta t \frac{a+\alpha}{t_{n-1}+\beta} \right) {\rm P}_{n-1}(H^i,a)\right. \qquad\qquad\qquad \notag \\
  & \left. \qquad\qquad\qquad + \Delta t\frac{a+\alpha-1}{(N-1)(t_{n-1}+\beta)} \sum_{j\neq i}{\rm P}_{n-1}(H^j,a-1) \right].
%  \label{Hptwoup_newnotation_SDEmulti} 
\end{align*}
%The initial and boundary conditions are given by: $\PP_n(H^i,a)=0$ for $a<0$, $\PP_0(H^i,a)=0$ for $a>0$, and $\PP_0(H^i,0)=\PP_0(H^i)$. To avoid discontinuities at $t=0$, we consider $\PP_0(a)$ as in the case of $N=2$. \Kcomment{Again, I suggest not mentioning the boundary conditions until later.}

%the initial conditions of Eq.~\eqref{Hptwoup_newnotation_SDEmulti} at $t=t_{n_0}=T$ to be 
%\begin{align*}
%\PP_{n_0}(H^i, a) = 
%% \frac{f^i(\xi_{n_0})}{P(\xi_{n_0})} P_{n_0}(a).
%{\rm P}_{n_0}(H^i|a) P_{n_0}(a).
%\end{align*}

Dividing by ${\rm P}_{n-1}(H^i,a)$, taking logarithms, and denoting $x^i_{t_n}(a):=\ln {\rm P}_{n}(H^i,a)$ we obtain
\begin{align*}
\Delta x^{i}_{t_n}(a) \propto \ln f^i_{\Delta t}(\xi_n) +
\ln \left[ 
 1 - \Delta t \frac{a+\alpha}{t_{n-1}+\beta} 
 + \Delta t \frac{a+\alpha-1}{(N-1)(t_{n-1}+\beta)} 
\sum_{j\neq i} e^{x^{j}_{t_{n-1}}(a-1)-x^{i}_{t_{n-1}}(a)}
 \right].
% \label{Hptwoup_newnotation_log_multi} 
\end{align*}
Using the approximation $\ln (1+z)\approx z$ valid for small $z$ yields
\begin{align*}
\Delta x^{i}_{t_n}(a) \propto \ln f^i_{\Delta t}(\xi_n) +
  \Delta t \left( 
 \frac{a+\alpha-1}{(N-1)(t_{n-1}+\beta)}  \sum_{j\neq i} e^{x^j_{t_{n-1}}(a-1)-x^i_{t_{n-1}}(a)}
 -\frac{a+\alpha}{t_{n-1}+\beta} \right).
% \label{log_discrete_multi} 
\end{align*}
Similar to the $N=2$ case, we may then take the continuum limit to yield Eq.~(\ref{eq:SDEmulti}).

\subsection{Consistency of the moment hierarchy equations}
\label{momentconsist}

We begin by taking the SDE given by Eq.~(\ref{sdeknown}) for $N=2$ states, and the known rate $\ep$, and changing variables to $\bar{\PP}_t^{\pm} = \e^{x_t^{\pm}}$, so
\begin{align*}
\d \bar{\PP}_t^{\pm} = \bar{\PP}_t^{\pm} \left[ g^{\pm}_t \d t + \d W^{\pm} \right] + \epsilon \left[ \bar{\PP}_t^{\mp} - \bar{\PP}_t^{\pm} \right] \d t.
%\label{likeknownsde}
\end{align*}
Furthermore, note that in the limit  $t \to \infty$, the ${\mc O}(1)$ terms and ${\mc O}(t^{-1})$ terms vanish in Eq.~(\ref{likeSDE}):
\begin{subequations} \label{limSDE}
\begin{align}
\d \overline{\PP}^{\pm}_t &= \overline{\PP}^{\pm}_t  \left[ \left( g^{\pm}_t + \frac{1}{2} \right) \d t + \d W^{\pm} \right] + \left[ \bar{A}^{\mp}_t - \bar{A}^{\pm}_t \right] \d t \label{limSDEa} \\
\d \bar{A}^{\pm}_t &=  \bar{A}^{\pm}_t \left[  \left( g^{\pm}_t + \frac{1}{2} \right) \d t + \d W^{\pm} \right] +  \left( \bar{A}^{\mp}_t - \bar{A}^{\pm}_t \right) \left( \bar{A}^{\mp}_t + \bar{A}^{\pm}_t \right) \d t. \label{limSDEb}
\end{align}
\end{subequations}
Therefore, in the event that $\bar{A}^{\pm}_t \to \epsilon \bar{\PP}_t^{\pm}$ in the long time limit ($t \to \infty$), we find the truncated system, Eq.~(\ref{limSDE}), becomes
\begin{subequations} \label{limSDE2}
\begin{align}
\d \bar{\PP}^{\pm}_t &= \bar{\PP}^{\pm}_t  \left[ \left( g^{\pm}_t + \frac{1}{2} \right) \d t + \d W^{\pm} \right] + \ep \cdot \left[ \bar{\PP}^{\mp}_t - \bar{\PP}^{\pm}_t \right] \d t \label{limSDE2a} \\
\ep \d \bar{\PP}^{\pm}_t &=  \ep \bar{\PP}^{\pm}_t \left[  \left( g^{\pm}_t + \frac{1}{2} \right) \d t + \d W^{\pm} \right] +  \ep^2 \cdot \left( \bar{\PP}^{\mp}_t - \bar{\PP}^{\pm}_t \right) \left( \bar{\PP}^{\mp}_t + \bar{\PP}^{\pm}_t \right) \d t. \label{limSDE2b}
\end{align}
\end{subequations}
Dividing by $\ep$ and noting that $\bar{\PP}_t^+ + \bar{\PP}_t^- = 1$,  Eq.~(\ref{limSDE2b}) becomes
\begin{equation*} 
\d \bar{\PP}^{\pm}_t = \bar{\PP}^{\pm}_t \left[  \left( g^{\pm}_t + \frac{1}{2} \right) \d t + \d W^{\pm} \right] +  \ep \cdot \left[ \bar{\PP}^{\mp}_t - \bar{\PP}^{\pm}_t \right]  \d t,
\end{equation*}
which is consistent with Eq.~\eqref{limSDE2a}, and 
indicates the truncated moment hierarchy is consistent with the case of known rates and two choices ($N=2$) in the SDE, Eq.~(\ref{sdeknown}).

\subsection{Noise-free limit of the neural population model}
\label{lownoiselim}

Consider the neural population Eq.~(\ref{rateSDEa}) for the evolution of $u^-_t$ in the event of environmental state $H_t = H^+$ and no observation-noise $f^{\pm}(\xi|H^+) = \delta(\xi - \xi^{\pm})$. As a result, the drift terms diverge $g^{\pm}_t : = \lim_{\Delta t \to 0} \frac{1}{\Delta t} {\rm E}\left[ \ln f^{\pm}_{\Delta t} (\xi)| H^+ \right] \to \pm \infty$ and the covariance matrix $\Sigma^{ij}_t := \lim_{\Delta t \to 0} \frac{1}{\Delta t} {\rm Cov} \left[ \ln f^{i}_{\Delta t} (\xi), \ln f^{j}_{\Delta t} (\xi) \right | H^+] \to 0$. Thus, the dominant terms on the right hand side of Eq.~(\ref{rateSDEa}) for $u^-_t$ come from the input so $\d u^-_t = - u^-_t |g^-| \d t$, and the population activity immediately decays to  $u^-_t = 0$. As a result, since $u^+_t + u_t- = 1$, we expect $u^+_t = 1$, when $H_t = H^+$.

%Consider the neural population Eq.~(\ref{rateSDEa}) for the evolution of $u^-_t$ in the event of low observation noise, and environmental state $H_t = H^+$. For example, if we assume normally distributed observations $\displaystyle f^{\pm}(\xi) = \frac{\e^{-(\xi \mp \mu)^2/(2 \sigma^2)}}{\sqrt{2 \pi \sigma^2}}$ where $\sigma \to 0$ and $\mu$ remains fixed, then when $H_t = H^+$, we have $g^+ := {\rm E}\left[ \ln f^+(\xi)| H^+ \right] = - \ln \left[ \sqrt{2 \pi} \sigma \right]$ and $g^- := {\rm E}\left[ \ln f^-(\xi)| H^+ \right] = - \mu^2/(2 \sigma^2) - \ln \left[ \sqrt{2 \pi} \sigma \right]$. Thus, for $\mu$ fixed and $\sigma \to 0$, we have $g^{\pm} \to \pm \infty$. Furthermore $\Sigma^{++} : = {\rm Var}\left[ \ln f^+(\xi), \ln f^+(\xi) | H^+ \right] = 1/2 $ whereas $\Sigma^{--} : = {\rm Var}\left[ \ln f^-(\xi), \ln f^-(\xi) | H^+ \right] = 4 \mu^2/\sigma^2$. In this case, the dominant terms on the right hand side of Eq.~(\ref{rateSDEa}) for $u^-_t$ come from the input so $\d u^-_t \approx u^-_t \left[ g^- \d t + \d W_t^- \right]$, which will decay on average to an absorbing state at $u^-_t = 0$. As a result, since $u^+_t + u_t- = 1$, we expect $u^+_t$ to saturate to unity ($u^+_t = 1$), when $H_t = H^+$. Note also, it is not essential that the observation distributions $\displaystyle f^{\pm}(\xi)$ be normally distributed, since the key feature of this result is that ${\rm E}\left[ \ln f^{\pm}(\xi)| H^+ \right]  \to 0$, which will occur for any peaked distribution.

\subsection{Performance of the neural population model}
\label{nperform}

\begin{figure}
\begin{center} \includegraphics[width=6cm]{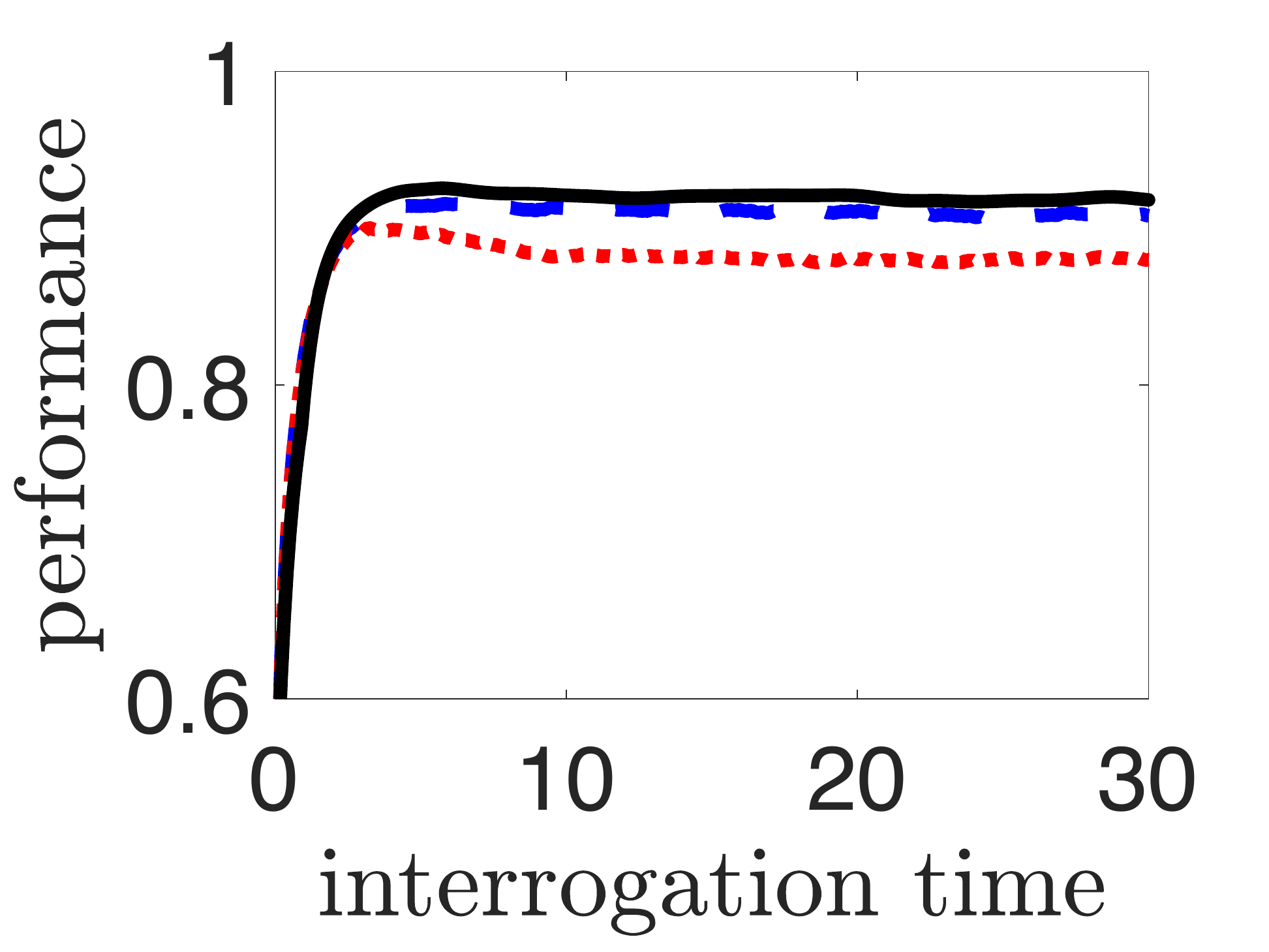} \end{center}
\vspace{-5mm}
\caption{The performance of the neural population model, Eq.~(\ref{plasticnet}), in comparison to the full SDE model, Eq.~(\ref{fulllikeSDE}), and the moment closure approximation, Eq.~(\ref{likeSDE}). Performance is tested under the interrogation paradigm, and defined as the percentage of correct responses at the interrogation time. Here $\ep = 0.05$, and SNR$=1$. The black curve represents the performance of the full SDE model given by Eq.~(\ref{fulllikeSDE}), an ideal observer who takes measurements continuously to infer the change rate. The blue dashed curve represents the performance of the moment closure model, Eq.~(\ref{likeSDE}), where have truncated as shown in Section \ref{momhier}. Lastly, the performance of the neural population model, Eq.~(\ref{plasticnet}), is represented by the red dotted curve. Simulation parameters are given in Section \ref{simappend}.}
\label{fig7}
\end{figure}

In Fig. \ref{fig7}, we compare the performance (percentage of correct responses) of the full SDE model, Eq.~(\ref{fulllikeSDE}), to the moment closure approximation, Eq.~(\ref{likeSDE}), and the neural population model, Eq.~(\ref{plasticnet}), where we have made a weak noise approximation. As in Fig. \ref{fig3}A, we employ the interrogation protocol, where the observer reports their predicted state of the environment at a fixed time. In both the full model (black solid line) and the moment closure model (blue dashed line), performance increases with time. On the other hand, there is a slight decrease in the performance of the neural population model (red dotted line) with time, suggesting that its estimate of the change rate may be corrupted by noise in a way that is not captured by our truncation. Regardless, all three models have relatively similar performance.

\subsection{Neural populations corresponding to log probabilities}
\label{neurlogprob}

In \cite{velizcuba15}, we derived a neural population model for optimal evidence accumulation when the environmental changerate is known. In contrast to our population rate model, this set of equations described neural population rates in terms of the log-probability of an environmental state, rather than the probability. For comparison with \cite{velizcuba15} and other previous neural population models of evidence accumulation in static environments~\citep{bogacz06,mcmillen06}, we map our equations to an equivalent system where the population rates correspond to log-probabilities. To do so, we make the change of variables $u^{\pm}_t = \e^{x^{\pm}_t}$ so that $\d u^{\pm}_t =  \e^{x^{\pm}_t} \d x^{\pm}_t = u^{\pm}_t \d x^{\pm}_t$. As in Section 5.1, It\^{o}'s  change of coordinates rules~\citep{gardiner04} imply our population model is equivalent to:
\begin{align*}
\d x^{\pm}_t &= I^{\pm}_t \d t + \d W^{\pm}_t + \left[ w^{\mp}_t \e^{x^{\mp}_t - x^{\pm}_t} - w^{\pm}_t \right] \d t \\
\d w^{\pm}_t &= \left[ \delta (x^+_t - x^-_t) - w^{\pm}_t \right] \cdot C_t \d t.
\end{align*}
Although the delta distribution $\delta \hspace{-1mm} \left( x^+_t-x^-_t \right)$ should technically be rescaled to account for its composition with $\e^{x}$, we ignore this transformation~\citep{keener88}, since the increment produced from the delta distribution models the changepoint counting process. Thus, each event where $x_t^+=x_t^-$ (equivalently $u^{+}_t = u^-_t$) should be counted the same in the log-probability equations as in the probability equations.

\subsection{Numerical simulations of SDE models}
\label{simappend}

Stochastic differential equation (SDE) models of evidence accumulation in symmetric environments changing between two states are simulated using a standard Euler-Maruyama integration algorithm~\citep{higham01}.  Eq.~(\ref{fulllikeSDE}) describes the evolution of an infinite number of SDEs over the changepoint vector $a \in \mathbb{Z}_{\geq 0}$, state vector $H_t \in \{ H^+, H^- \}$, and time $t \in [0,T]$, so we truncate this space to $a \in \{0,1,2,....,1000\}$, which is sufficient for transition rates $\ep$ and total simulation times $T$ not too large. We compared our results to cases with longer state vectors $a \in \{0,1,2,....,a_{\rm max}\}$ and the changes were negligible. Simulations shown in Fig. \ref{fig5} had a transition rate of $\ep = 0.1$ and total run time of $T = 1000$ with timestep ${\rm dt} = 10^{-3}$. Observations were sampled from a normal distribution $\displaystyle f^{\pm}(\xi) = \frac{\e^{-(\xi \mp \mu)^2/(2 \sigma^2)}}{\sqrt{2 \pi \sigma^2}}$ with mean $\mu = 0.5$ and variance $\sigma^2 = 1$, so the signal-to-noise ratio was $2 \mu / \sigma = 1$.
%\Adcomment{I used a different definition for SNR. I used SNR=(difference between means)/(stdev), we should use the same everywhere. With the definition in this section, what would we do if $\mu^-=0, \mu^+=1$?}
Initial conditions were chosen so that $\PP_0^{\pm} = 0.5$ and $\PP_0(a) = \frac{\alpha^a \e^{- \alpha}}{a!}$ where $\alpha = 1$ and $\beta=5$. A similar approach was used to numerically simulate the neural population model Eq.~(\ref{plasticnet}) and its variants to produce Fig. \ref{fig6} ($\mu = 1$ and $\sigma = 0.1$) and Fig. \ref{fig7} ($\mu = 0.5$ and $\sigma^2 = 1$, for consistency with the full Eq.~(\ref{fulllikeSDE}) and Eq.~(\ref{likeSDE})).

\bibliographystyle{spbasic} 
\bibliography{leaky}

\begin{thebibliography}{45}
\providecommand{\natexlab}[1]{#1}
\providecommand{\url}[1]{{#1}}
\providecommand{\urlprefix}{URL }
\expandafter\ifx\csname urlstyle\endcsname\relax
  \providecommand{\doi}[1]{DOI~\discretionary{}{}{}#1}\else
  \providecommand{\doi}{DOI~\discretionary{}{}{}\begingroup
  \urlstyle{rm}\Url}\fi
\providecommand{\eprint}[2][]{\url{#2}}

\bibitem[{Adams and MacKay(2007)}]{adams07}
Adams RP, MacKay DJ (2007) Bayesian online changepoint detection. arXiv
  preprint arXiv:07103742

\bibitem[{Beck et~al(2008)Beck, Ma, Kiani, Hanks, Churchland, Roitman, Shadlen,
  Latham, and Pouget}]{beck08}
Beck JM, Ma WJ, Kiani R, Hanks T, Churchland AK, Roitman J, Shadlen MN, Latham
  PE, Pouget A (2008) Probabilistic population codes for bayesian decision
  making. Neuron 60(6):1142--1152

\bibitem[{Bogacz et~al(2006)Bogacz, Brown, Moehlis, Holmes, and
  Cohen}]{bogacz06}
Bogacz R, Brown E, Moehlis J, Holmes P, Cohen JD (2006) The physics of optimal
  decision making: a formal analysis of models of performance in
  two-alternative forced-choice tasks. Psychol Rev 113(4):700--65,
  \doi{10.1037/0033-295X.113.4.700}

\bibitem[{Brunton et~al(2013)Brunton, Botvinick, and Brody}]{brunton13}
Brunton BW, Botvinick MM, Brody CD (2013) Rats and humans can optimally
  accumulate evidence for decision-making. Science 340(6128):95--8,
  \doi{10.1126/science.1233912}

\bibitem[{Churchland et~al(2008)Churchland, Kiani, and Shadlen}]{churchland08}
Churchland AK, Kiani R, Shadlen MN (2008) Decision-making with multiple
  alternatives. Nat Neurosci 11(6):693--702, \doi{10.1038/nn.2123}

\bibitem[{Cisek and Pastor-Bernier(2014)}]{Cisek:2014ei}
Cisek P, Pastor-Bernier A (2014) {On the challenges and mechanisms of embodied
  decisions.} Philosophical transactions of the Royal Society of London Series
  B, Biological sciences 369(1655)

\bibitem[{Deneve(2008)}]{deneve08}
Deneve S (2008) Bayesian spiking neurons i: inference. Neural Comput
  20(1):91--117, \doi{10.1162/neco.2008.20.1.91}

\bibitem[{Djuric and Huang(2000)}]{djuric2000}
Djuric PM, Huang Y (2000) Estimation of a bernoulli parameter p from imperfect
  trials. IEEE Signal Processing Letters 7(6):160--163

\bibitem[{Franks et~al(2002)Franks, Pratt, Mallon, Britton, and
  Sumpter}]{franks02}
Franks NR, Pratt SC, Mallon EB, Britton NF, Sumpter DJ (2002) Information flow,
  opinion polling and collective intelligence in house--hunting social insects.
  Philosophical Transactions of the Royal Society of London B: Biological
  Sciences 357(1427):1567--1583

\bibitem[{Gardiner(2004)}]{gardiner04}
Gardiner CW (2004) Handbook of stochastic methods for physics, chemistry, and
  the natural sciences, 3rd edn. Springer-Verlag, Berlin

\bibitem[{Glaze et~al(2015)Glaze, Kable, and Gold}]{glaze15}
Glaze CM, Kable JW, Gold JI (2015) Normative evidence accumulation in
  unpredictable environments. Elife 4:e08,825

\bibitem[{Gold and Shadlen(2007)}]{gold07}
Gold JI, Shadlen MN (2007) The neural basis of decision making. Annu Rev
  Neurosci 30:535--74, \doi{10.1146/annurev.neuro.29.051605.113038}

\bibitem[{G{\"u}tig et~al(2003)G{\"u}tig, Aharonov, Rotter, and
  Sompolinsky}]{gutig2003}
G{\"u}tig R, Aharonov R, Rotter S, Sompolinsky H (2003) Learning input
  correlations through nonlinear temporally asymmetric hebbian plasticity. The
  Journal of neuroscience 23(9):3697--3714

\bibitem[{Higham(2001)}]{higham01}
Higham DJ (2001) An algorithmic introduction to numerical simulation of
  stochastic differential equations. SIAM review 43(3):525--546

\bibitem[{Huk and Shadlen(2005)}]{huk05}
Huk AC, Shadlen MN (2005) Neural activity in macaque parietal cortex reflects
  temporal integration of visual motion signals during perceptual decision
  making. J Neurosci 25(45):10,420--36, \doi{10.1523/JNEUROSCI.4684-04.2005}

\bibitem[{Johnson et~al(2013)Johnson, Blumstein, Fowler, and
  Haselton}]{johnson13}
Johnson DD, Blumstein DT, Fowler JH, Haselton MG (2013) The evolution of error:
  Error management, cognitive constraints, and adaptive decision-making biases.
  Trends in ecology \& evolution 28(8):474--481

\bibitem[{Keener(1988)}]{keener88}
Keener JP (1988) Principles of applied mathematics. Addison-Wesley

\bibitem[{Kira et~al(2015)Kira, Yang, and Shadlen}]{kira15}
Kira S, Yang T, Shadlen MN (2015) A neural implementation of wald's sequential
  probability ratio test. Neuron 85(4):861--873

\bibitem[{Krajbich and Rangel(2011)}]{krajbich11}
Krajbich I, Rangel A (2011) Multialternative drift-diffusion model predicts the
  relationship between visual fixations and choice in value-based decisions.
  Proceedings of the National Academy of Sciences 108(33):13,852--13,857

\bibitem[{Kuehn(2016)}]{kuehn16}
Kuehn C (2016) Moment closure---a brief review. In: Control of Self-Organizing
  Nonlinear Systems, Springer, pp 253--271

\bibitem[{Lange and Dukas(2009)}]{lange09}
Lange A, Dukas R (2009) Bayesian approximations and extensions: optimal
  decisions for small brains and possibly big ones too. Journal of theoretical
  biology 259(3):503--516

\bibitem[{Machens et~al(2005)Machens, Romo, and Brody}]{machens05}
Machens CK, Romo R, Brody CD (2005) Flexible control of mutual inhibition: a
  neural model of two-interval discrimination. Science 307(5712):1121--1124

\bibitem[{McGuire et~al(2014)McGuire, Nassar, Gold, and Kable}]{mcguire14}
McGuire JT, Nassar MR, Gold JI, Kable JW (2014) Functionally dissociable
  influences on learning rate in a dynamic environment. Neuron 84(4):870--881

\bibitem[{McMillen and Holmes(2006)}]{mcmillen06}
McMillen T, Holmes P (2006) The dynamics of choice among multiple alternatives.
  Journal of Mathematical Psychology 50(1):30--57

\bibitem[{Miller(1994)}]{miller1994}
Miller KD (1994) A model for the development of simple cell receptive fields
  and the ordered arrangement of orientation columns through activity-dependent
  competition between on-and off-center inputs. J Neurosci 14:409--441

\bibitem[{Niv et~al(2015)Niv, Daniel, Geana, Gershman, Leong, Radulescu, and
  Wilson}]{Niv:2015}
Niv Y, Daniel R, Geana A, Gershman SJ, Leong YC, Radulescu A, Wilson RC (2015)
  {Reinforcement Learning in Multidimensional Environments Relies on Attention
  Mechanisms}. Journal of Neuroscience 35(21):8145--8157

\bibitem[{Olberg et~al(2000)Olberg, Worthington, and Venator}]{olberg00}
Olberg R, Worthington A, Venator K (2000) Prey pursuit and interception in
  dragonflies. Journal of Comparative Physiology A 186(2):155--162

\bibitem[{Pearson et~al(2011)Pearson, Heilbronner, Barack, Hayden, and
  Platt}]{pearson11}
Pearson JM, Heilbronner SR, Barack DL, Hayden BY, Platt ML (2011) Posterior
  cingulate cortex: adapting behavior to a changing world. Trends in cognitive
  sciences 15(4):143--151

\bibitem[{Petit et~al(2009)Petit, Gautrais, Leca, Theraulaz, and
  Deneubourg}]{petit09}
Petit O, Gautrais J, Leca JB, Theraulaz G, Deneubourg JL (2009) Collective
  decision-making in white-faced capuchin monkeys. Proceedings of the Royal
  Society of London B: Biological Sciences 276(1672):3495--3503

\bibitem[{Pfister and Gerstner(2006)}]{Pfister2006}
Pfister JP, Gerstner W (2006) Triplets of spikes in a model of spike
  timing-dependent plasticity. J Neurosci 26(38):9673--9682

\bibitem[{Portugues and Engert(2009)}]{portugues09}
Portugues R, Engert F (2009) The neural basis of visual behaviors in the larval
  zebrafish. Current opinion in neurobiology 19(6):644--647

\bibitem[{Ratcliff and McKoon(2008)}]{ratcliff08}
Ratcliff R, McKoon G (2008) The diffusion decision model: theory and data for
  two-choice decision tasks. Neural computation 20(4):873--922

\bibitem[{Redner(2001)}]{Redner:2001}
Redner S (2001) A guide to first-passage processes. Cambridge University Press

\bibitem[{Salinas and Abbott(1996)}]{salinas96}
Salinas E, Abbott L (1996) A model of multiplicative neural responses in
  parietal cortex. Proceedings of the national academy of sciences
  93(21):11,956--11,961

\bibitem[{Shvartsman et~al(2015)Shvartsman, Srivastava, and
  Cohen}]{shvartsman15}
Shvartsman M, Srivastava V, Cohen JD (2015) A theory of decision making under
  dynamic context. In: Advances in Neural Information Processing Systems, pp
  2476--2484

\bibitem[{Smith and Ratcliff(2004)}]{smith04}
Smith PL, Ratcliff R (2004) Psychology and neurobiology of simple decisions.
  Trends Neurosci 27(3):161--8, \doi{10.1016/j.tins.2004.01.006}

\bibitem[{Socha(2007)}]{socha07}
Socha L (2007) Linearization methods for stochastic dynamic systems, vol 730.
  Springer Science \& Business Media

\bibitem[{Sugrue et~al(2004)Sugrue, Corrado, and Newsome}]{sugrue04}
Sugrue LP, Corrado GS, Newsome WT (2004) Matching behavior and the
  representation of value in the parietal cortex. science 304(5678):1782--1787

\bibitem[{Veliz-Cuba et~al(2016)Veliz-Cuba, Kilpatrick, and
  Josic}]{velizcuba15}
Veliz-Cuba A, Kilpatrick ZP, Josic K (2016) Stochastic models of evidence
  accumulation in changing environments. SIAM Rev 58:264--289

\bibitem[{Wald and Wolfowitz(1948)}]{wald48}
Wald A, Wolfowitz J (1948) Optimum character of the sequential probability
  ratio test. The Annals of Mathematical Statistics 19(3):326--339

\bibitem[{Whittle(1957)}]{whittle57}
Whittle P (1957) On the use of the normal approximation in the treatment of
  stochastic processes. Journal of the Royal Statistical Society Series B
  (Methodological) pp 268--281

\bibitem[{Wilson and Niv(2011)}]{wilson11}
Wilson RC, Niv Y (2011) Inferring relevance in a changing world. Frontiers in
  human neuroscience 5:189

\bibitem[{Wilson et~al(2010)Wilson, Nassar, and Gold}]{wilson10}
Wilson RC, Nassar MR, Gold JI (2010) Bayesian online learning of the hazard
  rate in change-point problems. Neural Comput 22(9):2452--76

\bibitem[{Wong et~al(2007)Wong, Huk, Shadlen, and Wang}]{wong07}
Wong KF, Huk AC, Shadlen MN, Wang XJ (2007) Neural circuit dynamics underlying
  accumulation of time-varying evidence during perceptual decision making.
  Front Comput Neurosci 1:6, \doi{10.3389/neuro.10.006.2007}

\bibitem[{Wyart et~al(2012)Wyart, De~Gardelle, Scholl, and
  Summerfield}]{wyart12}
Wyart V, De~Gardelle V, Scholl J, Summerfield C (2012) Rhythmic fluctuations in
  evidence accumulation during decision making in the human brain. Neuron
  76(4):847--858

\end{thebibliography}
\end{document}